%% file: SM14.tex
\documentclass[useAMS,usenatbib]{mn2e}

\usepackage[T1]{fontenc}
 \usepackage{pslatex}
\usepackage{amsmath}
\usepackage{amsfonts}
\usepackage{color}
\usepackage{url}
\usepackage{graphicx}
\usepackage{subfigure}
\usepackage{amssymb}
{\newif\ifnotend
\notendtrue
\def\veclist{ABCDEFGHIJKLMNOPQRSTUVWXYZabcdefghijklmnopqrstuvwxyz.}
\def\top#1#2.{#1}
\def\tail#1#2.{#2.}
\loop\expandafter\xdef\csname v\expandafter\top\veclist\endcsname%
{{\noexpand\bf\expandafter\top\veclist}}
\edef\veclist{\expandafter\tail\veclist}
\if\veclist.\notendfalse\fi\ifnotend\repeat}
\mathchardef\mhyphen="2D

\title[Fingerprinting $(l,v)$ distributions]{Recognizing the
  fingerprints of the Galactic bar: a
  quantitative approach to comparing model $(l,v)$ distributions to observations}
\author[Sormani \& Magorrian]{Mattia C. Sormani and John Magorrian \\
Rudolf Peierls Centre for Theoretical Physics, 1 Keble Road, Oxford
OX1 3NP}
\begin{document}

\date{}

\def\p{\partial}
\def\Omegap{\Omega_{\rm p}}

\newcommand{\di}{\mathrm{d}}
\newcommand{\bfx}{\mathbf{x}}
\newcommand{\vlos}{\mathrm{v}_{\rm los}}
\newcommand{\Tspin}{T_{\rm s}}
\newcommand{\Tb}{T_{\rm b}}
\newcommand{\degree}{\ensuremath{^\circ}}
\newcommand{\Th}{T_{\rm h}}
\newcommand{\Tc}{T_{\rm c}}
\newcommand{\bfr}{\mathbf{r}}
\newcommand{\bfv}{\mathbf{v}}
\newcommand{\pc}{{\rm pc}}
\newcommand{\kpc}{{\rm kpc}}
\newcommand{\kms}{{\rm km\, s^{-1}}}
\newcommand{\de}[2]{\frac{\partial #1}{\partial {#2}}}

\maketitle

\begin{abstract}
  We present a new method for fitting simple hydrodynamical models to
  the $(l,v)$ distribution of atomic and molecular gas observed in the
  Milky Way.
  The method works by matching features found in models and
  observations.
  It is based on the assumption that the large-scale features seen in
  $(l,v)$ plots, such as ridgelines and the terminal velocity curve,
  are influenced primarily by the underlying large-scale Galactic
  potential and are only weakly dependent on local ISM heating and
  cooling processes.
  In our scheme one first identifies by hand the features in the
  observations: this only has to be done once.
  We describe a procedure for automatically extracting similar features
  from simple hydrodynamical models and quantifying the ``distance''
  between each model's features and the observations.
  Application to models of the Galactic Bar region ($|l|<30^\circ$)
  shows that our feature-fitting method performs better than $\chi^2$
  or envelope distances at identifying the correct underlying galaxy
  model.

\end{abstract}

\section{Introduction}
\input{Introduction.tex}
%
\section{Observations} \label{sec:observations}
\input{Observations.tex}
%
\section{Models} \label{sec:models}
\input{Models.tex}
%
\section{Comparing Models and Observations} \label{sec:compare}
\input{Comparing.tex}

\section{Tests with Mock Data} \label{sec:tests}
\input{Tests.tex}

\section{Application to Real Data} \label{sec:realdata}
\input{RealData.tex}

\section{Discussion}\label{sec:discussion}
\input{discussion.tex}
\section{Conclusion} \label{sec:conclusion}
\input{conclusion.tex}

%
%
%
\def\aap{A\&A}\def\aj{AJ}\def\apj{ApJ}\def\mnras{MNRAS}\def\araa{ARA\&A}\def\aapr{Astronomy \&
  Astrophysics Review}\def\apjs{ApJS}
\bibliographystyle{mn2e}
\bibliography{2d}

%
\appendix
\section{A Potential} \label{appendix1}

Here are the details of the potential used in the generation of the mock
data in Section~\ref{sec:humanintuition}.
The potential is inspired by \cite{DehnenBinney1998} models.
It is made by 3 components: Bar, Disk, Halo. Table \ref{table:1} shows the value of the parameters used.
The potential is steady in a frame that rotates pattern speed $\Omega= 48 \, \kms \kpc^{-1}$.

\subparagraph{\textbf{Bar}}
The density distribution generating the potential of the bar is given by
\begin{equation}\rho(a) = \rho_0 \left( \frac{a}{a_0}\right)^{-\alpha} \exp\left(- a^2/a_0^2\right) \end{equation}
where  $$a = \sqrt{x^2 + (y^2 + z^2)/q^2}.$$ 
To fully specify the bar potential we therefore need 4 parameters: the central concentration $\rho_0$, inner slope $\alpha$, major axis $a_0$ and axis ratio $q$.
Equivalently, we can specify the total mass $M$ instead of the central concentration $\rho_0$.

\subparagraph{\textbf{Disk}} The density distribution of the disk is exponential. It has zero thickness, and the surface mass density is given by
\begin{equation} \Sigma(R) = \Sigma_0 e^{-R/R_{\rm d}} \end{equation} 
To fully specify the disk potential we need 2 parameters: the radius $R_{\rm d}$ and the central surface mass density $\Sigma_0$. 

\subparagraph{\textbf{Halo}} The Halo potential is logarithmic. 
\begin{equation} \Phi_{\rm halo}(r) = \frac{1}{2} v_0^2 \log(r_{\rm h}^2 + r^2 ) \end{equation}
To fully specify the halo potential we need 2 parameters: the radius $r_{\rm h}$ and the circular velocity at infinity $v_0$.

\begin{table}
\begin{tabular}{| p{0.7cm} | p{2.2cm} | p{1cm} | p{1.9cm} | p{1cm} | p{1cm} |}
Bar 		& $M = 0.5 \times 10^{10}M_\odot$		& $\alpha = 1.8$ 							& $a_0 = 2.5 \, \kpc$ 		& $q = 0.4$ \\ [1.0ex]
Disk 		& \multicolumn{2}{| l |}{$\Sigma_0 = 0.07 \times 10^{10} M_\odot \kpc^{-2}$} 				& $R_{\rm d} = 2.5\, \kpc$ 	\\ [1.0ex]
Halo 	& $v_0 = 185\, \kms$				&										& $r_{h} = 5.0 \, \kpc$ \\ [1.0ex]
Omega	& \multicolumn{2}{| l |}{$\Omega_p = 48\, \kms \kpc^{-1}$}	 \\
Phi		& 35 \degree						
\end{tabular}
\caption{The parameters for the model used for the test in Sect. \ref{sec:humanintuition}.}
\label{table:1}
\end{table}

\section{Earth Mover Distance} \label{EMD}

\label{EMD}
The Earth-mover distance is
a way of quantifying the dissimilarity of two
distributions. Intuitively, given two distributions, one can be seen as
a collection of piles of earth spread in space, the other as a
collection of of holes in the same space. The amount of earth at each
point can be any positive real number.
The EMD measures the minimal amount of work needed to fill in the holes with earth taken
from the piles.  
A unit of work corresponds to transporting a unit of
earth by a unit of ground distance, which in our case would be a
metric suitably defined in the $(l,v)$ plane.
The earth contained in one pile can be shared among
many different holes if this solution requires less work than
other alternatives.
More details on this
distance can be found for example in \cite{EMD}.

As noted in Section \ref{calcdist}, EMD is an option that we initially
found intuitively appealing for comparing features.  In this case, the
idea is to apply the EMD to binary images such as panel (h) in
Fig. \ref{fig:comparison1}.  The amount of earth is 1 at pixels
corresponding to features, and zero otherwise. The dissimilarity
between model features and data features is quantified by the minimal
amount of work needed to turn the model features into data features
(or vice versa). When used in this way, the EMD turned out to
underperform the much simpler SMHD; if anything, the EMD was actually too
clever in matching features from one image to the other, with the
result that the variation of EMD with $\phi$ and~$\Omega$ had
spikes and false minima, and was much noisier than either the SMHD or
envelope distances.

Nevertheless, if one were faced with carrying out full
radiative-transfer modelling, the EMD used in a qualitatively
different way might be reconsidered as an alternative to $\chi^2$.
The idea in this case would be to use the Earth Mover Distance to
compare two full $(l,v)$ distributions, such as panel (a) in
Fig. \ref{fig:comparison1}.  The amount of earth would then be the
brightness temperature at each pixel, which would {\it not}
constrained to be either 0 or 1.  The potential advantage of EMD used
in this way would be that it avoids one of $\chi^2$ main problems,
namely that of ignoring cross-bin information.  We therefore suspect
that it might be useful in cases where one were trying to fit the
details of the chemistry and radiative transfer as well as the
potential.  In exploratory tests we found that EMD applied to the full
distributions performed well in retrieving parameters when the
rule~\eqref{eq:rhogrid} used to project the models was identical to
that used to generate the mock data.  As the EMD used in this way is
strongly dependent on the intensities at each point, it {\it requires}
that a full radiative-transfer model be included, however.

\end{document}

%% file: Introduction.tex
The Sun is located in the periphery of the Galactic Disk,
approximately 8 kpc away from the Galactic Centre (GC).
Observations towards the GC are obscured by dust at many wavelengths,
and, because of our position within the disk, it is complicated to unravel
what our Galaxy would look like if seen face-on. 
The structure and morphology of the Central Disk -- the region inside
Galactocentric Radius $r \simeq 3 \, \kpc$ -- is particularly complex,
but there is now a solid body of evidence that it contains a bar
(see \citealt{AthanassoulaBarReview, FuxBarReview, MerrifieldBarReview,
  GerhardBarReview} for reviews).

The gas kinematics in the Central Disk is mainly observed in spectral
lines of HI \citep{HIdata}, ${}^{12}$CO \citep{COdata,Sawada2001},
${}^{13}$CO and CS \citep{Bally1987} and most commonly represented
through longitude-velocity $(l,v)$ plots.  It has long been known
\citep[see for example][]{rougooroort1} that these plots have some
features that cannot be explained under the assumption of gas moving
in circular orbits in an axisymmetric potential, the most obvious
being the emission in the regions $(l >0, v < 0)$ and $(l<0, v >0)$
close to the GC, which are forbidden to pure circular motion. It is
now considered most likely that $(l,v)$ plots can be understood in
terms of gas flow driven by a bar, an hypothesis first suggested by
\cite{devaucouleurs1964}.  This hypothesis has received strong
independent confirmation by photometric evidence for a Galactic bar
\citep{Blitz1991,Dwek1995,Binney1997}, while other hypotheses put
forward in the early days, for example involving explosive phenomena
\citep{oort1977}, are now considered very unlikely.

There has been a long tradition of work trying to understand $(l,v)$
plots in terms of a non-axisymmetric gas flow driven by a bar.  The
first models were purely kinematical \citep[see for
example][]{peters1975, lisztburton1980}, in that they assumed that gas
follows closed streamlines, but without a physical model for the
origin of the assumed streamlines.  These were followed by ballistic
models, in which the gas streamlines were approximated by closed
orbits in an assumed underlying potential
\citep{gerhardvietri,binneyetal1991}.  \cite{binneyetal1991} convinced
the community that our Galaxy contains a bar arguing only on kinematic
grounds.

More recently, the interpretation of $(l,v)$ plots has relied on full
hydrodynamical calculations.
These involve finding quasi-steady solutions and/or running
hydrodynamical simulations, usually under the assumption that the
dynamics of the gas is governed by Euler's equation, complemented by the
equation of state of an ideal isothermal gas \cite[for a discussion of
this approximation see for example][]{ShuBookVol2}.
Many simulations have been carried out with a variety of methods:
sticky-particle codes \citep{jenkinsbinney,combesrodriguez2008},
Eulerian grid-based codes \citep{weinersellwood1999} and 
smoothed particle hydrodynamic simulations (SPH) \citep{leeetal1999,
  englmaiergerhard1999, bissantzetal2003}
in externally imposed potentials undergoing rigid rotation;
SPH simulations coupled to
self-consistent 3D N-body barred models of the Galaxy \citep{fux1999};
SPH codes that include
phenomenological terms to model heating and cooling processes such as
radiative cooling, heating caused by UV radiation, star formation and
SN feedback \citep{babaetal2010,Pettitt2014}.  \cite{mulderliem} found numerically a
quasi-steady solution of gas-dynamical equations in a given barred
gravitational potential.

An important limitation of all the works cited above is the lack of a
fully satisfactory way of comparing the models' predicted $(l,v)$
distributions against the observed ones.  The ballistic models tried
only to superimpose the projected traces of orbits to ridges and edges
identified by eye in the observed $(l,v)$ distribution, without any
consideration about the intensity produced by such traces.  Authors
who studied hydrodynamical simulations have mostly tried to take
snapshots of the gas density projected to the $(l,v)$ plane and then
qualitatively tried to identify by visual inspection counterparts of features found in observations. 
Some works \citep{fux1999} have given a very
detailed and coherent interpretation of many features.  Others have
carried out a quantitative comparison based only the envelope of the
$(l,v)$ distribution \citep{weinersellwood1999,englmaiergerhard1999},
ignoring the extra information contained in the internal structure of
the data.  More recently, \cite{Pettitt2014} have proposed a fit
statistic akin to $\chi^2$ that makes use of the full $(l,v)$
distribution.  Although this is clearly the correct approach in
principle, in that it makes full use of the available data, we believe
that it would be prohibitively expensive to use such a method to
constrain Galactic potential and ISM structure simultaneously.  First,
the need to include ISM cooling and chemistry and full radiative
transfer calculations, in addition to modelling the gas dynamics,
makes it very expensive computationally -- we note that
\cite{Pettitt2014} only applied their fit statistic to models using a
simplified version of the radiative transfer calculations.  Second,
when one is unsure about the Galactic potential, their method shares
with $\chi^2$ some serious drawbacks that we discuss in detail in
Section \ref{sec:tests}.

In this paper we argue that the problem of interpreting $(l,v)$ plots
is best split into two steps: one should first constrain the gross
distribution of gas and the overall Galactic potential by fitting
simple dynamical models to ``features'' in the observed $(l,v)$
distribution; then, once this gross structure has been found, the
model can be refined by including more detailed treatments of gas
chemistry, radiative transfer and so on, along the lines proposed by
\cite{Pettitt2014}.  We focus on the first step: the problem of
fitting features.  We present a new method that can be seen as an
automated way of performing the task that has been previously done by
visual inspection: comparing broad scale features in synthetic $(l,v)$
plots against those in the observed $(l,v)$ plots.  In our method,
features in the models are identified automatically by a computer,
while features in observations are identified by the astronomer, but
only once.  The method returns a single number that measures the
dissimilarity between a synthetic $(l,v)$ plot and the observations.
The approach takes inspiration from human face- and
fingerprint-recognition algorithms.

The paper is organised as follows.
In Section~\ref{sec:observations} we review the observations and
explain what we mean by a ``feature'' in the $(l,v)$ distribution.
In Section~\ref{sec:models} we describe our simple hydrodynamical
models and show how they reproduce the same kinds of features.
Section~\ref{sec:compare} is the core of the paper, where we describe
in detail our method for defining a ``distance'' between observed
and model features.
Section~\ref{sec:tests} presents a range of tests on mock data to
assess the performance of the method, including how it compares to
other methods, such as envolope fitting or~$\chi^2$. 
In Section~\ref{sec:realdata} we show an example of application to
real data, before summing up and discussing future plans in
Sections \ref{sec:discussion} and~\ref{sec:conclusion}.
%

%% file: Observations.tex
The $(l,v)$ distributions of HI and CO spectral lines
(Fig. \ref{fig:obs1}, panels (a) and (b), from data in \citealt{HIdata,COdata})
contain an incredibly rich and diverse amount of information.
They exhibit clumpiness and complicated structure on small scales, but
also coherent, broad features on large scales. 
In the present section we review the observational data, including the
various large-scale features that have been identified in the literature.
The most obvious such feature is the envelope of the emission.
Internal features consist mostly of bright ridges, some of which
are thought to correspond to spiral arms in the Galaxy.
Our fitting method requires that features in the observations (but not in the models) 
are identified by a human being and not by a computer. 
The advantage of this subjective identification is that it allows to include our own insight when identifying features; 
this can be tested by selectively omitting features or adding new ones.

\subsection{Envelope}
To extract the envelope of the $(l,v)$ plot we use the HI brightness
temperature measured by \cite{HIdata}, which has a spatial resolution of 
$0.5\degree$.  We average over $|b|\leq4^\circ$
and convolve with a Gaussian of $\sigma=5\,\kms$ in velocity to obtain
the smoothed temperature map, $T_B(l,v)$, shown in Fig.~\ref{fig:obs1}(b).
For each $l$ in the range $-5\degree \leq l \leq 30 \degree$ we determine
the upper envelope, $v^D_+(l)$, (i.e., the most positive line-of-sight
velocity for each~$l$) by examining $T_B(v)\equiv T_B(l,v)$ as follows.
\begin{enumerate}
\item Given~$l$, find the velocities $v$ at which $\partial T_B(v)
  / \partial v$ peaks.
  This amounts to finding points of high gradient, which is a general definition of an edge.
  To avoid fitting noise, discard any edge that has
  $T_B(v)<0.125\,\rm K$: this is slightly larger than the RMS
  noise in $T_B$ quoted by \cite{HIdata}.
  Let $v_{\rm pk}$ be the location of the highest velocity edge
  that remains for this~$l$.
\item Following \cite{Shane1966}, assign
\begin{equation} v^D_+ = v_{\rm pk} + \frac{1}{T_B(v_{\rm
      pk})}\int_{v_{\rm pk}}^\infty T_B(v)
  \, \di v.
\end{equation}
As this makes use of the full profile $T_B(v)$, it has the advantage
over other methods of being robust against systematic effects due to
noise in~$T_B(v)$.
\end{enumerate}
For $l<0$, one expects that most of the $v<0$ emission will be caused
by foreground material well outside the bar (see, e.g., Figure~9.3 of
\cite{BM}).  As our interest in the present paper is
restricted to the bar region, after visual inspection of the CO data,
we define the envelope to fall linearly from its value found using the
method above at $l = -5\degree$ to zero at $l=-12\degree$.

The lower envelope (the most negative line-of-sight velocities),
$v^D_-(l)$, is obtained in the same way, but reflecting $l\to -l$ and
$v\to -v$.
We make an exception for the five points
$l=\{0.5,\degree,1.0\degree,1.5\degree,2\degree,2.5\degree\}$, which
we corrected manually to account for extra absorption features.
The final result, superimposed on HI and CO data, is shown in
Fig. \ref{fig:obs1} panels (d) and (c).

\subsection{Internal Features} 
\begin{figure*}
    \subfigure[]
    {
        \includegraphics[width=0.48\textwidth]{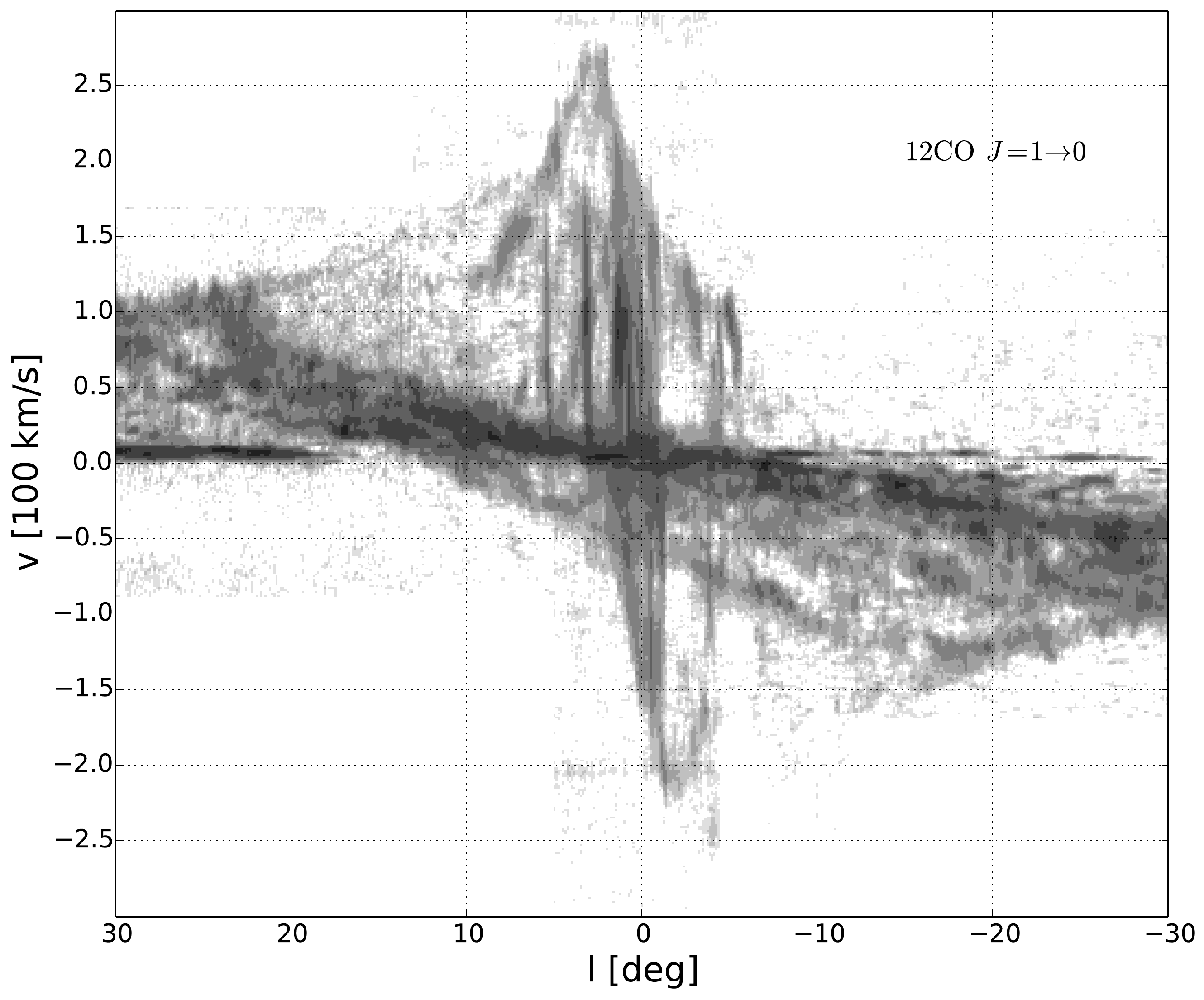}
        \label{sub:c1}
    }
    \subfigure[]
    {
        \includegraphics[width=0.48\textwidth]{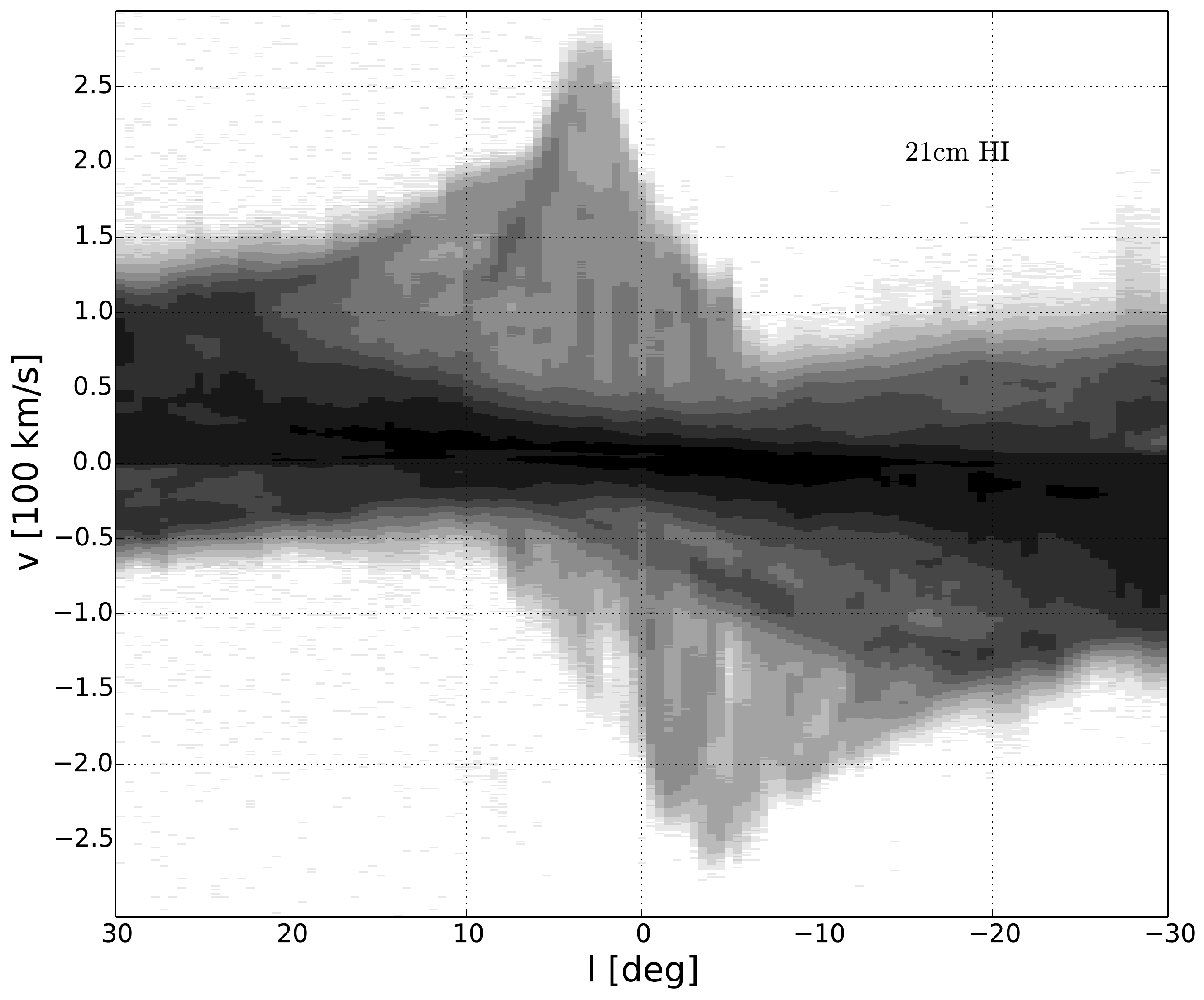}
        \label{sub:c2}
    }    
    \\
        \subfigure[]
    {
        \includegraphics[width=0.48\textwidth]{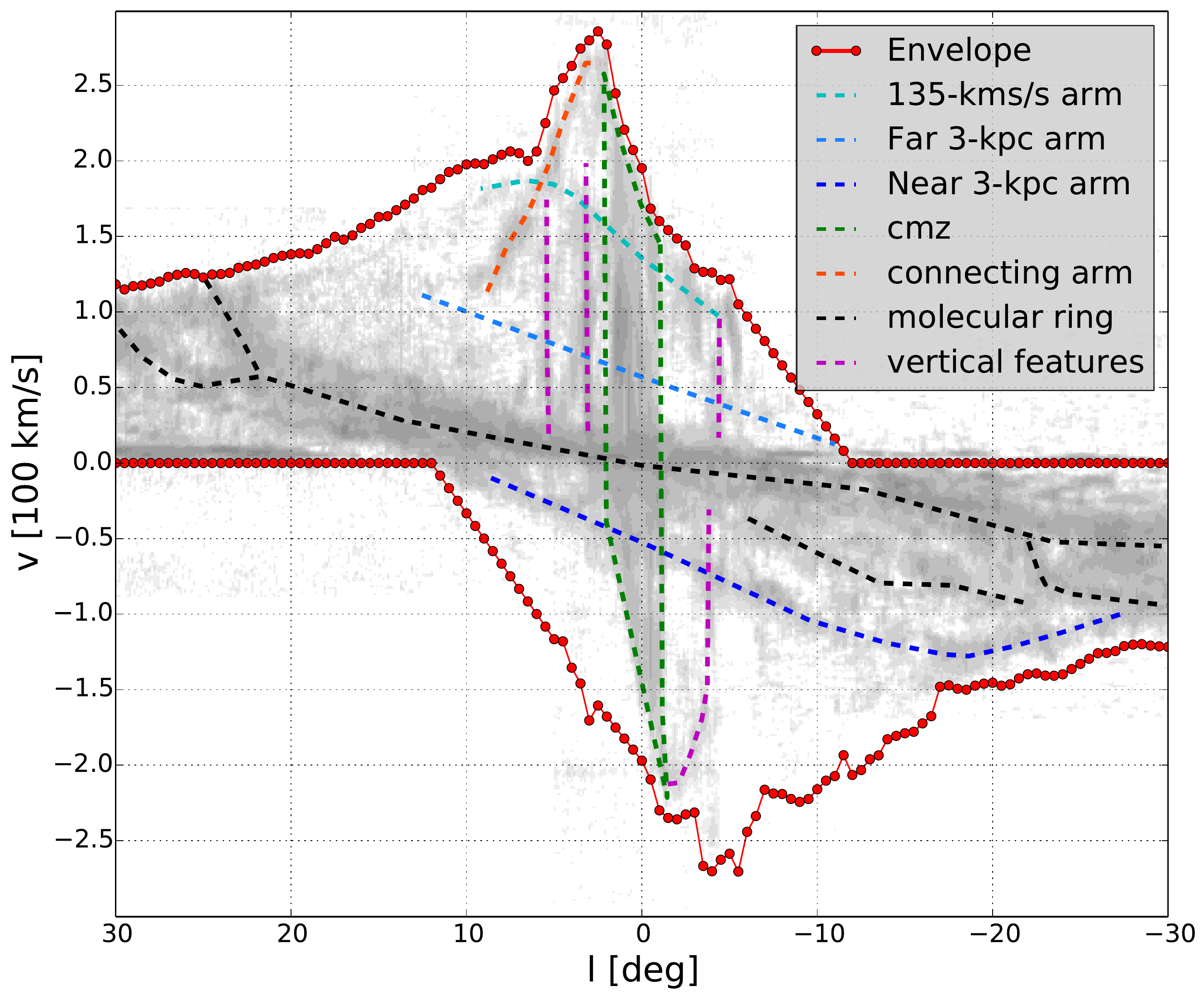}
        \label{sub:c5}
    }
    \subfigure[]
    {
        \includegraphics[width=0.48\textwidth]{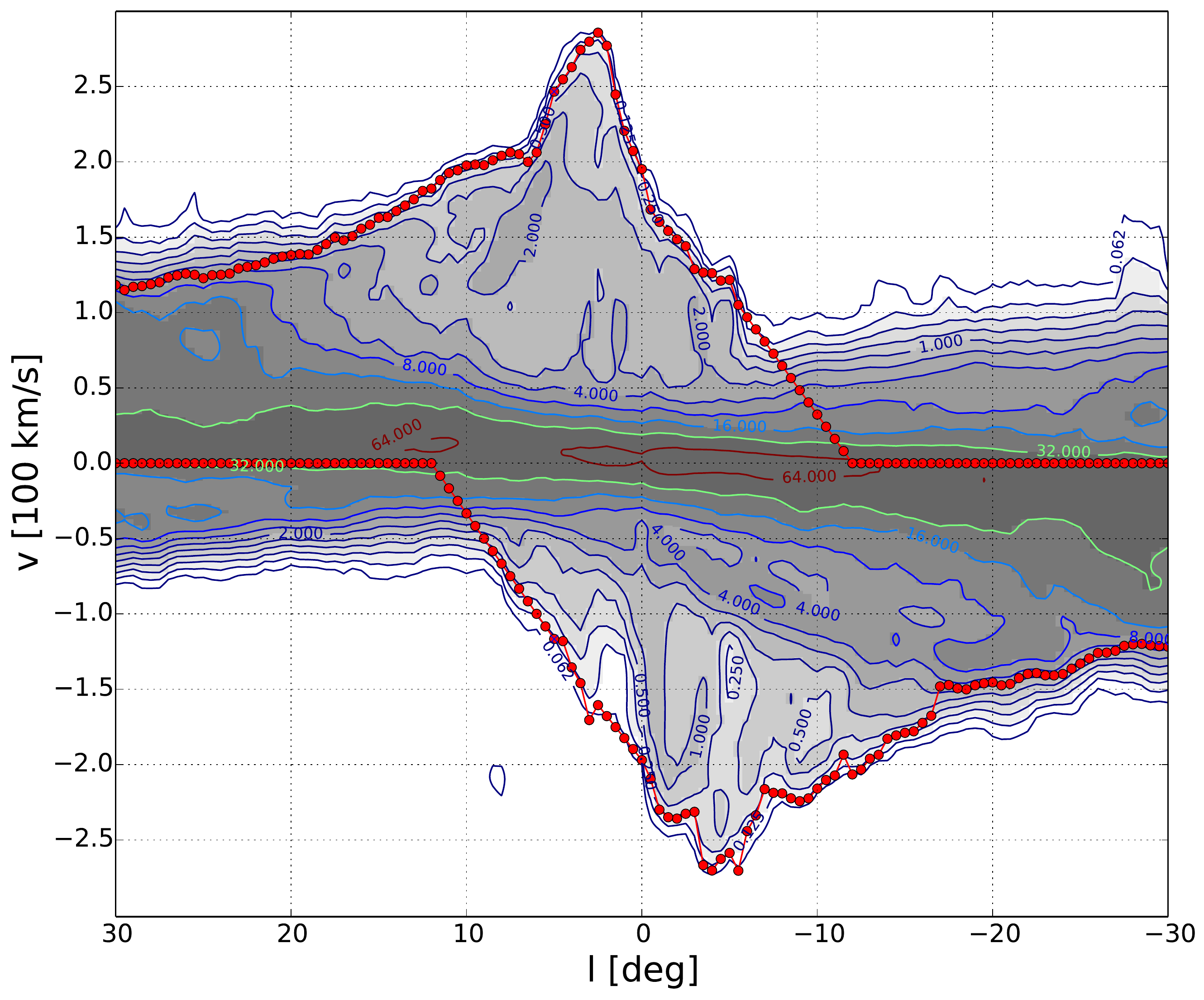}
        \label{sub:c4}
    }
    \\
    \subfigure[]
    {
        \includegraphics[width=0.48\textwidth]{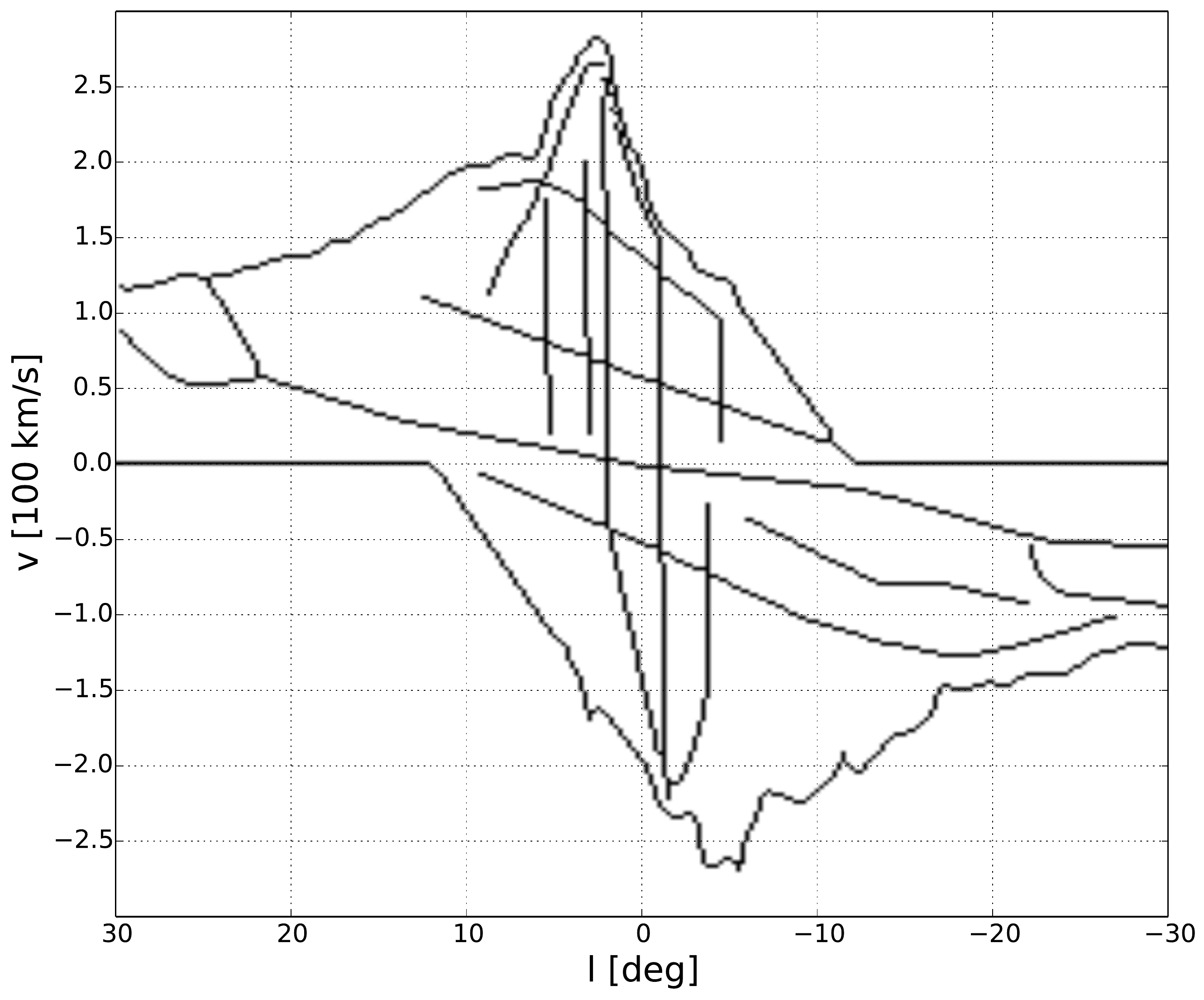}
        \label{sub:c6}
    }
    \caption{(a) CO observations integrated over $|b| \leq 4
      \degree$. (b) HI observations integrated over $|b| \leq 4
      \degree$. (c) CO observations with envelope (determined from HI data) and internal
      features superimposed. (d) HI observations with superimposed
      envelope and brightness temperature contours indicated. Contours are spaced by factors of 2 in $T_B$. (e) all
      the features used in comparison with models in the final format. This is the data input for the SMHD.}
\label{fig:obs1}
\end{figure*}

Many internal features can be identified in the $(l,v)$
diagrams in the region $| l | < 30 \degree$.
Most of them consist of bright ridges and often bear a name that can
be confusing, as it refers to an original, now discredited, interpretation which may
no longer be accurate.
Lists of features can be found for example in
\cite{Rougoor1964,Kruit1970,Cohen1975,Bania1977,Bally1988,fux1999}.
Here, our goal is to focus on those that are most likely to trace the
large-scale gravitational potential of the Galaxy, avoiding any
that are most likely due to stellar feedback processes.
Possible candidates, shown in Fig. \ref{fig:obs1}, are the following
features:
\begin{itemize}
\item[-] \textit{3-kpc arm}. This is the most apparent and coherent
  feature. It is a bright ridge that can be traced over a large range
  in longitude and and crosses $l=0$ with a velocity of $-53\,
  \kms$. Absorption against radio continuum emission from the Galactic
  centre shows that it lies in front of the Galactic centre. It is
  probably associated with a spiral arm.
\item[-] \textit{Far side 3-kpc arm}. This recently discovered feature
  \citep{Dame2008} lies beyond the Galactic centre and is thought to
  be the far-side counterpart of the 3-kpc arm, a role that was
  sometimes previously assigned to the $135\,\kms$ arm.  It can be
  seen clearly only in sufficiently high-resolution data, so it is not
  identifiable in our figures.
\item[-] \textit{$135\,\kms$ arm}. This is a high-velocity arm that
  crosses $l=0$ at a velocity of $135 \, \kms$.  As it is not seen in
  absorption against Sgr A, it is most probably caused by gas that
  lies beyond the Galactic centre \citep[e.g.,][]{Cohen1975}.
\item[-] \textit{Connecting arm}. This very bright feature lies at
  very high velocity and touches the positive-velocity peak emission.
  There is no unanimous consensus on whether it corresponds to a real
  arm.
  Most often it is associated with a dust lane
  \citep{Marshall2009,Liszt2008}.
  An alternative interpretation considers it the edge of the nuclear
  $x_2$ ring \citep{lisztburton1980}.
  In the interpretation of \cite{Marshall2009} it lies in front of the
  Galactic Center.
\item[-] \textit{Central Molecular Zone}. This refers to the off-centered
  concentration of dense molecular gas in the region $-1\degree
  \lesssim l \lesssim 1.5\degree$. The parallelogram bounding this
  structure has received particular attention
  \citep{binneyetal1991}.
\item[-] \textit{Molecular Ring}. Despite being one of the most
  prominent structures in the CO data, its structure it is still
  debated. It could be either a ring-like structure, possibly
  associated with a resonance, or an intertwined structure of spiral
  arms. Sub-branches and bifurcations can be observed near the edges.
  
\item[-] \textit{Vertical Features}. We use this umbrella term to
  denote various features that span a large velocity range while being
  confined in a small longitude range. The vertical feature at
  $l\simeq 3\degree$ is called \textit{Clump2} after
  \cite{StarkBania1986}. These authors related it with an inner dust
  lane or spiral arm. \cite{Marshall2009} interpreted  the vertical features at negative longitudes as the far-side counterpart of the Connecting
  arm. Some of these vertical features might be due to magneto-hydrodynamic effects \citep{Fukui2009}.
  
\end{itemize}
Most interesting, \textit{all} the features listed above can be
identified both in both HI and CO data, though some of them (Central
Molecular Zone and Bania Clump2) are notably fainter in HI. All of
them are candidates for tracing the large-scale dynamics of the Milky
Way.
Some of them, like the Molecular Ring, the 3-kpc arm and its far side
counterpart, are most likely to be independent of small scale
physics, whereas others are more dubious, such as those that have been associated
with dust lanes like the Connecting Arm and the Vertical Features.

%% file: Models.tex
In this section we explain how we obtain synthetic $(l,v)$
distributions given a model for the Galactic potential.  We split the
process into two steps:
\begin{enumerate}
\item Construct snapshots of the density and velocity distribution of
  the gas (i.e., obtain $\rho(\bfx)$ and $\bfv(\bfx)$ for each point
  in the Galaxy).
\item Project the density and velocity distribution onto $(l,v)$ space
\end{enumerate}
Sect \ref{sec:hydrodescription} and Sect. \ref{sec:projection} deal respectively with these two steps. 

\subsection{Hydro Simulation Scheme}
\label{sec:hydrodescription}
We assume that the gas is a fluid governed by the Euler equation
complemented by the equation of state of a isothermal ideal gas.
Then we run 2D hydrodynamical simulations in an externally imposed
rigidly rotating barred potential.
The output of each simulation is snapshots of velocity and density
distributions $\rho(\bfx)$ and $\bfv(\bfx)$ at chosen times.

We use a grid-based, Eulerian code based on the second-order
flux-splitting scheme developed by \cite{vanAlbada+82} and later used
by \cite{Athan92b} and \cite{weinersellwood1999} to study gas dynamics
in bar potentials.
The sound velocity was chosen to be $c_s =10 \, \kms$; we verified for some sample simulations 
that the results are quite insensitive to the exact value in the range
$c_s = 5 \, \mhyphen 15 \, \kms$.
We used a $400\times400$ grid, with cells of side $50 \, {\rm pc}$. Thus the
total simulated area is a square of $20 \, \kpc$ side.

In each run the initial conditions are as follows.
We start with an axisymmetrized bar and, to avoid transients, turn on
the non-axisymmetric part of the potential gradually during the first
150 Myr, in such a way that the total mass of the Galaxy is conserved
in the process.
The gas starts with a uniform density profile and moves on circular
orbits at the local circular velocity.
For the models presented here, we verified that the gas reaches an
approximate steady-state, although this is not a requirement for the
application of our comparison method (section~\ref{sec:compare}
below).
We do not include any recycling or star-formation laws for the gas in
high-density regions.
As a consequence, in the central region our simulations reach very
high densities that should be taken as upper bounds on real
densities.\footnote{To test the importance of this we have run some
  models that do include a simple gas recycling law.
  The central spike vanishes in these models, but the rest of the
  features are unchanged.}
The boundary conditions are such that the gas can freely escape the
simulated region, after which it is lost forever.
The potential well is sufficiently deep that very little gas escapes
from our simulation box.

\begin{figure}
\centering
        \subfigure[]
    {
	\includegraphics[width=0.42\textwidth]{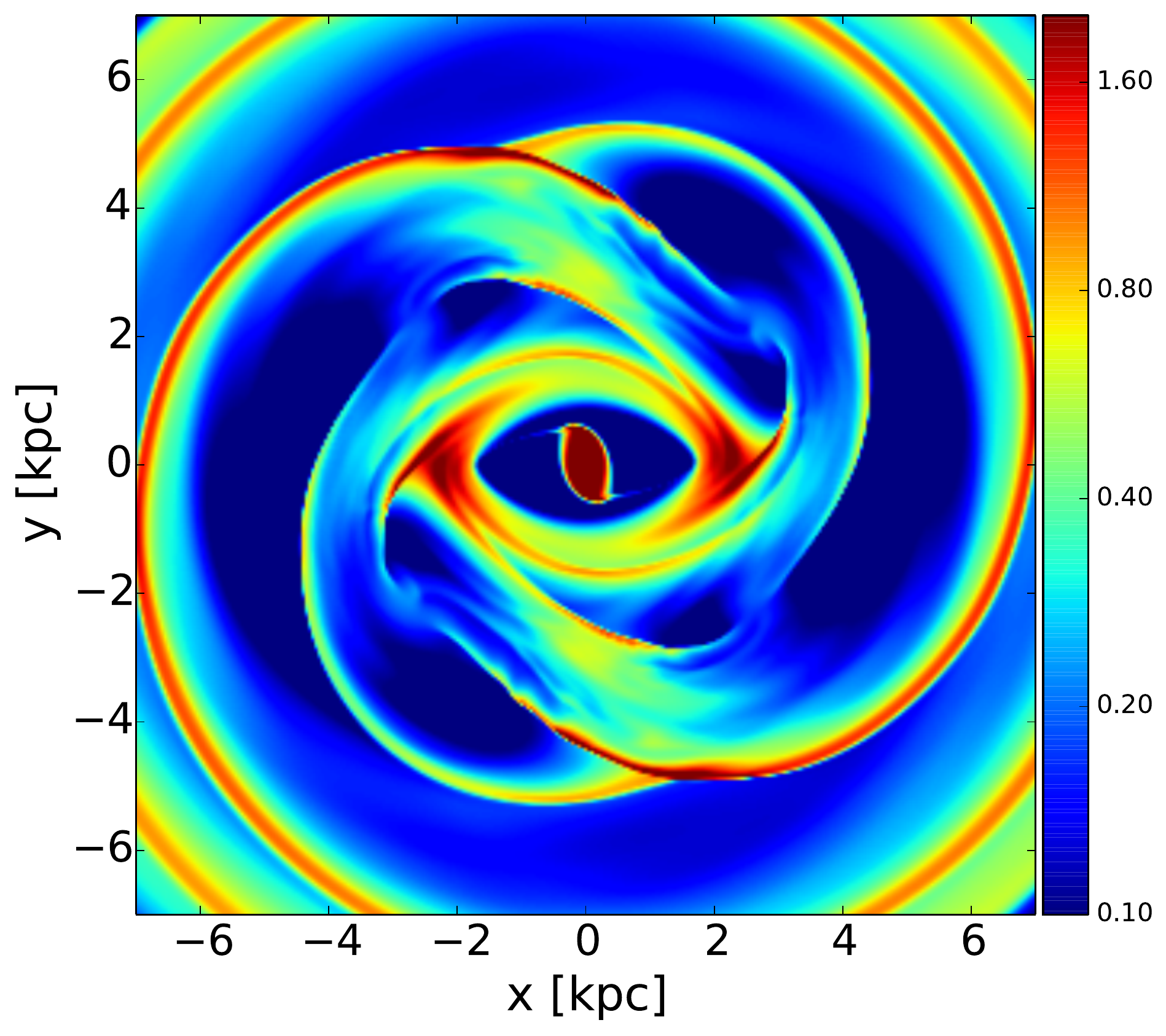} \label{fig:models2a}
    }
        \subfigure[]
    {
	\includegraphics[width=0.42\textwidth]{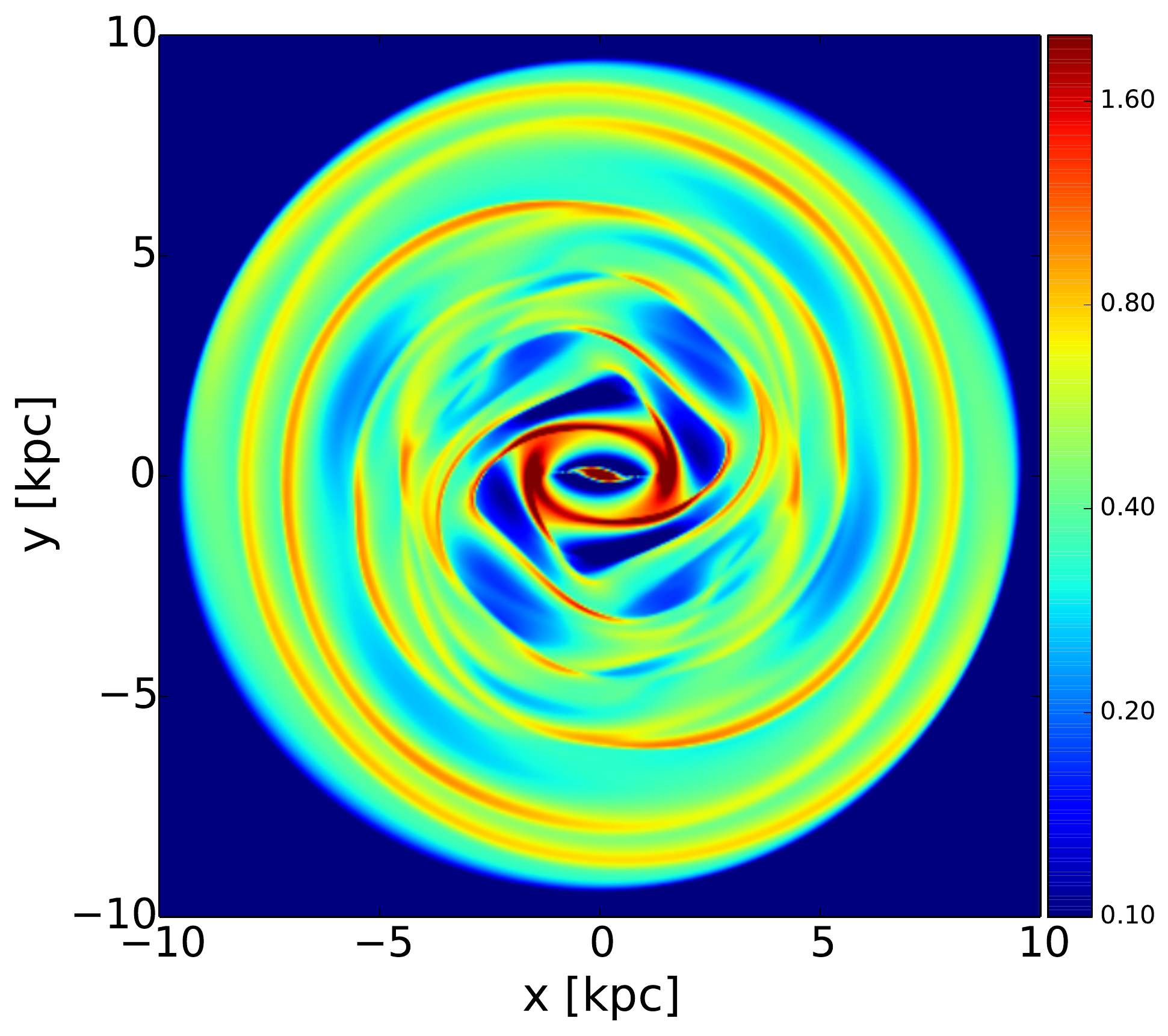}
    }
        \subfigure[]
    {
	\includegraphics[width=0.42\textwidth]{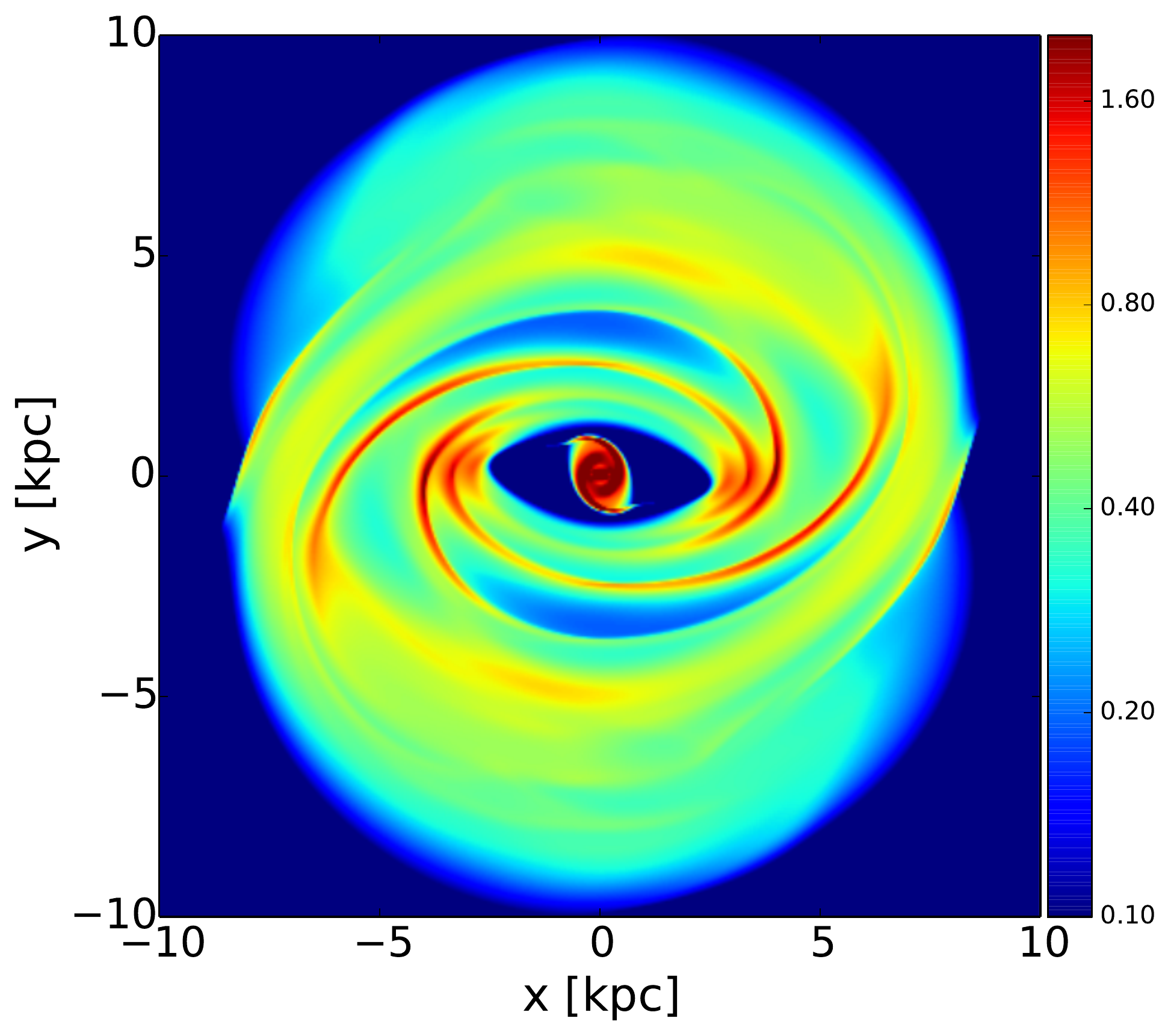}
    }
    \caption{Gas densities produced by our simulation scheme for some
      models from the literature.  In all panels the $x$-axis is
      coincident with the major axis of the
      bar. Density is in arbitrary units, and the initial density is uniform with value 1. (a) Englmaier \& Gerhard's (1999) standard
      potential with $\Omega_p=55 \, \kms \kpc^{-1}$ at taken
      evolutionary time $t=367 \rm Myr$.  Compare with Fig. 9 in their
      paper.  (b) Bissantz et al.'s (2003) standard potential at
      evolutionary time $t=310\rm Myr$. The bar pattern speed is
      $\Omega_p=58.6 \, \kms$. A spiral component with pattern speed
      $\Omega_{\rm spiral}=19.6 \, \kms \kpc^{-1}$ is also included in
      the potential. Compare with their Fig.~12. (c) The potential of
      Rodriguez-Fernandez \& Combes (2008) with $\Omega_p=30 \, \kms
      \kpc^{-1}$ at evolutionary time $t=367 \rm Myr$. Compare with
      their Fig. 8, second row.  }
    \label{fig:models2}
\end{figure}

For much of this paper we will use our reconstruction of the
``standard" potential of \cite{englmaiergerhard1999} from the
multipole moments they plot in their Figure~3; our reconstruction 
reproduces their rotation curve and effective
potential (their Figures 4 and~5, respectively) correctly.
To allow comparison with their results, in Fig.~\ref{fig:models2},
panel (a), we show the gas density obtained by our simulations in this
reconstructed potential for the bar pattern speed of
$\Omega_p=55\,\kms\,\kpc^{-1}$ that they assumed in their Fig.~9:
there is a very good match between the density produced by our
Eulerian simulations with that of their SPH scheme.
Panels (b) and (c) of Fig. \ref{fig:models2} show our reconstruction
of of two other models from the literature, \cite{bissantzetal2003}
and \cite{combesrodriguez2008}.
We discuss these models in Section~\ref{sec:realdata}.

\subsection{Calculating model $(l,v)$ distributions}
\label{sec:projection}
Throughout this paper, we follow the earler modelling work above in
assuming that the Sun is undergoing circular motion at a radius
$R_0=8\, \kpc$ with speed $v_\odot = 220\, \kms$.
The major axis of the Bar always lies along the $x$ axis in our
models.
Calling $\phi$ the angle between the major axis and the Sun--GC line,
the cartesian coordinates of the Sun are given by $x_\odot = R_0 \cos
\phi$, $y_\odot = R_0 \sin \phi$.

We adopt a very simple binning procedure to produce the predicted
$(l,v)$ distributions corresponding to each simulation snapshot
($\rho(\bfx),\bfv(\bfx)$).
We use an $(l,v)$ grid with spacing $\Delta l=0.25\degree$, $\Delta
v=2.5\kms$, whereas the simulations use a $50\,\pc\times50\pc$ grid.

First, we use linear interpolation to resample the
simulation's $\rho(\bfx)$ and $\bfv(\bfx)$ grids down to finer
$6.25\,\rm pc$ grids.  Then, given an assumed position and velocity of
the Sun, we calculate the Galactic longitude $l_i$ and line-of-sight velocity
$v_i$ corresponding to each resampled grid point and bin the resulting
$(l_i,v_i)$ onto the
$(l,v)$ grid 
with weight
\begin{equation}
w_i = \frac{ \rho_i^{\alpha} }{s_i^2 + (2\, \kpc)^2},
\label{eq:rhogrid}
\end{equation}
where $s_i$ is the distance of the grid point from the Sun.  The
$2\,\kpc$ softening term is used to avoid sampling artefacts from grid
points that lie close to the Sun's position.

\begin{figure}
        \subfigure[$\alpha=0$]
    {
        \includegraphics[width=0.25\textwidth]{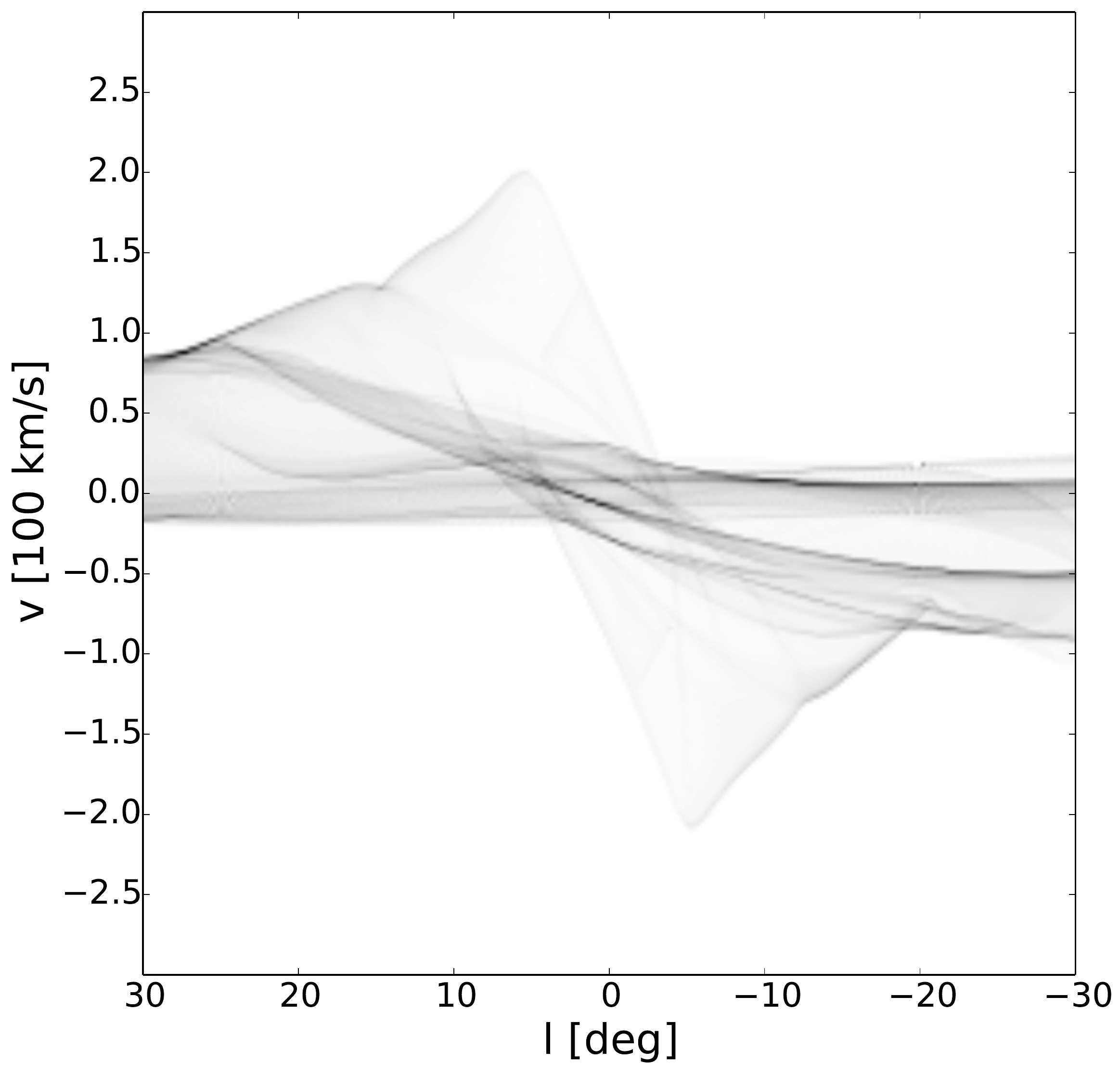}
        \includegraphics[width=0.25\textwidth]{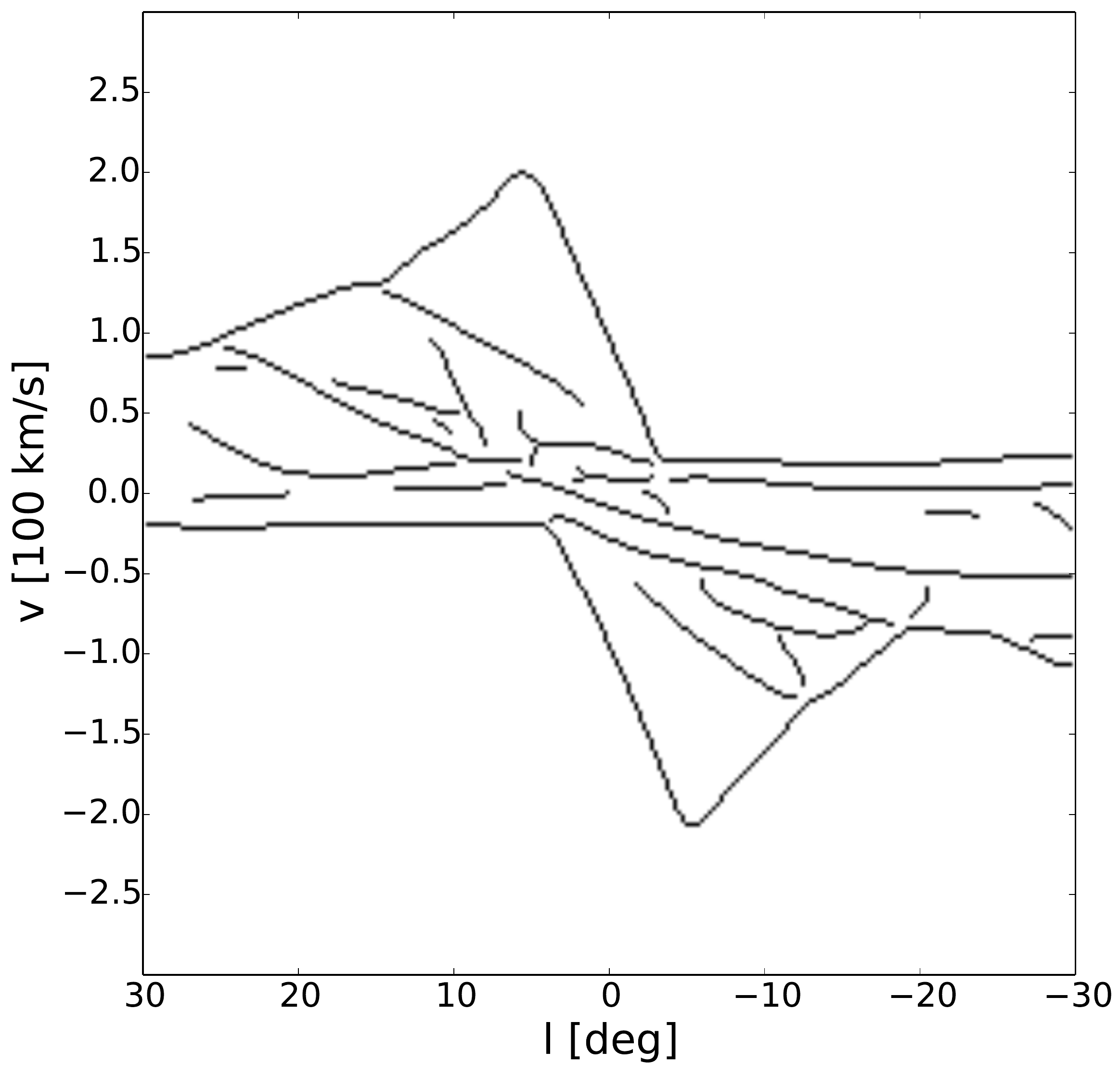}
    }
        \subfigure[$\alpha=1$]
    {
        \includegraphics[width=0.25\textwidth]{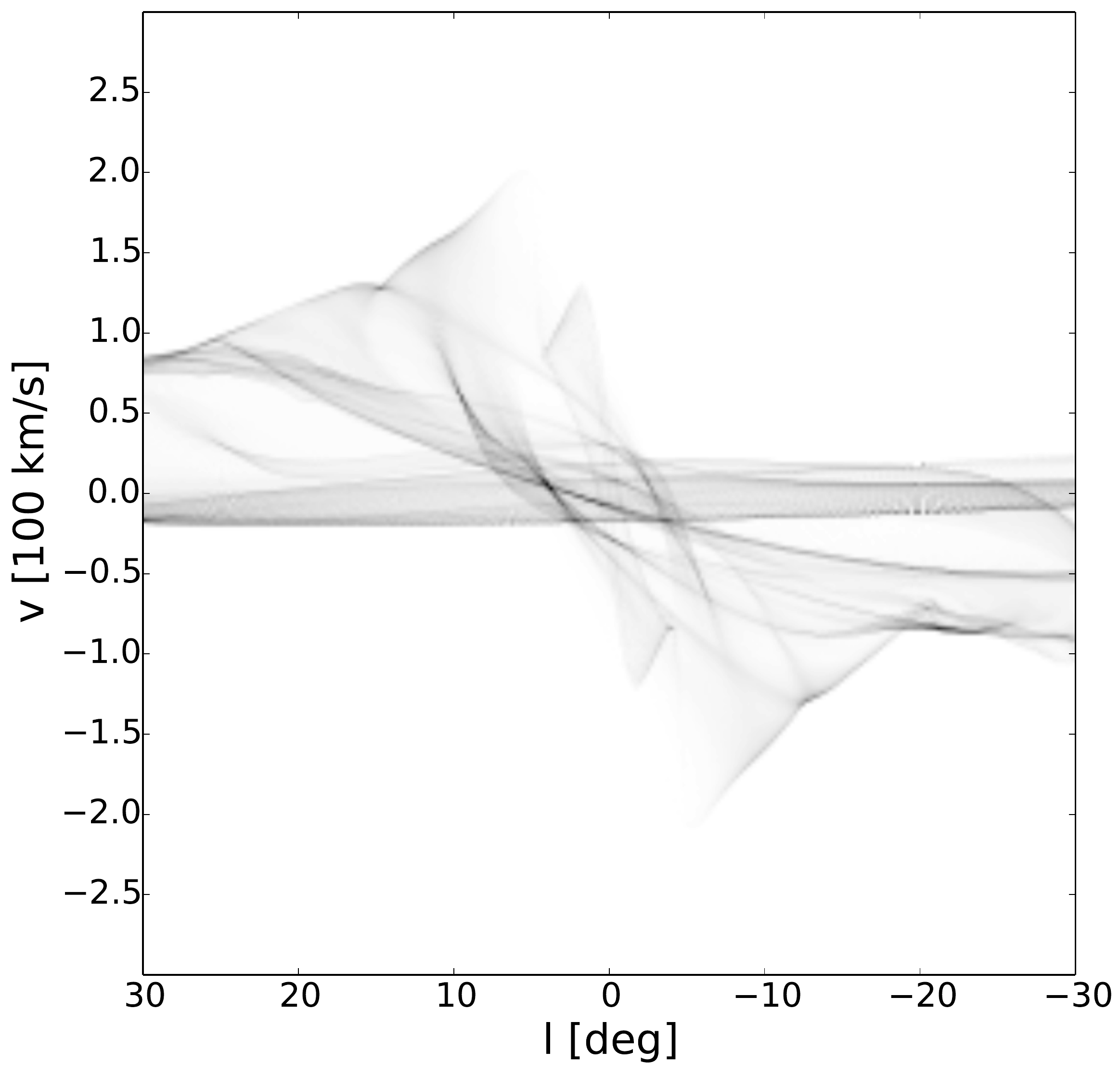}
        \includegraphics[width=0.25\textwidth]{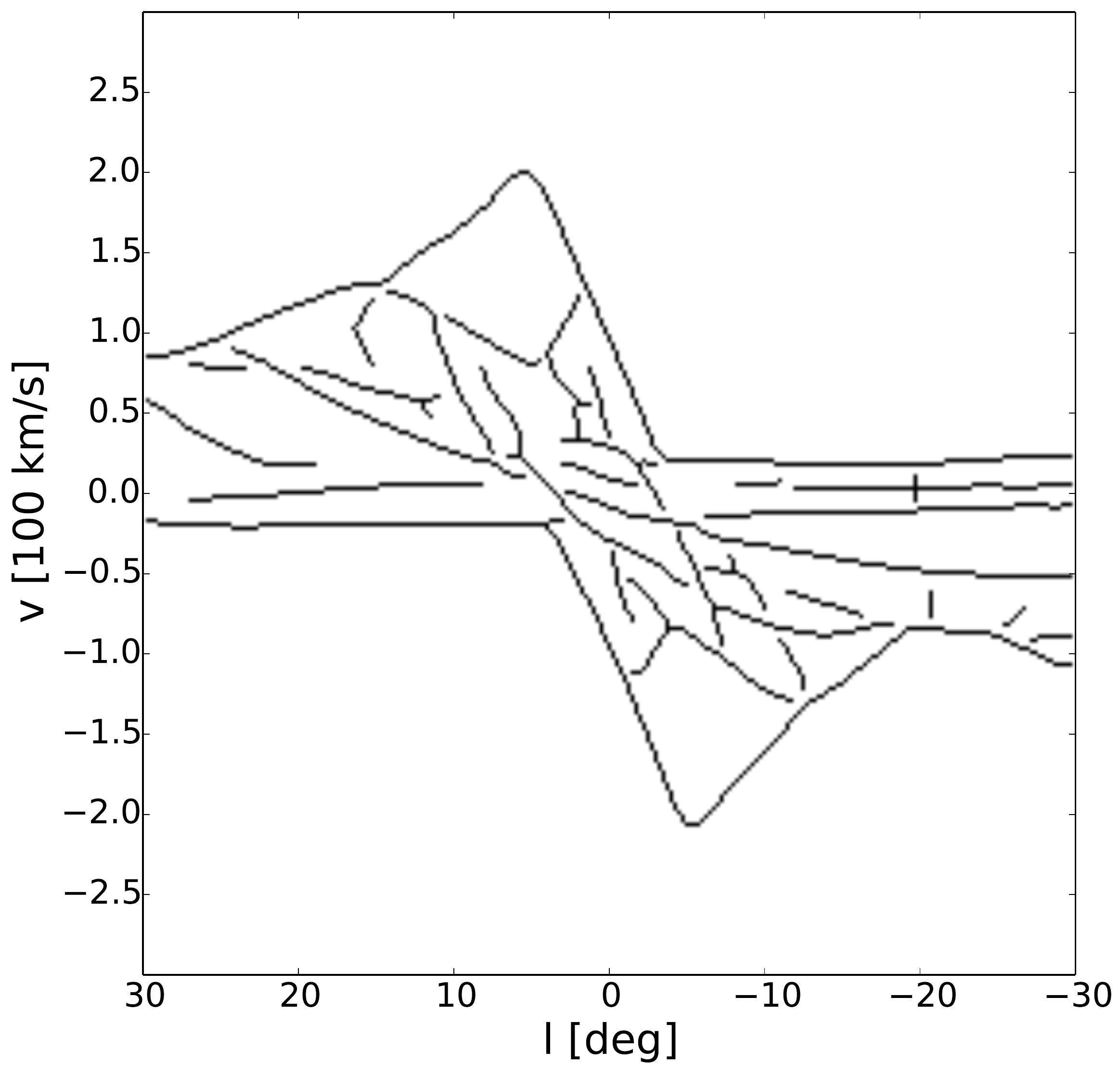}
    }
        \subfigure[$\alpha=2$]
    {
        \includegraphics[width=0.25\textwidth]{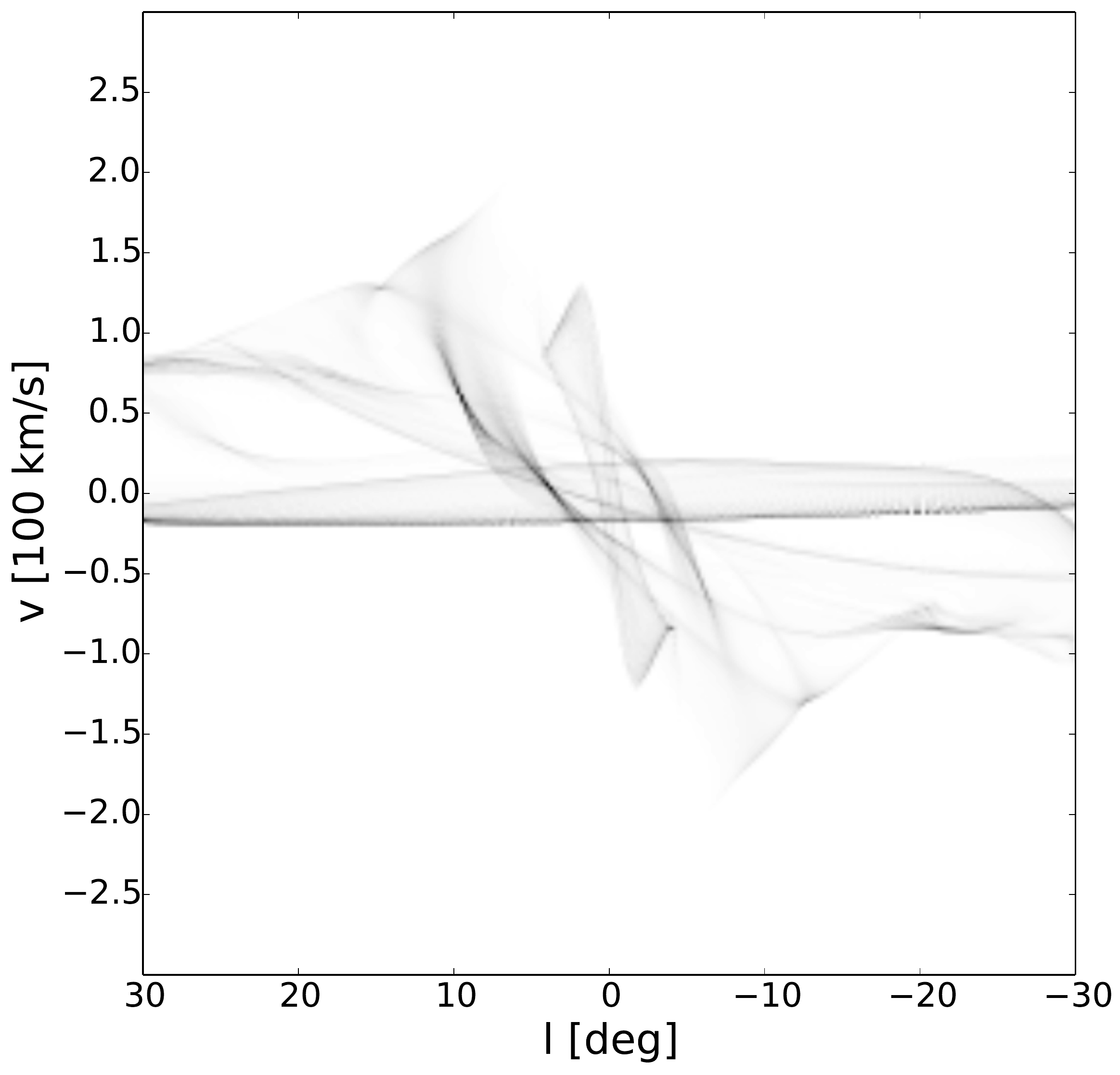}      
        \includegraphics[width=0.25\textwidth]{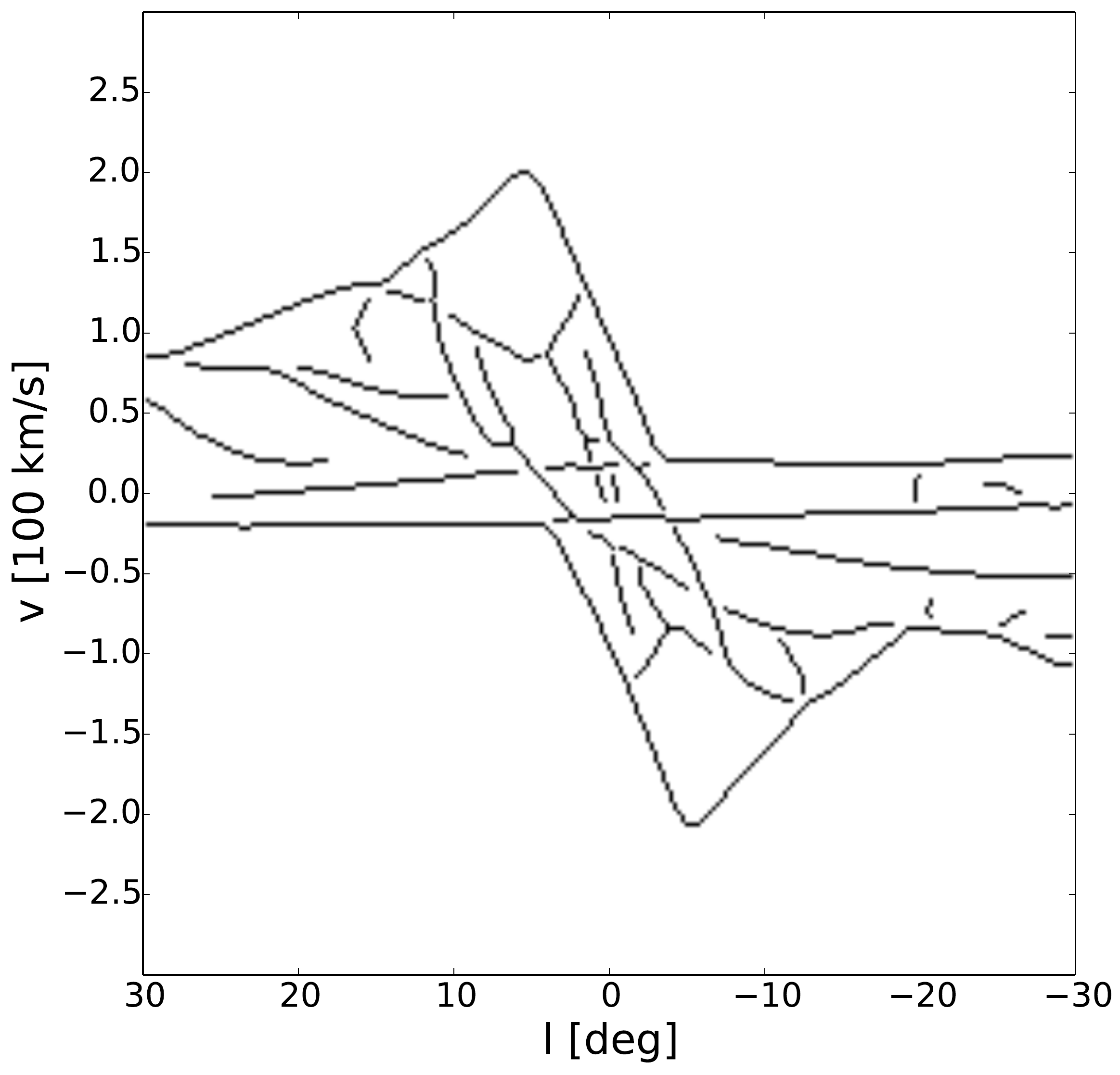}
    }
        \subfigure[$\alpha=3$]
    {
        \includegraphics[width=0.25\textwidth]{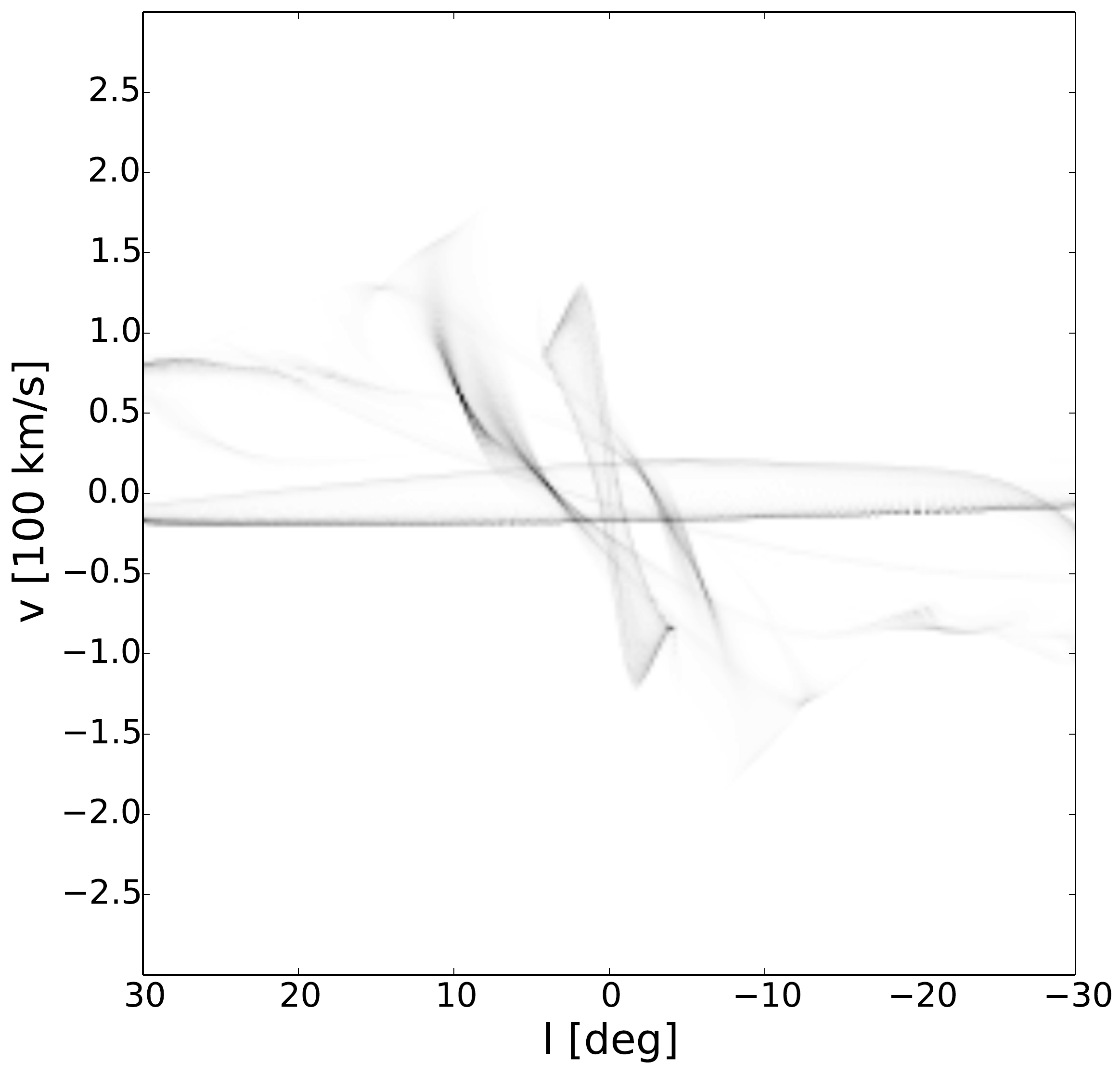}
         \includegraphics[width=0.25\textwidth]{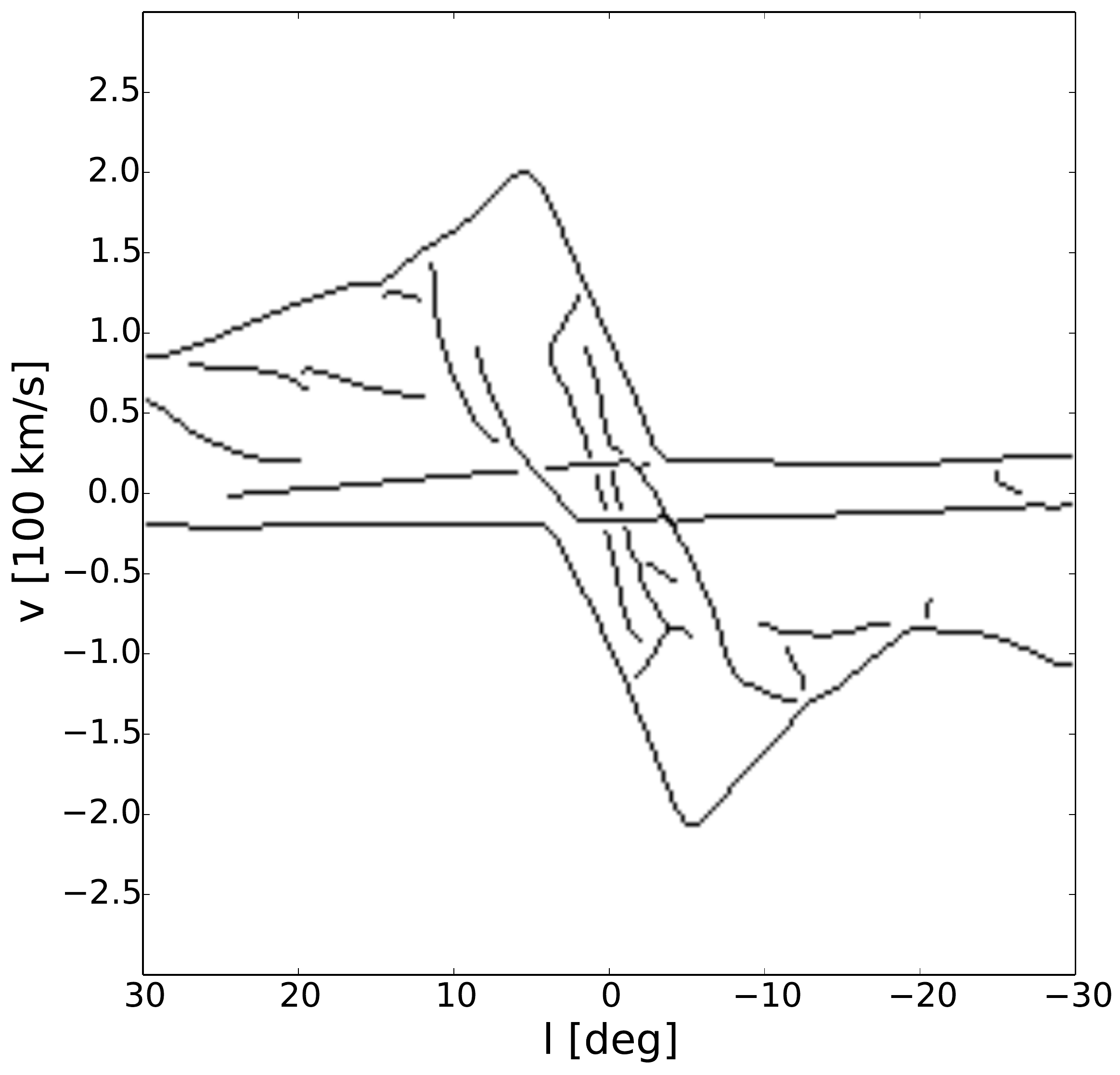}
    }
    \caption{The effect of the exponent $\alpha$ adopted in the
      projection law~\eqref{eq:rhogrid}.  
      The panels on the left show how the model's
      predicted $T_B(l,v)$ varies with~$\alpha$.
      The underlying model is our reconstruction of
      Englmaier \& Gerhard's (1999) potential with bar pattern speed
      $\Omega_{\rm p}=55\, \kms \kpc^{-1}$ viewed at time $t= 367 \rm Myr$
      with angle $\phi=20\degree$.
      The panels on the right show the corresponding
      features found using the algorithm of Section~\ref{sec:findfeat}.
      For $\alpha=0$, the gas density in the $(x,y)$ plane is a
      effectively uniform, and all information about density variation
      is washed away.
      Yet, many features in the $(l,v)$ plot remain visible,
      particularly those corresponding to spiral arms.
      This suggests that the velocity field is more important than the
      density distribution in producing $(l,v)$ features (see also
      Mulder \& Liem 1986).}
\label{fig:models3}
\end{figure}

We include the exponent $\alpha$ in~\eqref{eq:rhogrid} as a very
crude proxy for radiative transfer effects, which allows us to test
how sensitive model features are to how the projection is done.
Fig. \ref{fig:models3} shows projections made for different values of
$\alpha$.
We usually adopt $\alpha=1$, in which case our projection is linear in
the density and gives results equivalent to radiative transfer for
some simple cases.
In the case of HI, radiative transfer calculations show that the
optical depth is linear in the column density if the
temperature\footnote{that is, the spin temperature, which is
  associated with the relative population of the energy levels.} is
constant, and in the optically thin limit also the corresponding brightness temperature is linear.
The assumption of constant temperature is known to be a simplification
for Galactic HI, as it is often modelled as a medium made by two or
more phases at different temperatures \citep[see for
example][]{Ferriere2001}.
In the case of ${}^{12}$CO, the linearity is invalid when considering
a single cloud, but can be roughly recovered when shadowing is not
important between many clouds \citep[see, e.g.,][]{BM}.
  
A number of further minor comments are in order regarding this
projection.  As we do not have a recycling law that lowers the gas in
high-density regions, gas accumulates at the very center reaching
implausibly high densities.  So, before projecting the density, we
clip all gas with density more than four times higher than the initial
density to this maximum value.  
We also project only the gas inside $R<8\, \kpc$: this is justified by the fact as we focus only on the Central disk ($| l |<30\degree$) and exclude low-velocity emission from our comparisons.
Material outside the Solar circle would produce emission only at high
longitudes or low line-of-sight velocity, which we do not include in
our fits.

%% file: Comparing.tex
Given a set of synthetic $(l,v)$ distributions, we would like to judge
which one is ``closest'' to the observations.
This is actually a very challenging task, and the main focus of this
paper.

On small scales, the detailed intensities at each point are the result
of complicated radiative transfer physics, and therefore depend on
local density, temperature, and chemical composition of the gas. Also
the local clumpiness of the gas plays a major role in determining the
observed longitude-velocity intensities on small scales.
On the other hand, the presence of large-scale features, such as
bright ridges tracing spiral arms, is more robust to changes in the
radiative transfer physics and local physics, and more sensitive to
changes in the large-scale dynamics of the gas.
The evidence for this can be seen both in data and models:

\begin{itemize}
\item The main features believed to trace spiral arms are present both
  in CO and HI.  The CO and HI features are coincident with one
  another on the scale of the $(l,v)$ maps of Figure~\ref{fig:obs1}.
  The same cannot be
  said for the detailed intensities, as features have different
  relative intensities in the two species. This is to be expected,
  since the two species obey different radiative transfer physics and
  probe different density-temperature regimes; therefore, spatial
  densities are not completely correlated.
  
\item The data also show a clumpy sub-structure that can be modeled as
  the result of heating and cooling processes, such as stellar
  feedback and supernovae feedback; interestingly, if a smooth
  simulation is run turning off these processes, the main large-scale
  features stay the same \citep{babaetal2010}.
  
\item In Fig. \ref{fig:models3} we see the result of
  projecting the gas density for four different values of $\alpha$.
    Clearly, the same bright ridges can be identified at same locations
  for a wide range of values of $\alpha$, while the relative intensity
  of these ridges varies.  Only at extreme values ($\alpha=3$) does the
  difference start to be noticeable.\footnote{Nevertheless, our
    fitting scheme method still finds the
  correct parameters in this case; see Sect. \ref{sec:tests}.}
  Particularly striking is the case $\alpha=0$, where any information
  about overdensities along spiral arms has been washed out: the
  features stay more or less the same as long as \textit{some} gas is
  present at all points, suggesting that the large-scale velocity
  field is what mostly determines the features \citep[see
  also][]{mulderliem}.
  
\item \cite{Pettitt2014} produced a longitude-velocity diagram from a
  full radiative transfer calculation for CO, and compared it to the
  diagram obtained by a simple-minded projection approach akin to
  ours.
  Their results show that while detailed intensities are different,
  the overall morphology and large-scale features are identical in the
  two versions.
  
\end{itemize}

Therefore we argue that, loosely speaking, detailed intensities at
each point probe the radiative transfer physics and local physics,
while broad features trace the large-scale dynamics of the gas.

Proper modelling of radiative transfer physics, local physics
\textit{and} gas dynamics at the same time is certainly possible
\citep[e.g.,][]{ShettyOstriker2008,Dobbs11,Tasker11,Pettitt2014}, but
its computational expensive renders it unattractive when the parameter
space to explore is large \citep[see, e.g., Section~2.3.2
of][]{Pettitt2014}.
Nevertheless, the above considerations suggest that it is possible to
break the problem of understanding the Galaxy's ISM into two steps:
first constrain the large-scale gravitational potential and dynamics
by matching only the broad features in the $(l,v)$ distribution; then,
once the potential has been constrained, go back and use more
sophisticated models to understand the internal structure and
chemistry of the ISM in detail.

Here we present a quantitative comparison scheme to use in the first
step, focusing only on features that are believed to depend the most
on gas dynamical physics and the least on the radiative transfer and
local physics.
This ``feature comparison'' problem has much in common with the
problem of matching fingerprints, in which, given a smudged,
incomplete, contaminated print taken from a crime scene, the goal is
to find which members in a database of prints look most like it.
Here the Milky Way is the crime scene and we use a set of simple
hydrodynamical models as the database of potential culprits.

\subsection{Identifying features in model $(l,v)$ distributions}
\label{sec:findfeat}

\begin{figure*}
\centering
    \subfigure[Starting Image]
    {
        \includegraphics[width=0.23\textwidth]{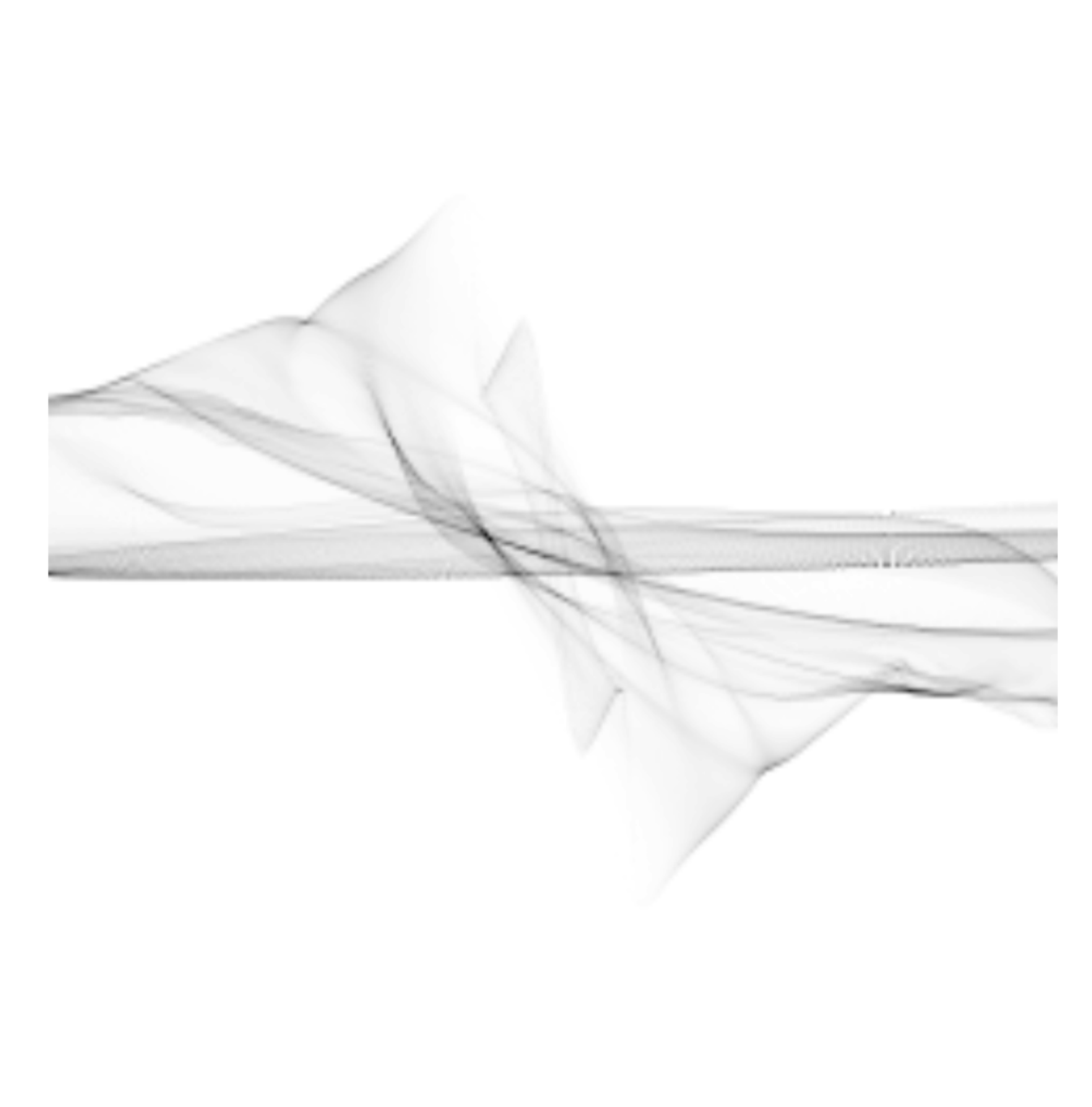}
        \label{sub:c1}
    }
    \subfigure[Enhanced Envelope]
    {
        \includegraphics[width=0.23\textwidth]{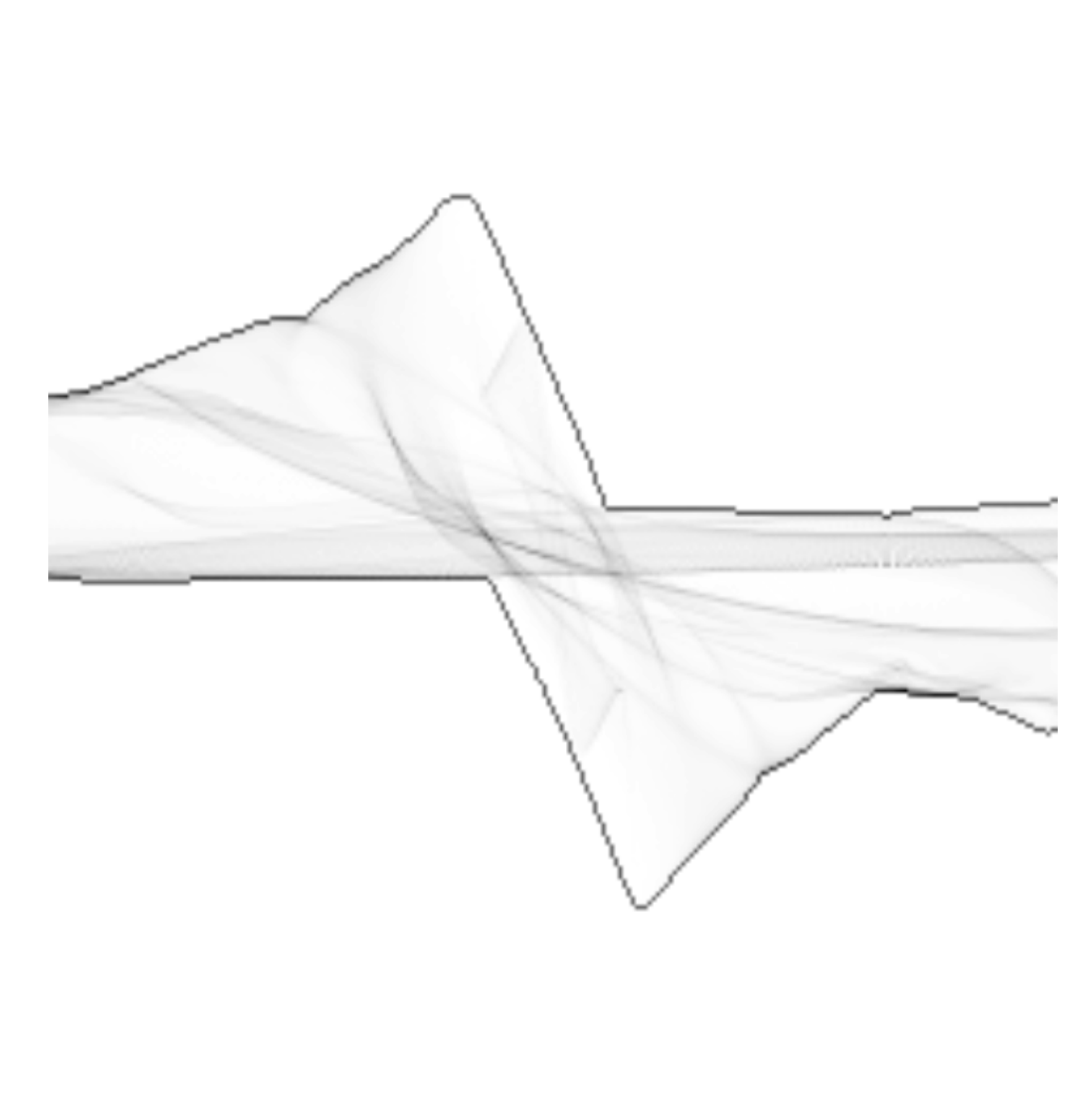}
        \label{sub:c2}
    }
    \subfigure[Smoothed with Gaussian filter]
    {
        \includegraphics[width=0.23\textwidth]{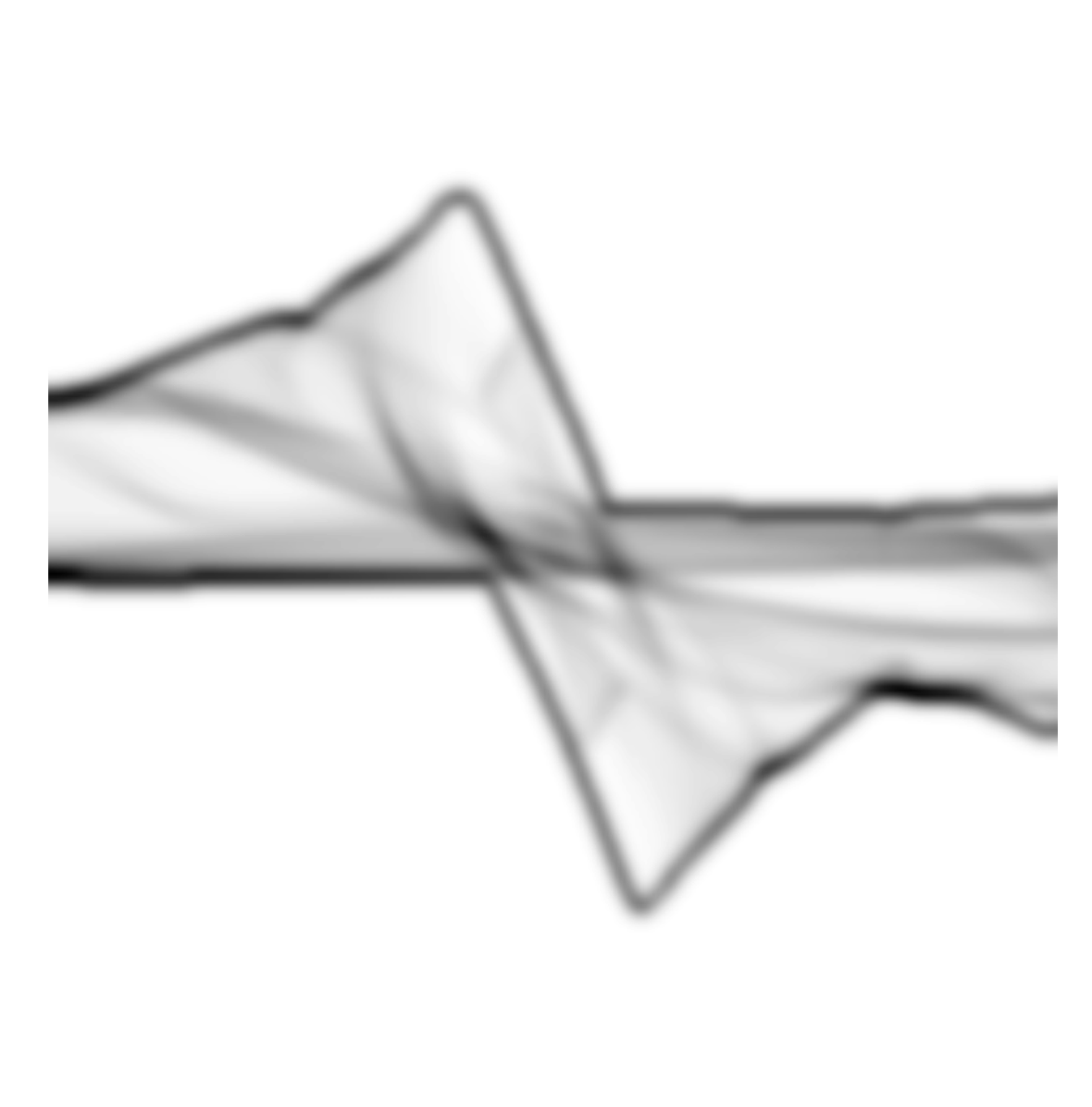}
        \label{sub:c3}
    }
    \subfigure[Ridge Filtered Image]
    {
        \includegraphics[width=0.23\textwidth]{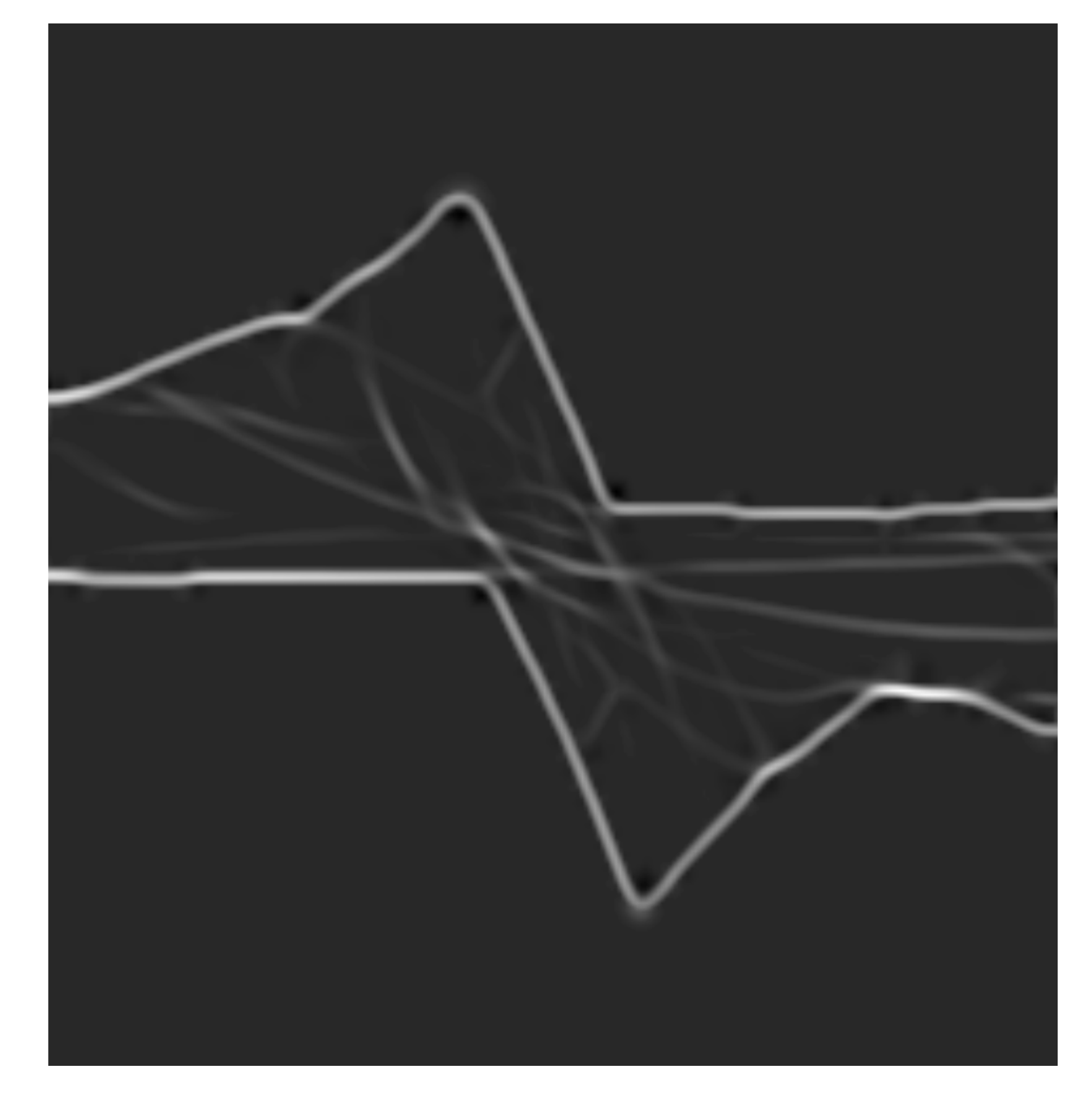}
        \label{sub:c4}
    }
    \\
    \subfigure[Points after Non-Maximal Suppression]
    {
        \includegraphics[width=0.23\textwidth]{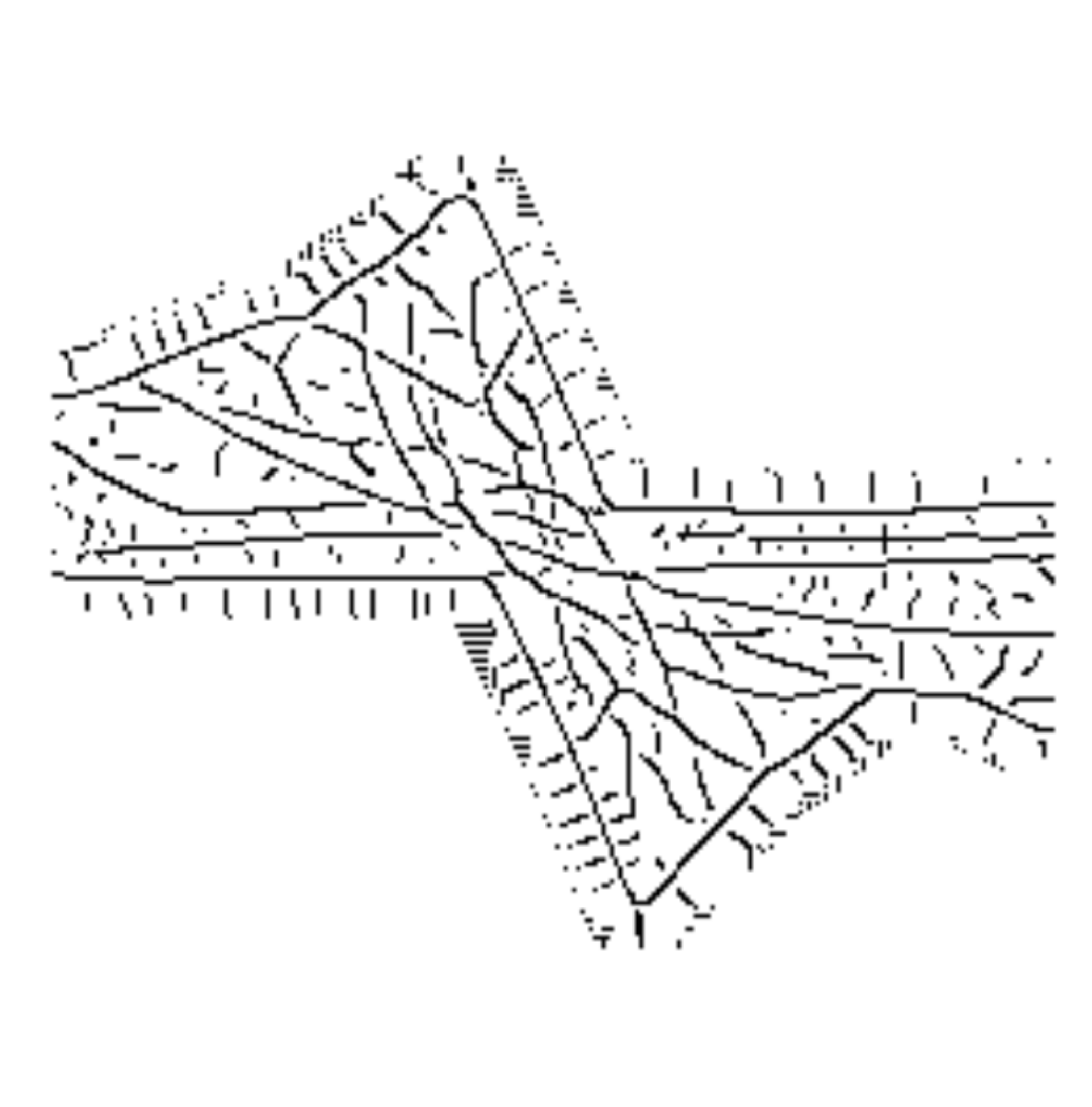}
        \label{sub:c5}
    }
    \subfigure[Points after Hysteresis Thresholding]
    {
        \includegraphics[width=0.23\textwidth]{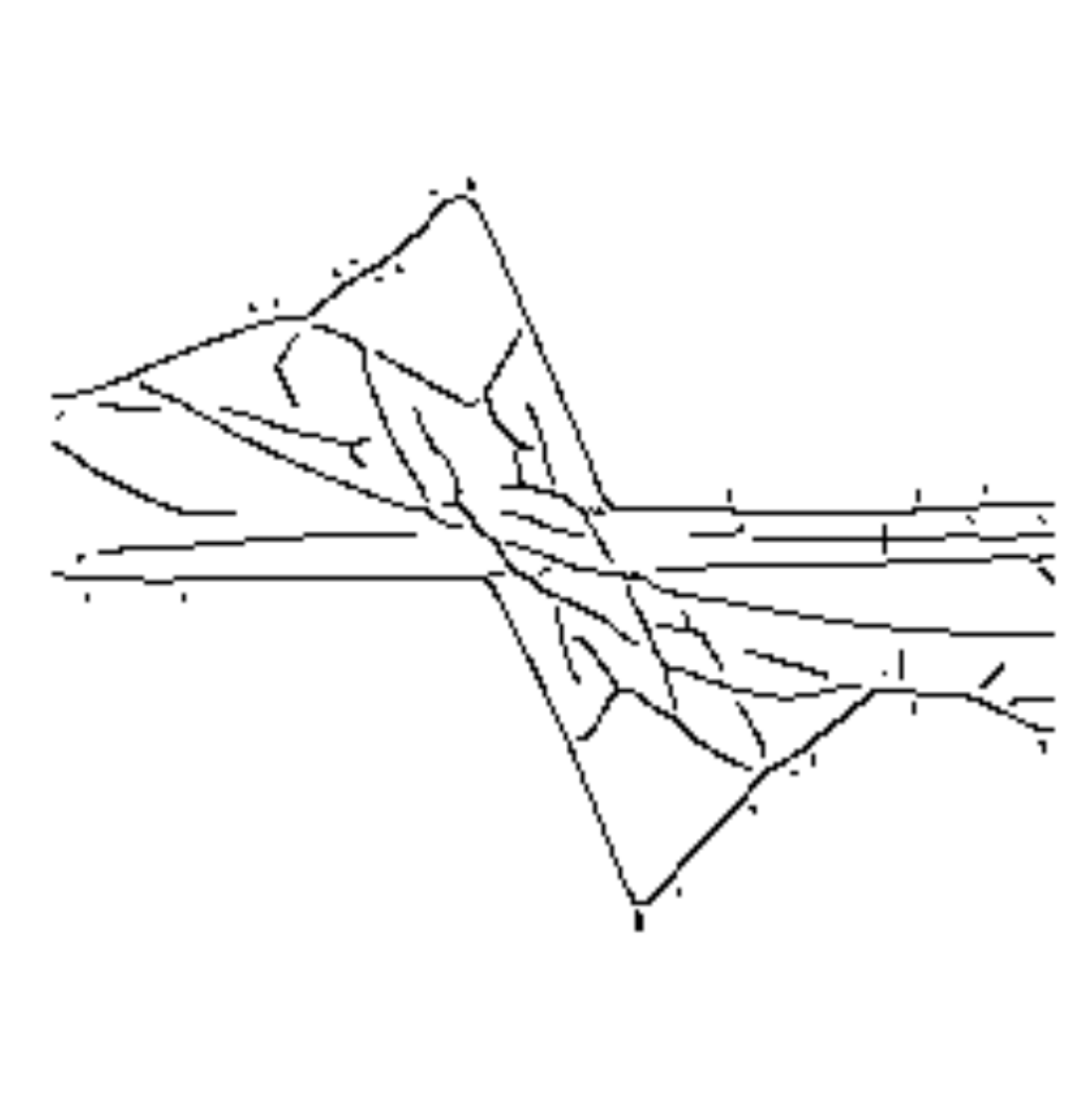}
        \label{sub.c6}
    }
    \subfigure[After Thinning]
    {
        \includegraphics[width=0.23\textwidth]{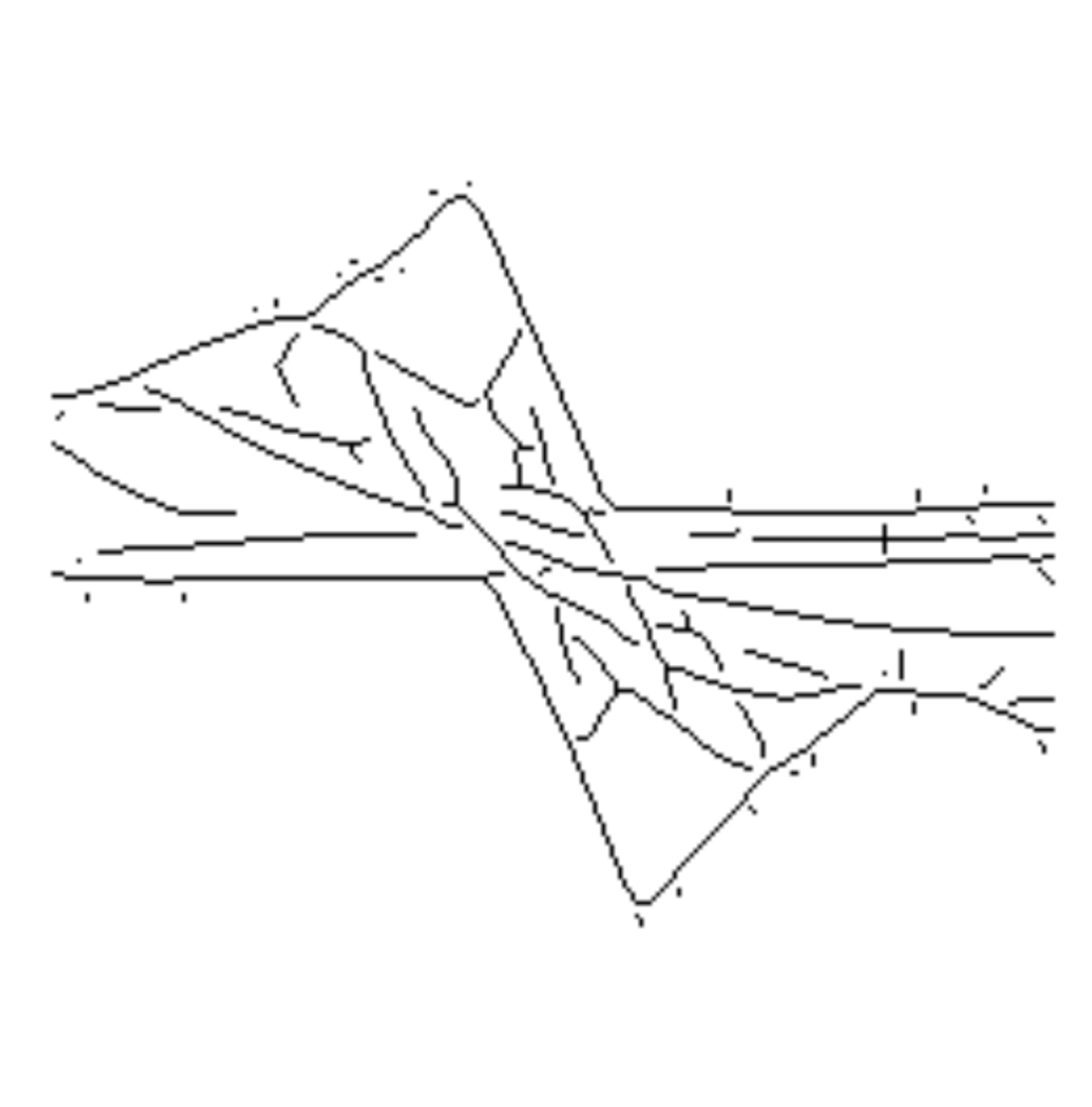}
        \label{sub:c7}
    }
    \subfigure[Final Output]
    {
        \includegraphics[width=0.23\textwidth]{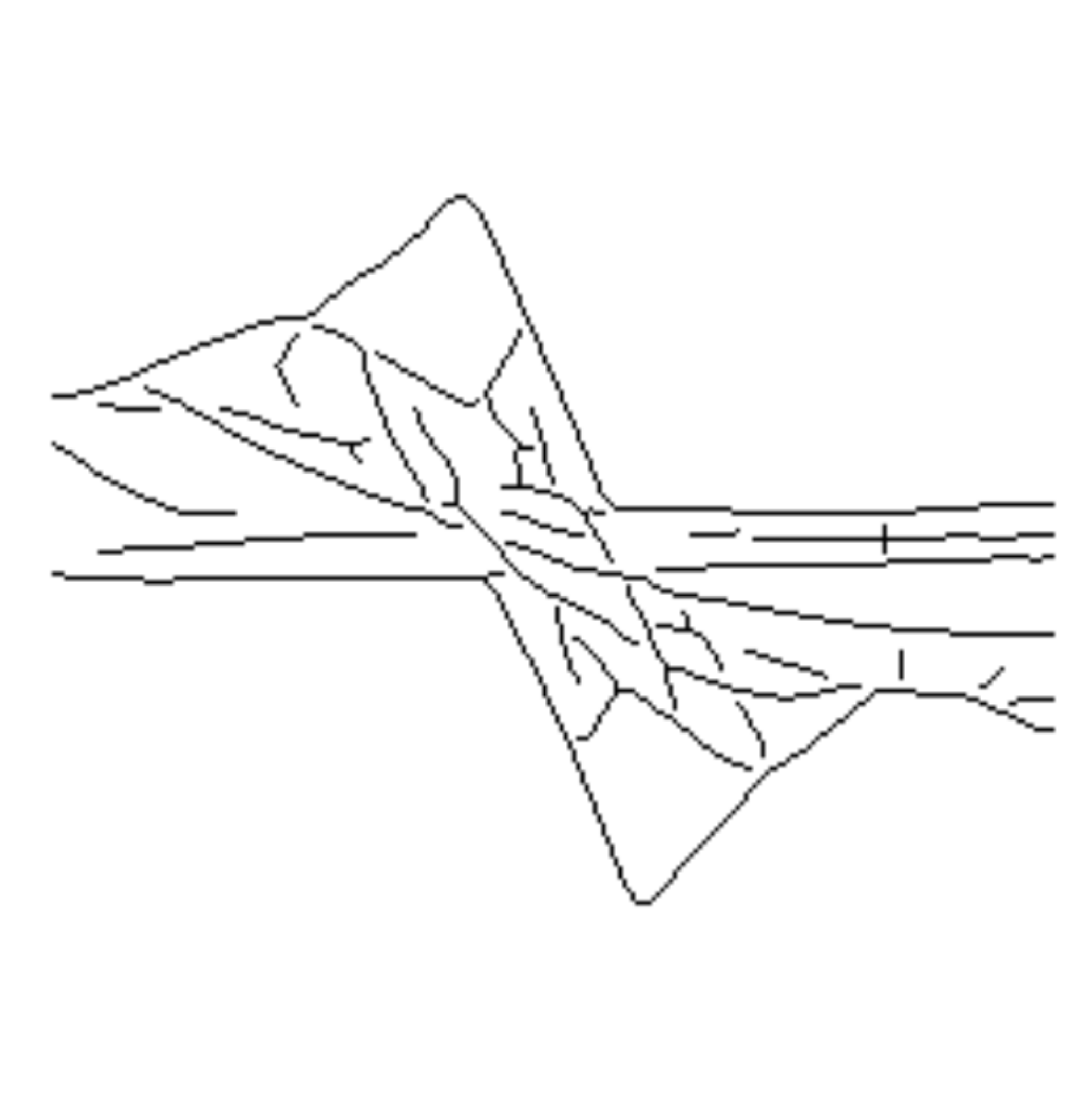}
        \label{sub:c8}
    }
    \caption{The steps of our feature-finding algorithm
      (section~\ref{sec:findfeat}), ordered from top left to bottom
      right. The final output represents the features detected in the
      model, constituted by the bright ridges and the
      envelope.  Features are 1-pixel wide lines at least 5 pixels
      long. The final output is then used as input for the SMHD.}
    \label{fig:comparison1}
\end{figure*}

As discussed above, features that mostly probe the large-scale
dynamics of the gas, besides the envelope, are generally bright ridges
in the longitude-velocity plane.
Therefore our first step is to identify such ridges: given a binned
model $(l,v)$ distribution (Section~\ref{sec:projection} above), the
task is to return a corresponding binary image in which each bin is 1
if the emission at that bin is part of a feature, 0 otherwise.
Features are lines 1-pixel wide and at least 5 pixels long.
Our procedure relies heavily on the widely used edge-detection
algorithm of \cite{Canny1986}, modified to detect ridges instead of
edges.
The steps of our procedure are the following, each illustrated in
Fig.~\ref{fig:comparison1}.

\subparagraph{\textbf{Envelope Enhancement}} In a typical model, some
parts of the envelope are brighter than others. Since we want our
algorithm to pick up the whole envelope, not only its brightest parts,
we increase the intensity of pixels lying on the envelope.
Thus we put the value of all pixels on the envelope equal to the value
of the brightest pixel in the image.
A pixel is defined to belong to the envelope if it has at least one
side in common with an empty pixel which is above the highest velocity
pixel with positive emission or below the lowest velocity pixel with
positive emission.

\subparagraph{\textbf{Smoothing}} We convolve the image (including the
enhanced envelope) with a Gaussian to remove noise. We used a standard
deviation $\sigma=2$ pixels.
(Recall that a pixel size is $\Delta l = 0.25 \degree$ in longitude
and $\Delta v = 2.5 \,\kms$ in velocity.)

\subparagraph{\textbf{Ridge Filter}} This step filters the image in
order to highlight ridges. Let $F(x, y)$ denote a two-dimensional
function and $H$ its Hessian Matrix,
$$H = \begin{bmatrix}
F_{xx} & F_{xy} \\  
F_{xy} & F_{yy}
\end{bmatrix}.
$$
A measure of the presence of a ridge at a point is the value of the
main negative eigenvalue of $H$ \citep{Lindeberg1996}.
The direction of the ridge is orthogonal to the direction of the
eigenvector associated with this main negative eigenvalue.
We therefore compute the lowest eigenvalue $\lambda_{\rm low}$ of the
Hessian Matrix and its associated direction $p$.
The greater the value of $R \equiv -\lambda_{\rm low}$, the stronger the ridge. To
compute the derivatives $F_{xx}$, $F_{xy}$ and $F_{yy}$ we use Sobel
operators \citep{Sobel1968} with a kernel size of 3 pixels.

\subparagraph{\textbf{Non-maximum suppression}} A search is carried
out to determine whether the ridge strength $R$ assumes a local
maximum in the direction $p$ orthogonal to the ridge.
At every pixel, we round the $p$ direction to the nearest $45 \degree$
(that is, the rounded direction points to one of the 8 nearest
neighbours).
Then we compare the ridge strength at the current pixel with the ridge
strength of the pixel in the positive and negative $p$ direction.
If $R$ at the current pixel is greater or equal, we mark the point as
a possible ridge, otherwise we suppress it, i.e., we declare that no
ridge goes through this point.

\subparagraph{\textbf{Hysteresis thresholding}} We need to decide
which points that are left unsuppressed by the previous step are
actual ridges.
Large values of $R$ are more likely to correspond to ridges.
It is in many cases difficult to specify a unique threshold at which
points switch from corresponding to ridges to not doing so.
For this reason, following \cite{Canny1986}, we use thresholding with
hysteresis.
Thresholding with hysteresis requires two thresholds, high and low.
All points above the high threshold are marked as certainly ridges.
All points below the lower threshold are marked as certainly
non-ridges.
The points between the two are marked as ridges only if they are
connected to a point above the high threshold.
This allows to follow a fainter section of a strong ridges, that is
likely to be a genuine ridge.
We use the mean value of~$R$ over the whole image as the high
threshold and one half of this value for the low threshold.
We have experimented with varying these by a
factor of two (but keeping the same high-to-low ratio) and find little
change in the resulting feature maps.

\subparagraph{\textbf{Thinning}} We make each detected line thinner,
reducing its width to one-pixel.
This is done according to the algorithm of \cite{Zhang1984}, which
preserves end points and pixel connectivity.

\subparagraph{\textbf{Remove small components}} As a final polishing,
we remove features that are too small and likely to be noise.
We therefore remove from the image all the connected components that
are less or equal than 4 pixels to obtain our final result.

\subsection{Comparing the features} \label{calcdist} Having two binary
images $A$ and $B$ representing the features of model and data
respectively, we need to produce a single number that quantifies their
dissimilarity.
For this purpose, we have tried different options.
The one we found intuitively most appealing involved applying the
``Earth-mover distance'' (Appendix \ref{EMD}).
Surprisingly, in practical tests this did not perform as well as a
much simpler alternative, the Modified Hausdorff Distance\footnote{
  However, as we discuss in more detail in Appendix \ref{EMD}, we
  believe that EMD used in a qualitatively different way could prove
  to be a useful alternative to $\chi^2$.}
\citep{Dubuisson94}.

Call $a = \{a_1, ... , a_N\}$ the set of pixels containing 1 in the
first image and $b = \{b_1, ... , b_M\}$ the same for the second
image.
Then the Modified Hausdorff distance is defined as
\begin{equation}
{\rm MHD}(a,b) \equiv\sum_i \min_j\left( d(a_i,b_j) \right)
\end{equation}
where $d(a_i,b_j)$ is a suitable metric between pixels.
In words, for each positive pixel in the first image, find the 
distance from the closest positive pixel in the second image, according
to the chosen metric~$d$.  Then the MHD is the sum of these distances over all positive pixels in the first image. 
An unattractive feature of the MHD is that it is not symmetric in its
arguments.  We define a symmetrized version as
\begin{equation}
  {\rm SMHD}(a,b)\equiv \frac{1}{2N} {\rm MHD}(a,b) +  \frac{1}{2M} {\rm MHD}(b,a)
  \label{eq:SMHD}
\end{equation}
Our recipe is that the dissimilarity between two binary images is
their SMHD.
The only thing left to decide is $d(a,b)$, the metric in the pixelated
$(l,v)$ plane.
After some experimentation, we decided to use the city-block distance,
\begin{equation}
  d(a,b) = \frac{|l_a-l_b|}{\Delta l} + \frac{|v_a-v_b|}{\Delta v},
\label{eq:distancefn}
\end{equation}
where $(l_a,v_a)$ are the coordinates of pixel $a$ in the first image,
and, similarly, $(l_b,v_b)$ are the coordinates of pixel $b$ in the second.
The city-block distance between two points is the sum of the absolute
differences of their Cartesian coordinates.
We found very little difference between this and the ``Euclidean''
distance obtained by squaring each of the terms
in~\eqref{eq:distancefn}.
The results of the fitting method depend also on the choice of the
ratio $\zeta = \Delta v / \Delta l$, which should be adapted to the
nature of features under consideration.
A natural choice is to take $\zeta$ equal to the ratio of the typical
velocity extension and the typical longitude extension of a feature,
which in our case is approximately $\zeta=10$.
This value could in principle be adjusted to better suit other
situations.

One might ask whether we need to symmetrize the MHD: we extract
features from data and models in different ways, so why should we
impose symmetry on the function we use to compare the resulting sets
of features?
In fact, variants of our method can be defined that do not symmetrize
the MHD.
Let $D$ be the features in data and $M$ the features in models.
If we measure the dissimilarity of data and observations using ${\rm
  MHD}(M,D)$, then models are penalized if a model feature is not
present in the data, but not viceversa.
Thus, ${\rm MHD}(M,D)$ is insensitive to contaminants in the data that
cannot be reproduced by the model.
On the other hand if we use ${\rm MHD}(D,M)$, the opposite would be
true: we are not penalized at all if our model predicts extra features
in addition to those present in the data.
An example of such a situation occurs in Figure~\ref{fig:retrieving1},
panel (b), below.
We decided to use the symmetrized version as giving a good compromise,
but one could choose to calculate both and process independently the
separate bits of information acquired.

%% file: Tests.tex
In this Section we use mock data generated from a variety of simulated
galaxies to asses the performance of our method.
By fitting the mock data with a family of parametrized simulated
galaxies, we test how well we can recover the correct parameters
describing the potential underlying the mock data.
We also compare the performance of our SMHD distance against two other
measures of goodness-of-fit that have been used previously, namely
$\chi^2$ and envelope distance.

Unlike our SMHD, which measures distance between model and observed
features, $\chi^2$ is a direct measure of the difference between model
and observed brightness temperatures.
It is defined as
\begin{equation}
  \label{eq:chisq}
  \chi^2\equiv \sum_n\left[\frac{T_B^D(l_n,v_n)-T_B^M(l_n,v_n)}{\Delta
      T_B^D(l_n,v_n)}
    \right]^2,
\end{equation}
where $T_B^M(l_n,v_n)$ is the model's prediction for the brightness
temperature at the point $(l_n,v_n)$ and $T_B^D(l_n,v_n)$ is the
corresponding ``measurement'' from the simulated dataset, with
measurement uncertainty $\Delta T_B^D(l_n,v_n)$.
For the tests here we take $T_B$ to be directly proportional to the
binned $(l,v)$ distribution constructed in section~\ref{sec:projection},
with $\Delta T_B=\hbox{constant}$.

The Envelope Distance (ED) is defined as
\begin{equation}
D_{\rm e}^2 = \frac{1}{N} \left[ \sum_{n=1}^N \left[
      v^{D}_+(l_n) - v^M_+(l_n) \right]^2
  + \sum_{n=1}^N \left[ v^D_-(l_n) - v^M_-(l_n) \right]^2 \right]
\label{eq:envelope}
\end{equation}
where $v^{D}_{\pm}(l)$ and $v^M_\pm(l)$ are the positive- and
negative-velocity envelopes of the ``data'' and ``model'' respectively.
We consider the range $-6 \leq l \leq 30\, \degree$ for the
positive-velocity envelope and $-30 \leq l \leq 6\, \degree$ for the
negative; we omit portions of the envelope that in the real observations
are seen to be heavily influenced by material outside the solar circle.
The envelope distance is closer in spirit to our SMHD than $\chi^2$, as
it involves measuring the distance between the terminal velocity
features of model and data.

For both $\chi^2$ and SMHD we exclude all data/features at low
velocities, $|v|<40\kms$.
This is done to simulate the modelling of real data, in which the low
velocity features are dominated by foreground emission.

The models we fit in this section are all based on our reconstruction of
the \cite{englmaiergerhard1999} potential.
They have only three free parameters: the pattern speed $\Omega_{\rm
  p}$, the angle $\phi$ between the major axis of the bar and the
Sun-Galactic Centre line and the evolutionary time $t$.
We shall see in section~\label{stationarity} below that, as these
particular models settle into an approximate steady state, we can
eliminate the time $t$, reducing the number of interesting parameters to
be fit to just the pair $(\phi,\Omega_{\rm p})$.
For each model, we evolve an initial axisymmetric state, as described in
Section~\ref{sec:hydrodescription} and project onto the $(l,v)$ plane
assuming an exponent $\alpha=1$ in equation~\eqref{eq:rhogrid}.
At this point we can calculate $\chi^2$ and the ED~\eqref{eq:envelope}.
For the SMHD~\eqref{eq:SMHD}, we use the algorithm of
Section~\ref{sec:findfeat} to find the features.

\subsection{On Stationarity} \label{stationarity}

\begin{figure}
\centering     
        \includegraphics[width=0.23\textwidth]{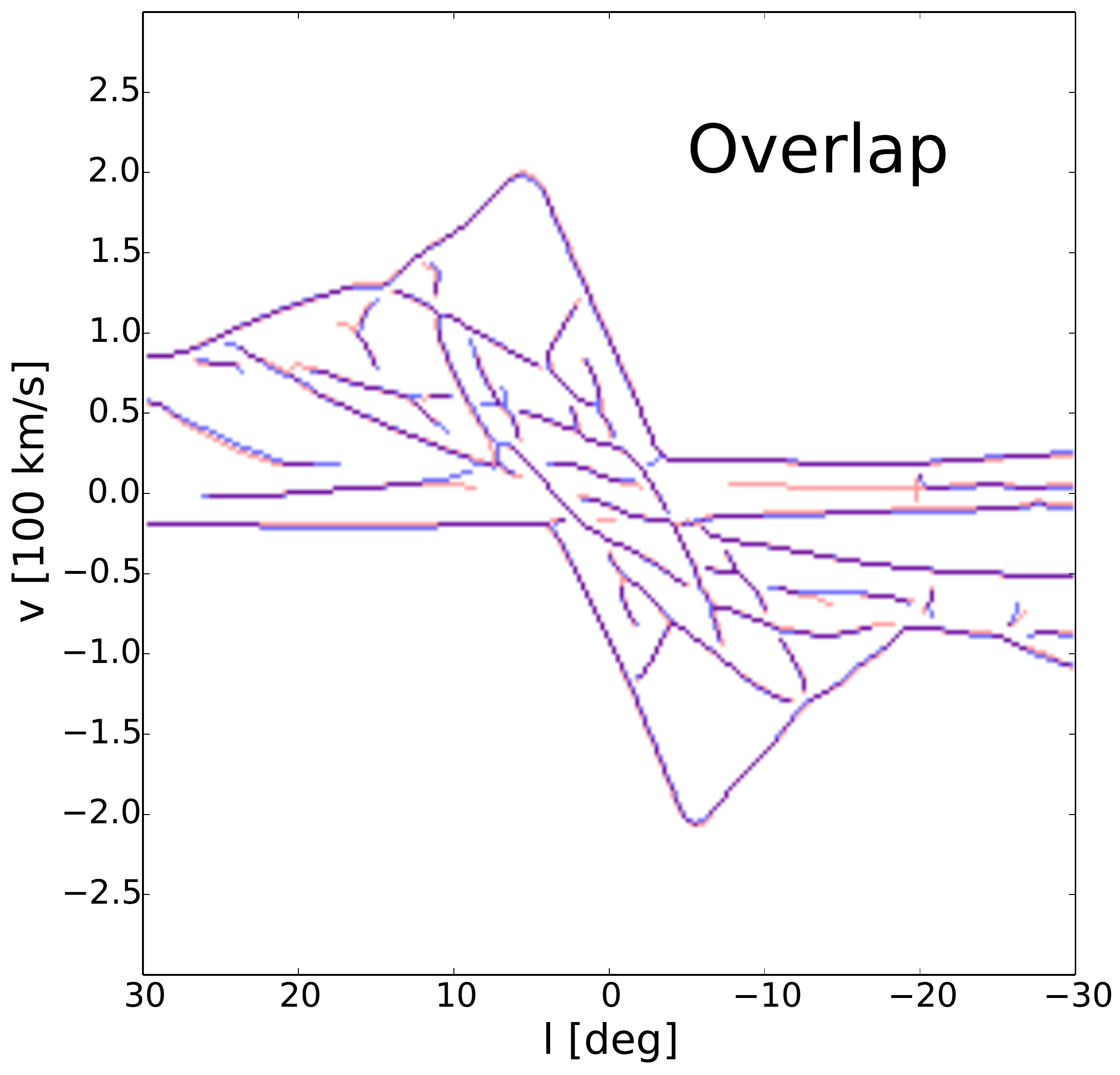}
        
        \includegraphics[width=0.23\textwidth]{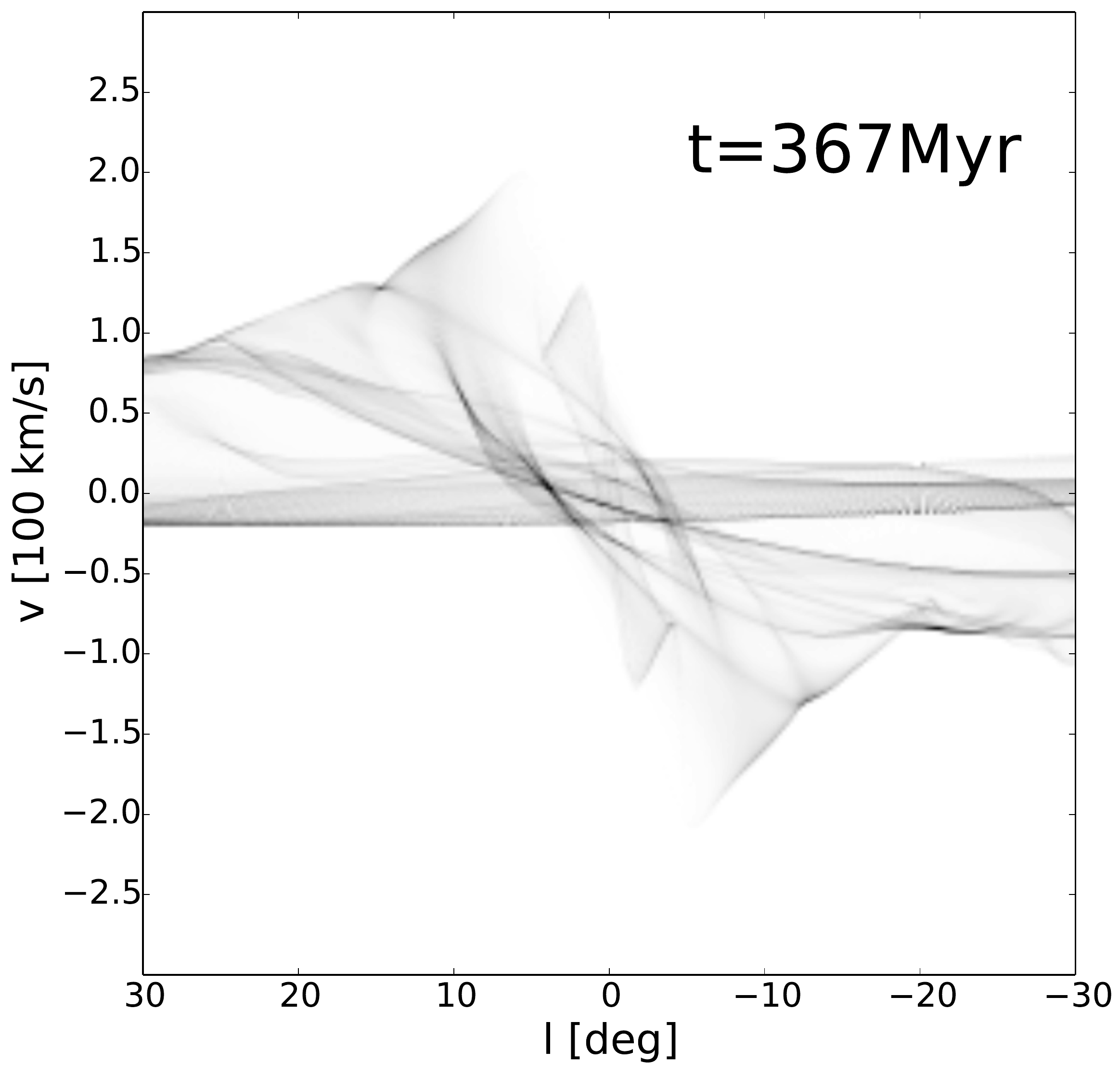} 
        \includegraphics[width=0.23\textwidth]{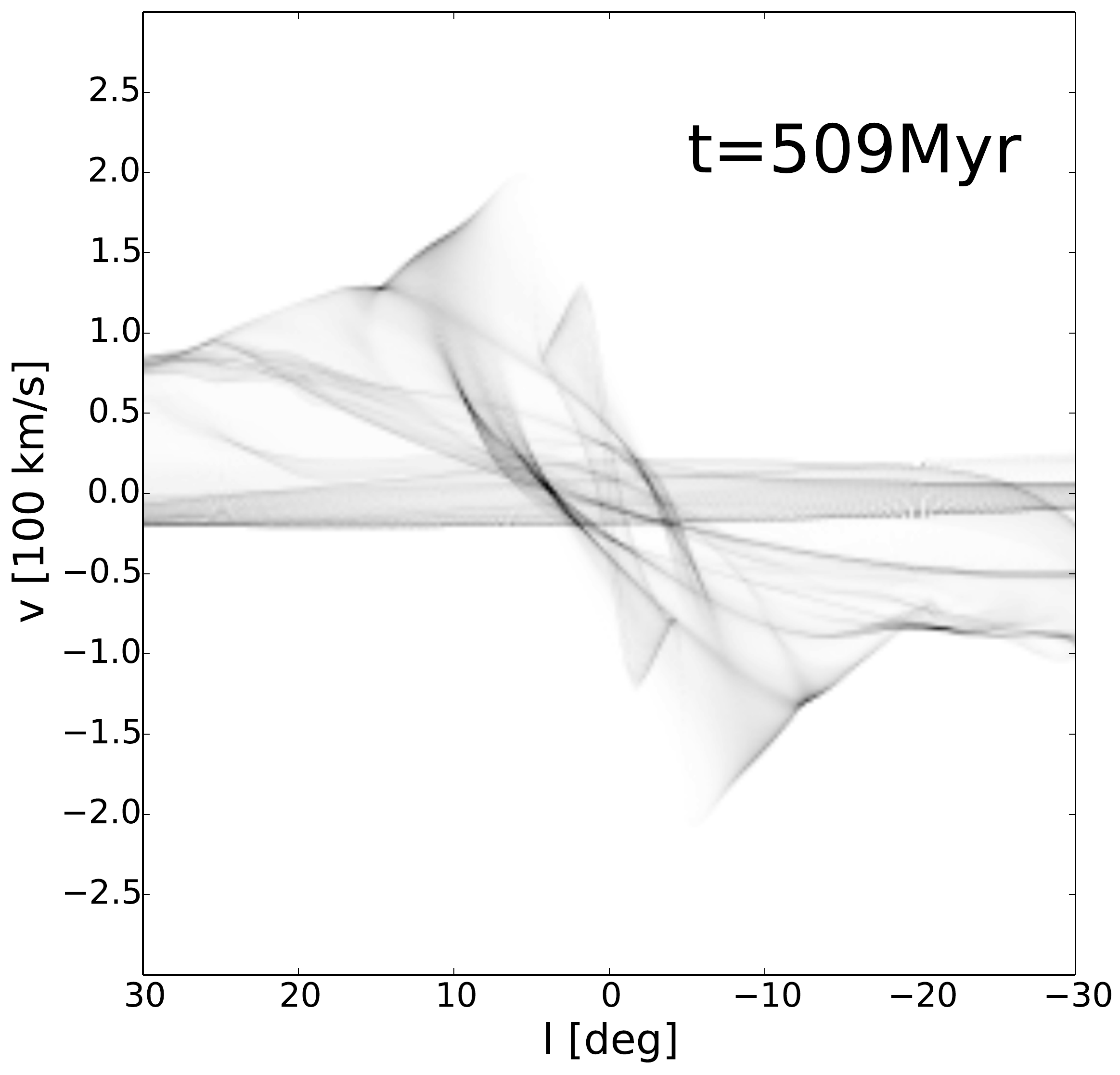}
    \caption{$(l,v)$ plots of the simulation with $\Omega_{\rm p} = 55 \, \kms
      \kpc^{-1}$ viewed from $\phi = 20 \degree$ at different evolutionary times.
      The top panel overlays the features extracted from the two snapshots.} 
    \label{fig:stationarity1}
\end{figure}

The \cite{englmaiergerhard1999} models we consider here quickly reach a
steady state.
Fig. \ref{fig:stationarity1} shows two snapshots of the model having
pattern speed $\Omega_{\rm p}=55\,\kms\kpc^{-1}$ viewed with
$\phi=20\degree$ taken at two different evolutionary times.
The features are almost identical, particularly outside the $|v|<40\kms$
band we exclude from our fitting procedure.

\begin{figure}
  \includegraphics[width=0.25\textwidth]{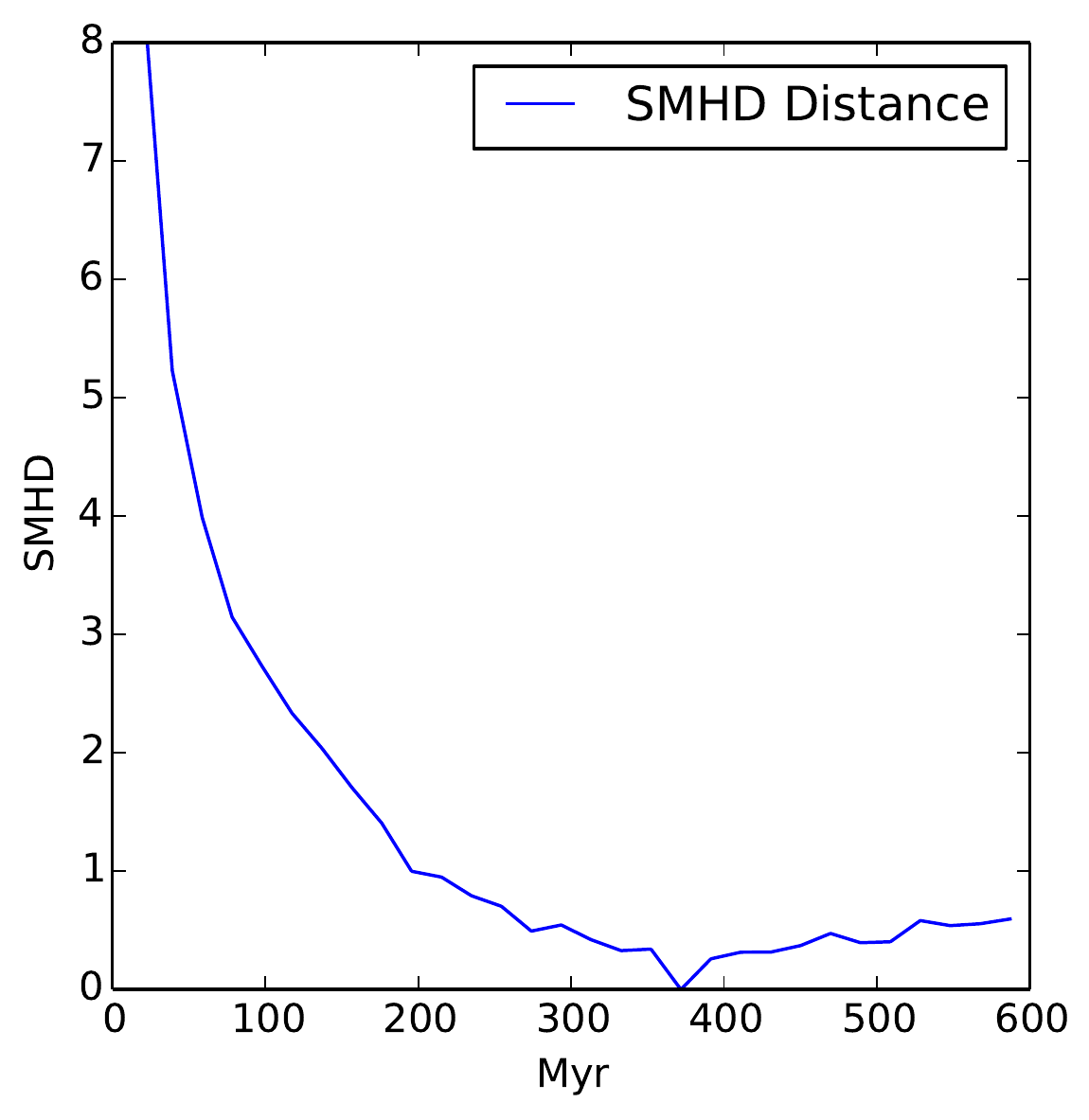}\includegraphics[width=0.25\textwidth]{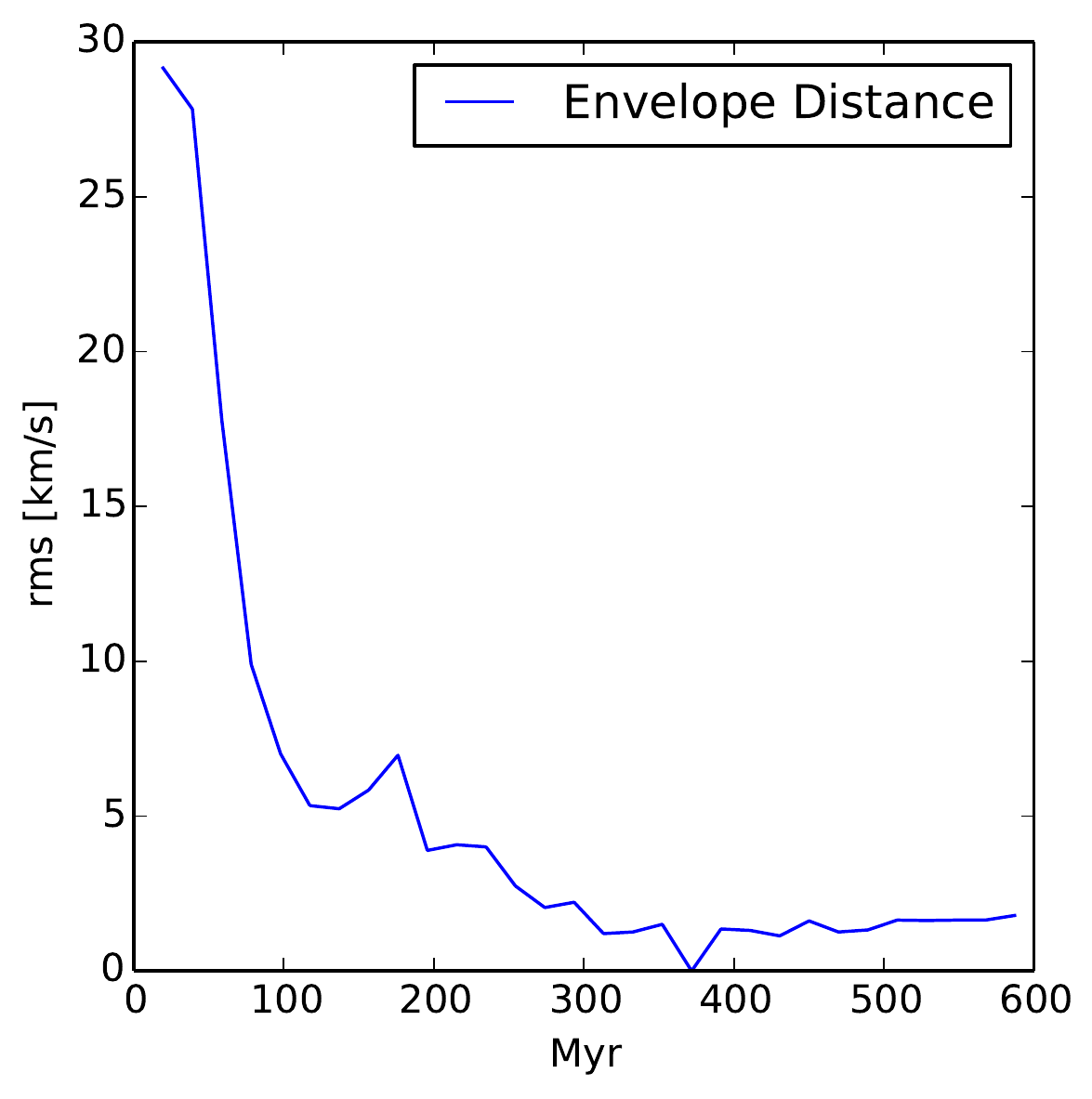}
\caption{Variation of SMHD distance (left) and envelope distance
  (right) versus time for models with $\Omega_{\rm p} = 55 \, \kms
      \kpc^{-1}$ and $\phi = 20 \degree$.  The mock ``data'' used
      are the snapshot taken at time $t=367\rm Myr$.  SMHD is calculated using all features, including envelope.}
\label{fig:stationarity2}
\end{figure}

To quantify this, we calculate the SMHD and ED between a synthetic
$(l,v)$ plot at a fiducial evolutionary time, $t=367\rm Myr$, with
synthetic $(l,v)$ plots at \textit{different} evolutionary times in the
\textit{same} simulation.
Thus all synthetic $(l,v)$ plots considered in this section come from exactly the same potential,
pattern speed~$\Omega_{\rm p}$ and viewing angle $\phi$; only the
time~$t$ is allowed to vary.
Fig. \ref{fig:stationarity2} shows the SMHD and ED between the fiducial
snapshot at $t=367\rm Myr$ and all other snapshots in
the same simulation.
At $t=0$ the gas moves on purely circular orbits and the SMHD and ED are
both very large.
As the bar is gradually turned on over the first 150 Myr of the run, the
gas settles into a distribution that approaches that of the fiducial
snapshot at $t=367\rm Myr$.
While this is happening, the SMHD and ED both decrease from their high
initial values to much lower values, reaching an approximate plateau
starting from 300 Myr.
Of course, at the special point $t=367\rm Myr$ both distances go to
zero, but the sizes of the SMHD and ED in the plateau region away from
this point provide lower bounds on the level of noise expected when we
fit only relaxed, evolved versions of our models to mock data.
Indeed, we will see below that varying one of the other parameters
$(\phi,\Omega_{\rm p})$ produces differences in SMHD and ED much greater
than the height of the plateau.

We emphasize that, although this shows that the particular models we
consider here do reach an effective steady state, the feature-fitting
algorithm and SMHD distance of Section~\ref{sec:compare} do not require
that such a steady state be reached.
In the following subsections we make use of this effective steady state
and compare mock data only to model snapshots taken at some time $t>300$
Myr long enough to allow the model to settle into a steady state; when
fitting more general models with more than one pattern
speed~\citep[e.g.,][]{fux1999} one would have to include~$t$ as a
parameter to be fit.

\begin{figure}
    \subfigure[Mock data for the \textbf{basic test}, generated using $\alpha=1$]
    {
        \includegraphics[width=0.25\textwidth]{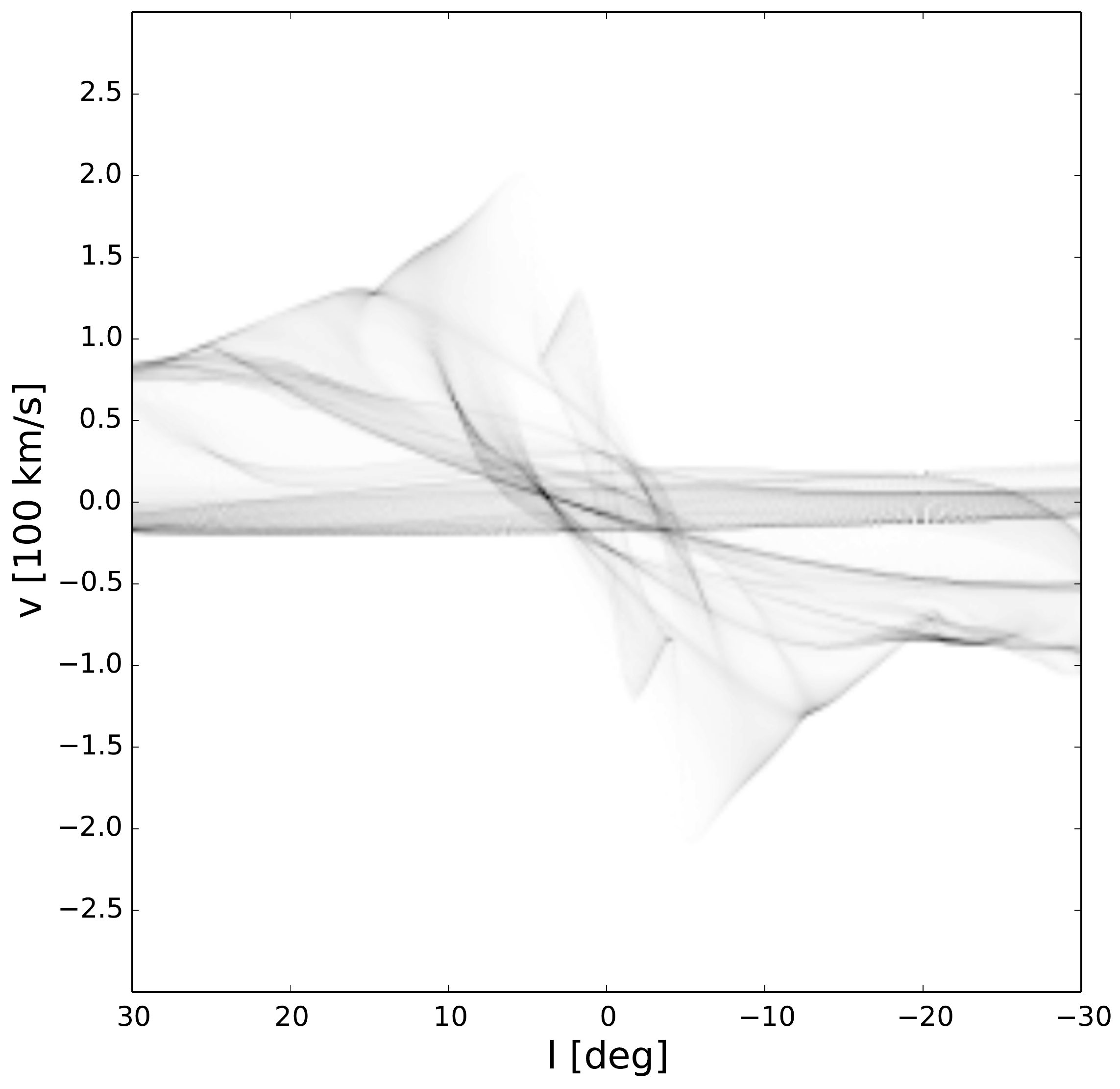}
        \includegraphics[width=0.25\textwidth]{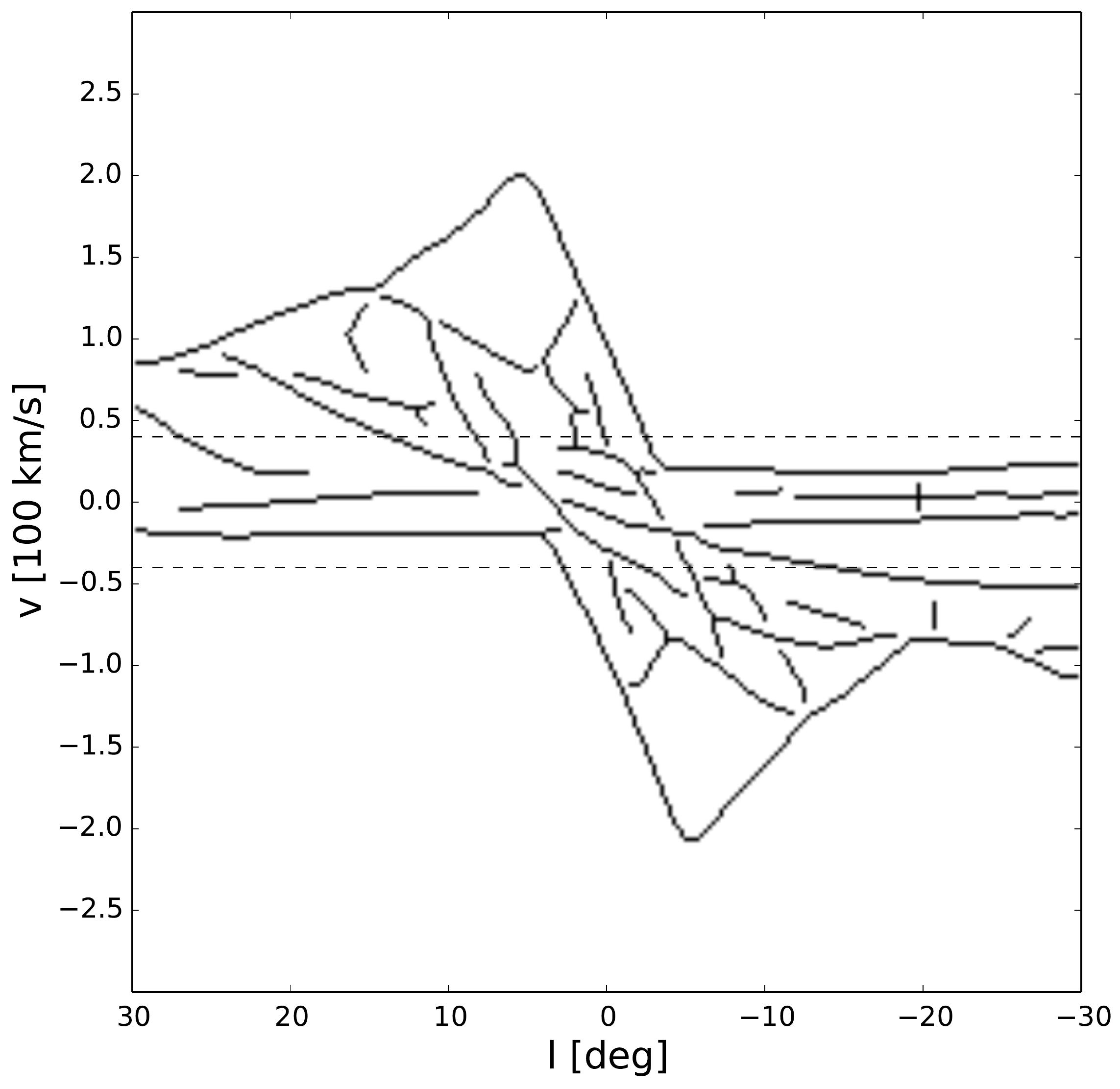} \label{fig:retrieving1a}
    }
     \subfigure[Mock data for the \textbf{alpha test}, generated using
     $\alpha=3$]
    {
        \includegraphics[width=0.25\textwidth]{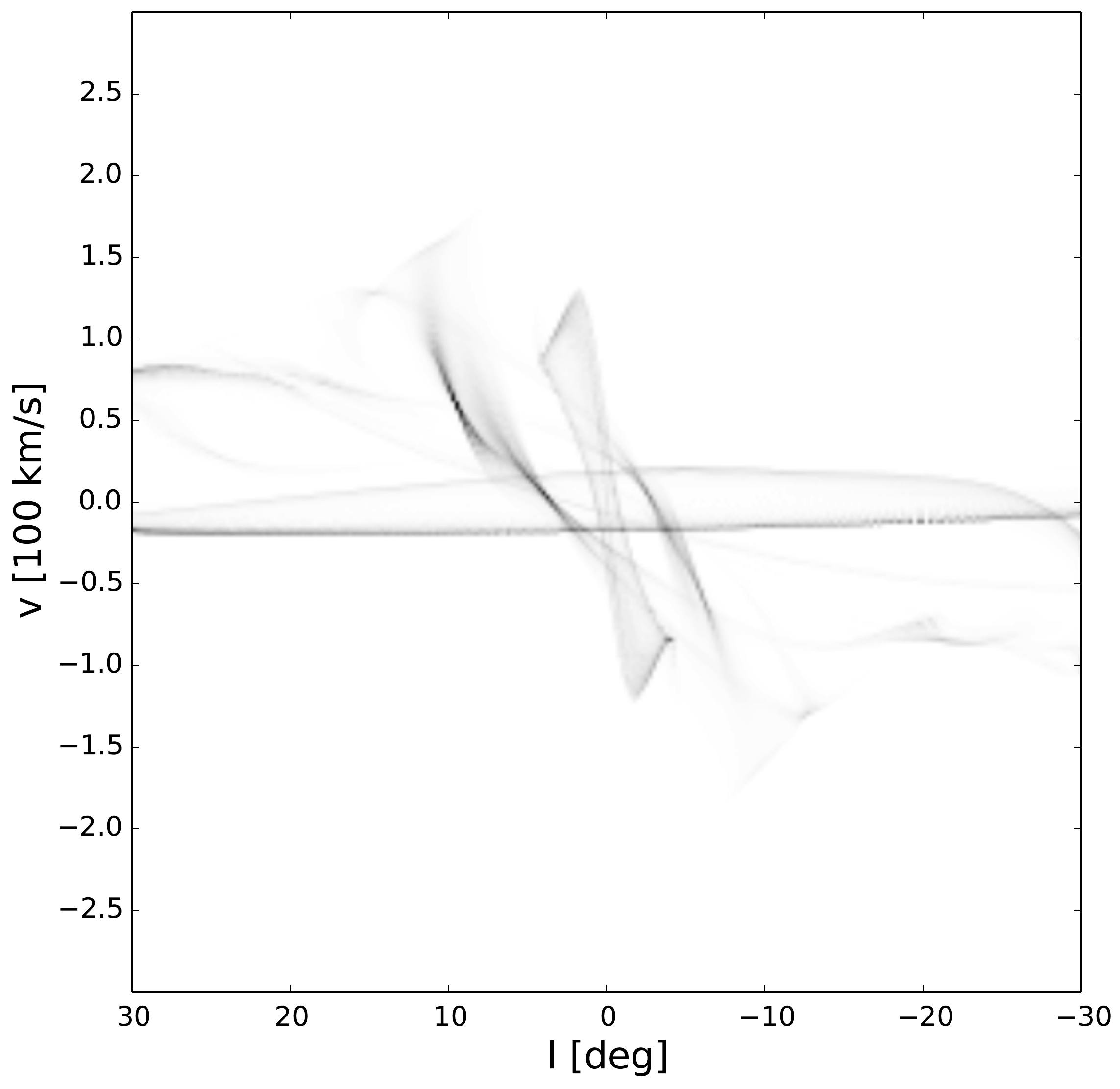}
        \includegraphics[width=0.25\textwidth]{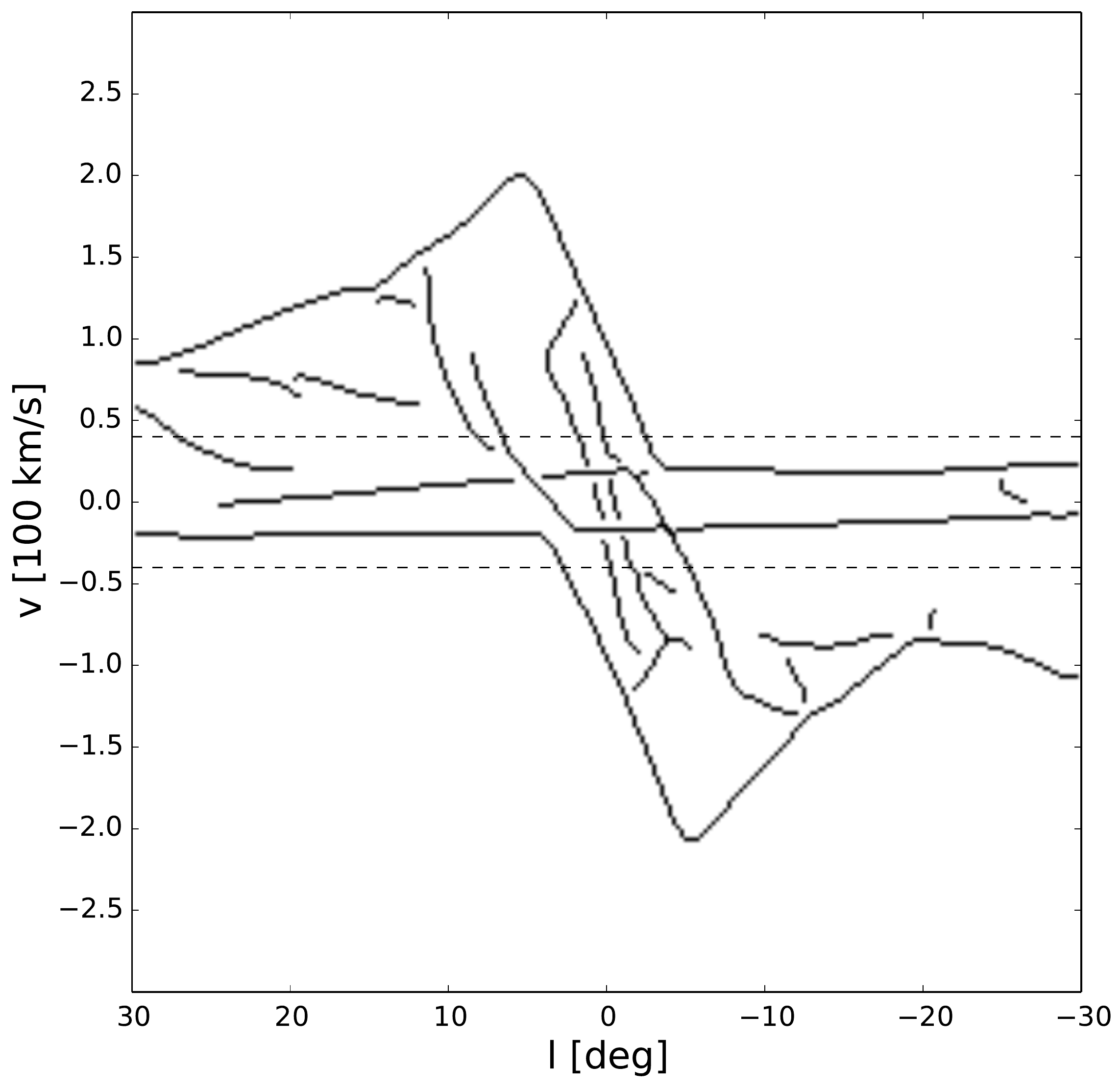} \label{fig:retrieving1b}
    }
    \subfigure[Mock data for the \textbf{contamination test}, using
    $\alpha=1$ and adding contaminating features]
    {
        \includegraphics[width=0.25\textwidth]{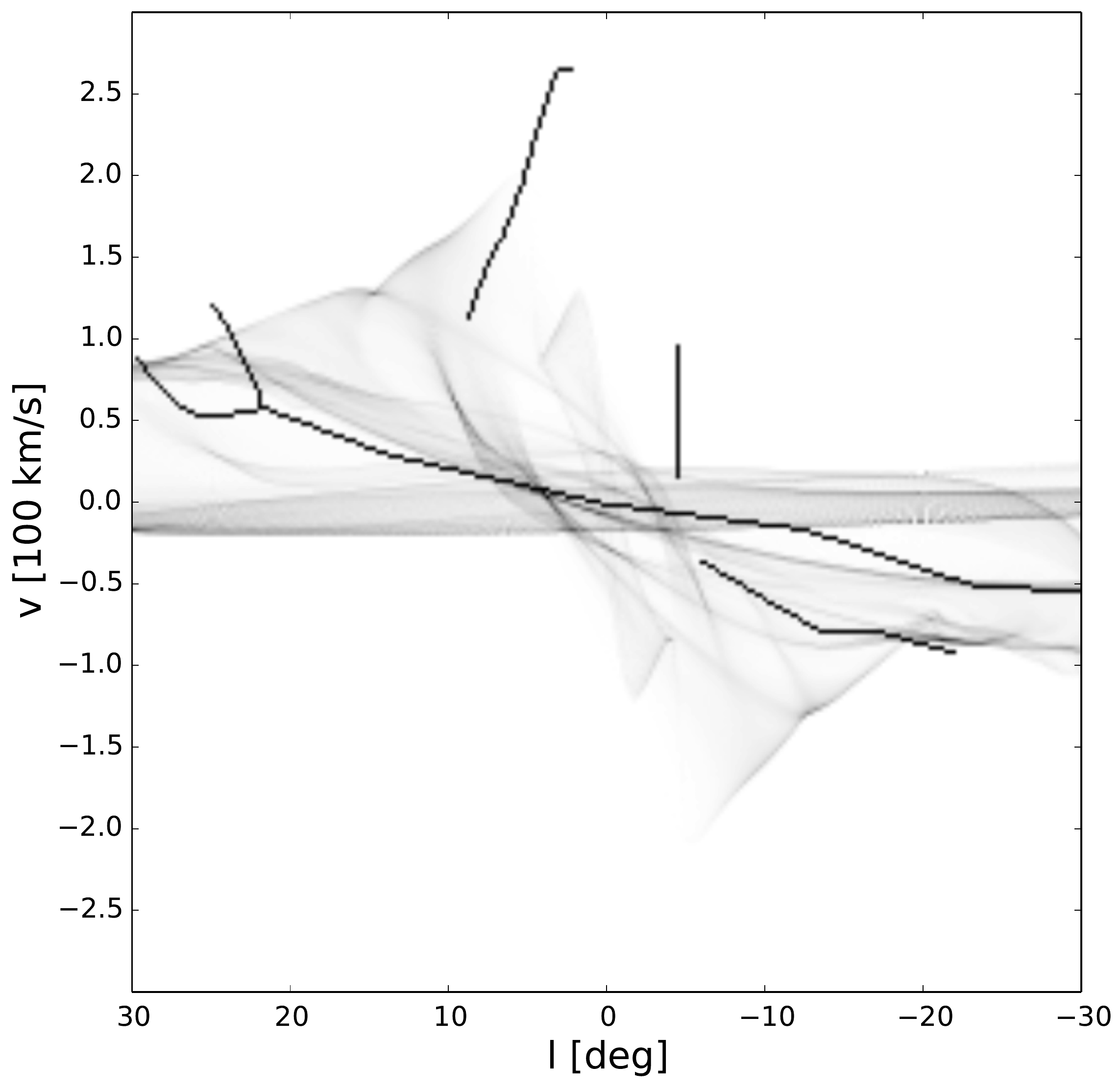}
        \includegraphics[width=0.25\textwidth]{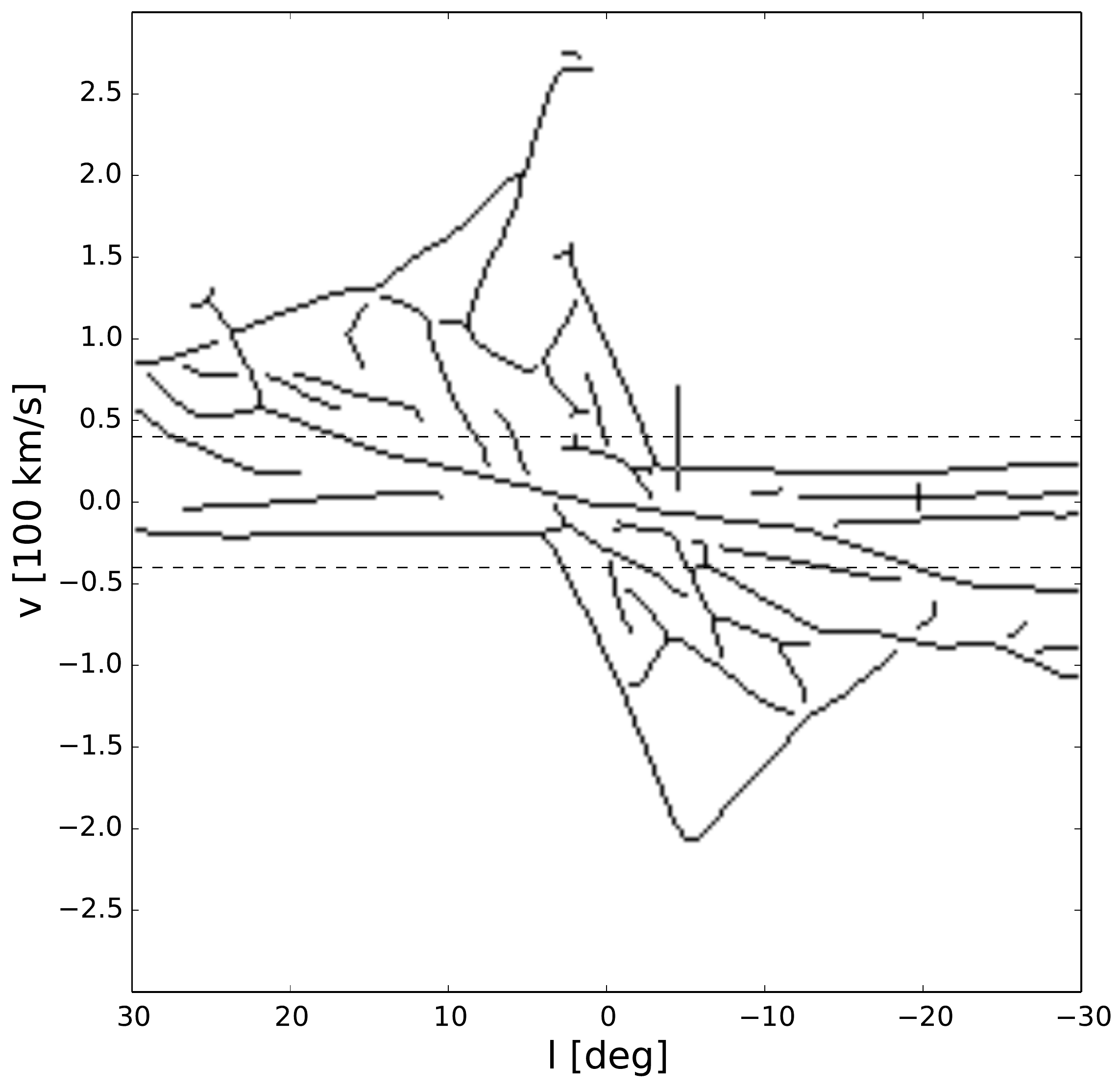} \label{fig:retrieving1c}
    }
    \caption{The mock data used in the tests of Section~\ref{sec:retrieving}.
      All data are based on our reconstruction of Englmaier \&
      Gerhard's (1999) potential with $\Omega_{\rm p} = 55 \, \kms \kpc^{-1}$,
      $\phi = 20 \degree$ taken at $t \simeq 310\,\rm Myr$.  The top row
      shows the results of projecting this model with $\alpha=1$
      in the projection law~\eqref{eq:rhogrid}.  The panel on the left shows the
      resulting $(l,v)$ distribution, while the panel on the right shows the full set of features
      extracted using the algorithm of Section~\ref{sec:findfeat}.
      The middle row shows the second dataset, projected with
      $\alpha=3$.  The bottom row shows the third dataset, projected
      with $\alpha=1$, but to which the additional contaminating
      features indicated in the panel on the left have been added.
      \label{fig:retrieving1}}
\end{figure}

\subsection{Recovering the correct model parameters}
\label{sec:retrieving}

Having eliminated time~$t$, we now turn to the more interesting question
of how reliably one can identify the model parameters $\Omega_{\rm p}$
and $\phi$.
We construct three different mock datasets, shown in
Figure~\ref{fig:retrieving1}.
All are generated from a fiducial model having pattern speed
$\Omega_{\rm p}=55\kms \rm Myr^{-1}$ viewed at $t=310$ Myr and angle
$\phi=20\degree$, but differ in how the mock data are generated from that
model.
The first is constructed by projecting the fiducial model with
$\alpha=1$ and using the algorithm of Section~\ref{sec:findfeat} to extract
features.
The second is identical to the first, but projected with $\alpha=3$.
The third is a modification of the first to which additional contaminating features
have been added by hand.

For each mock dataset we construct models having $\Omega_{\rm p}$ in the
range $20 \mhyphen 70 \, \kms \kpc^{-1}$ in steps of $2\,\kms\kpc^{-1}$
and $\phi$ in the range $0 \mhyphen 60 \degree$ in steps of $2 \degree$.
For each model we project a single snapshot taken at $t= 370\,\rm Myr$
with $\alpha=1$.
Notice that none of our models use the ``correct'' values of $t=310$ Myr
and $\Omega_{\rm p}=55\kms\rm Myr^{-1}$ from which the mock data are
generated.

\begin{figure*}
  \includegraphics[width=0.3\hsize]{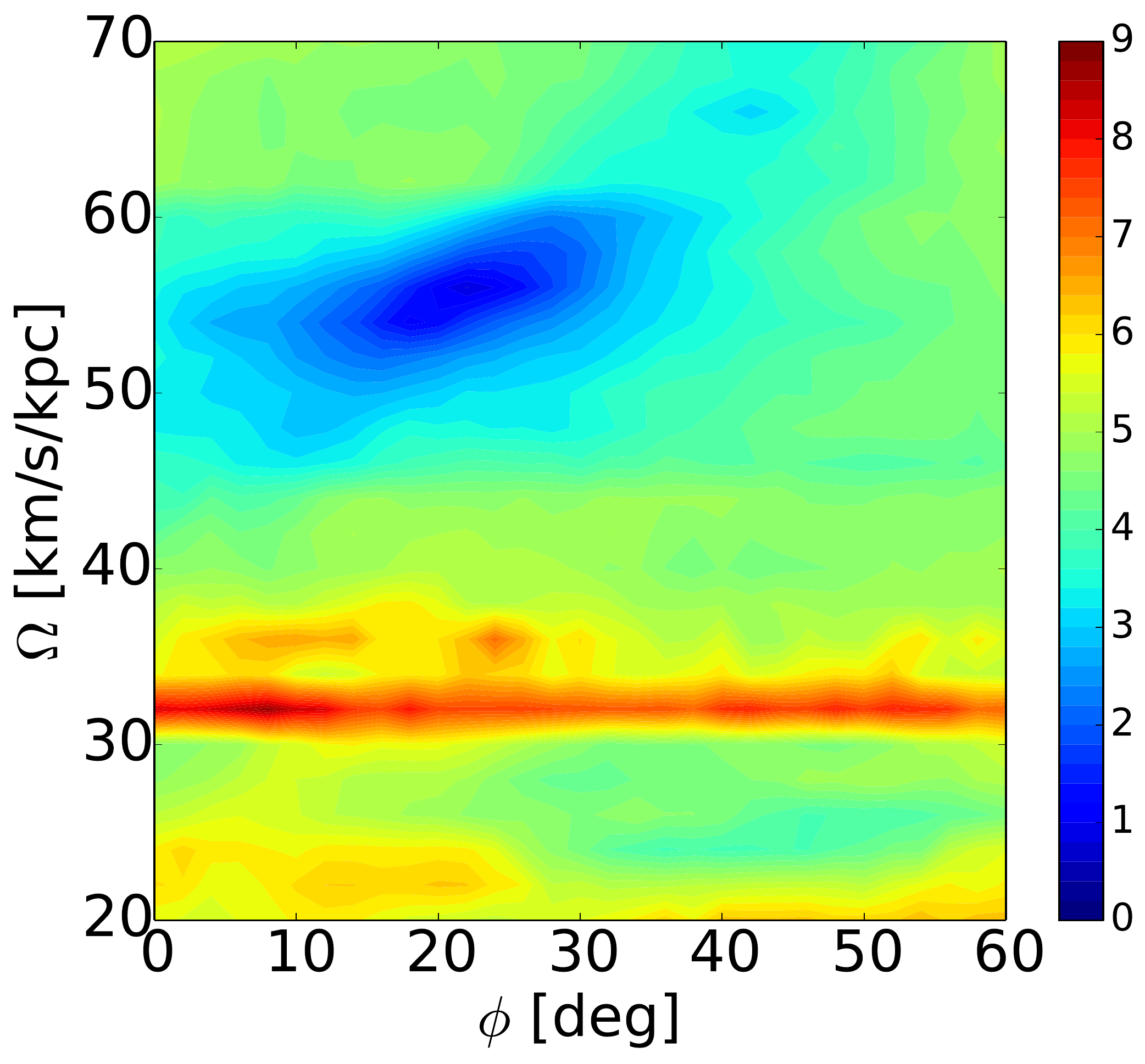}
  \includegraphics[width=0.6\textwidth]{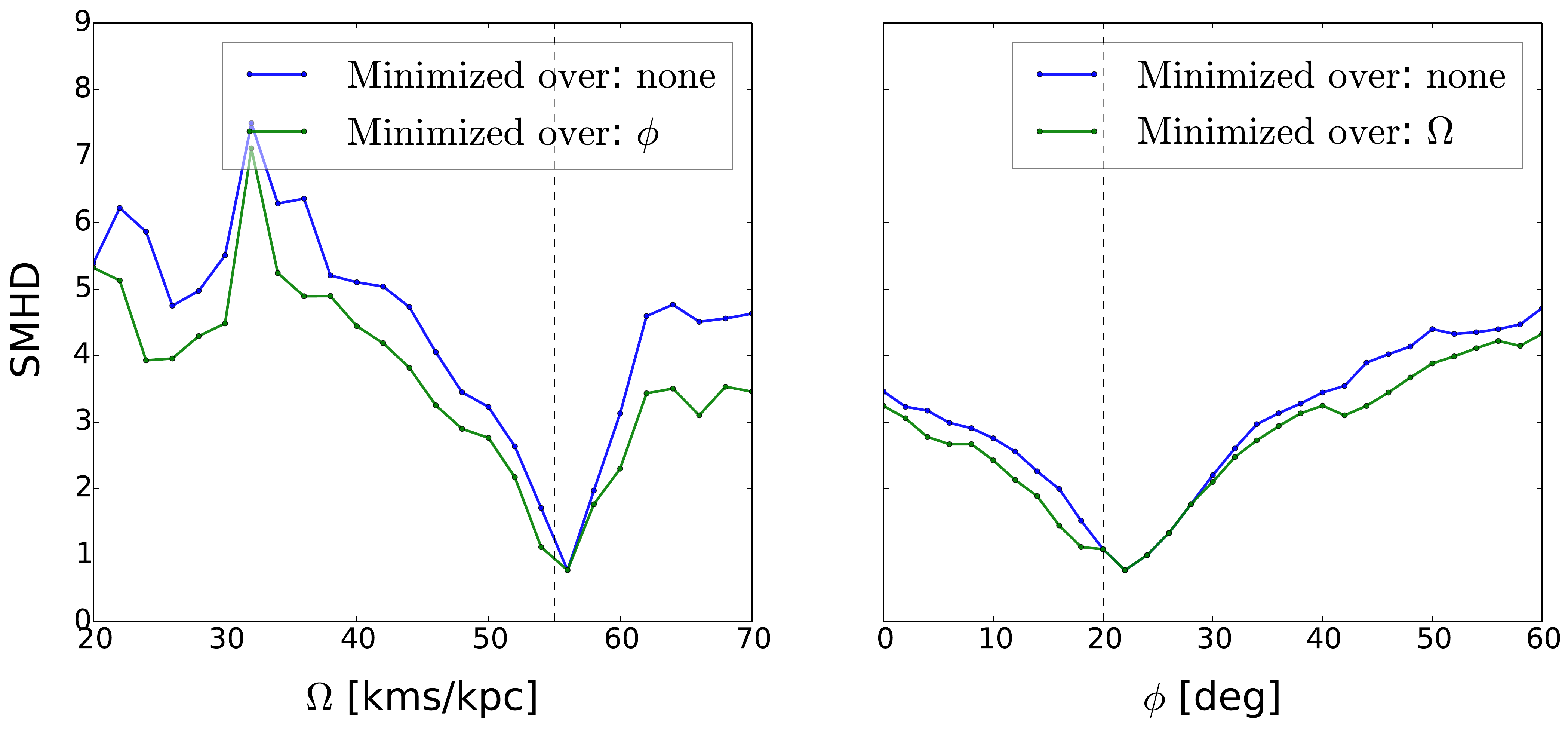} 

  \includegraphics[width=0.3\hsize]{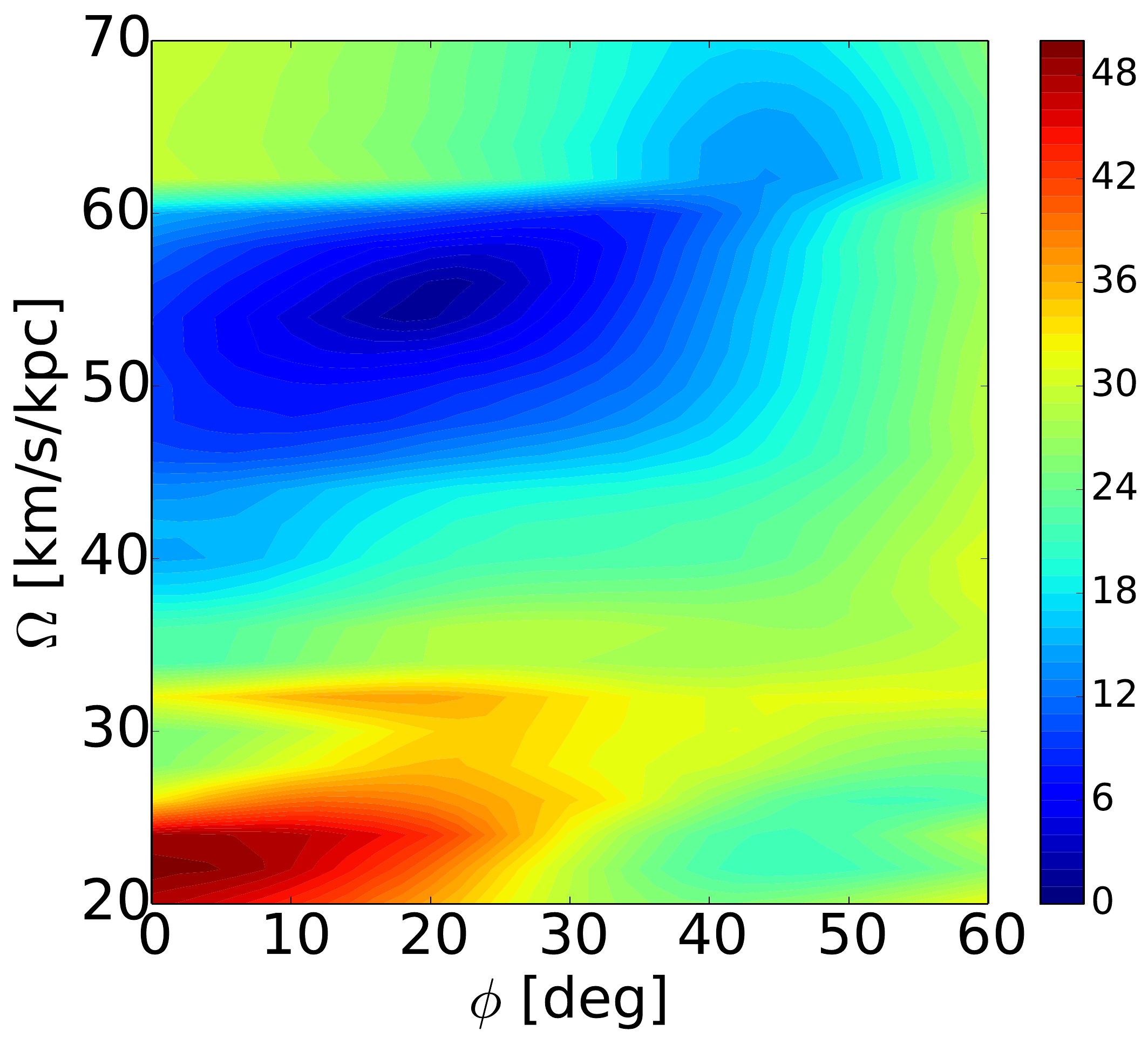}
  \includegraphics[width=0.6\textwidth]{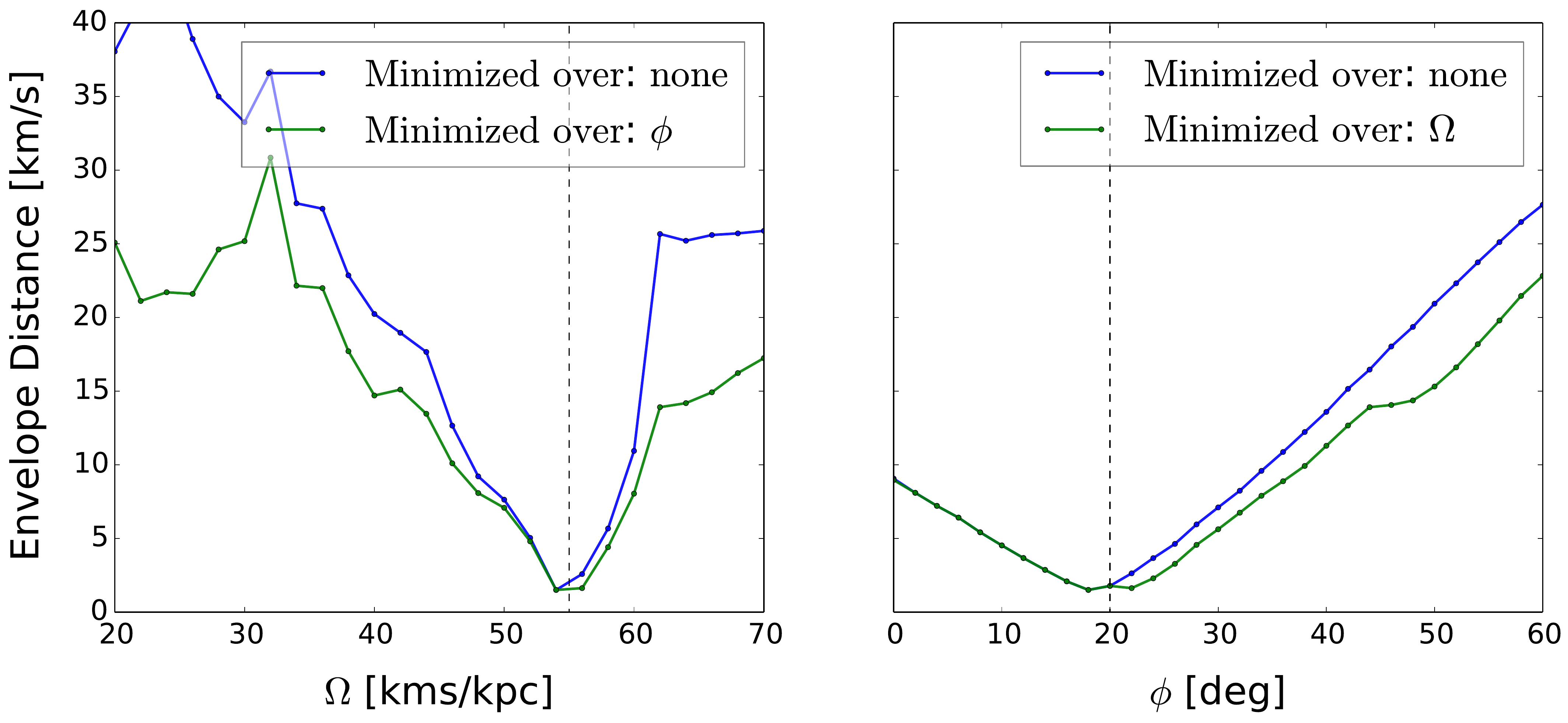} 

  \includegraphics[width=0.3\hsize]{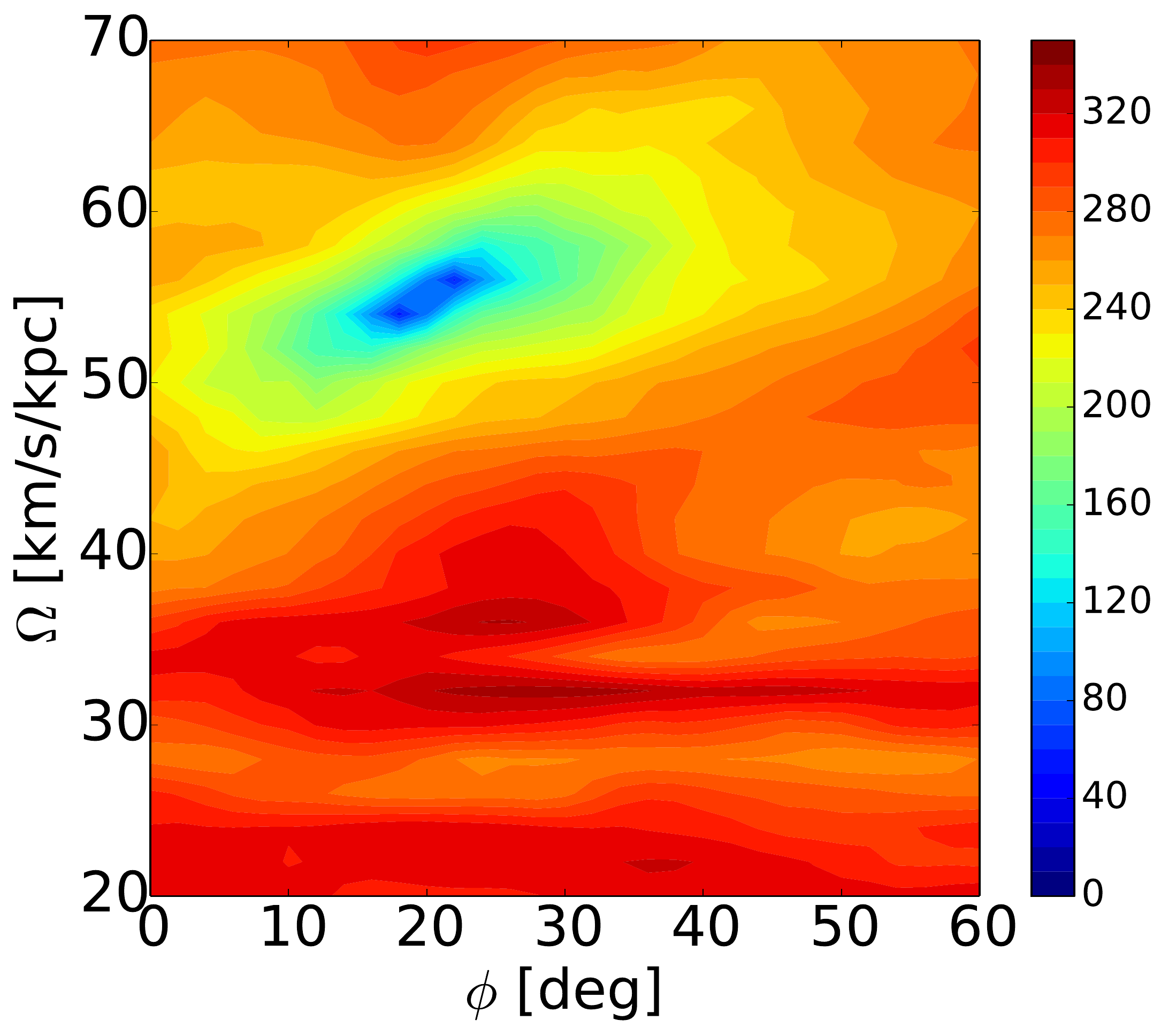}
  \includegraphics[width=0.6\textwidth]{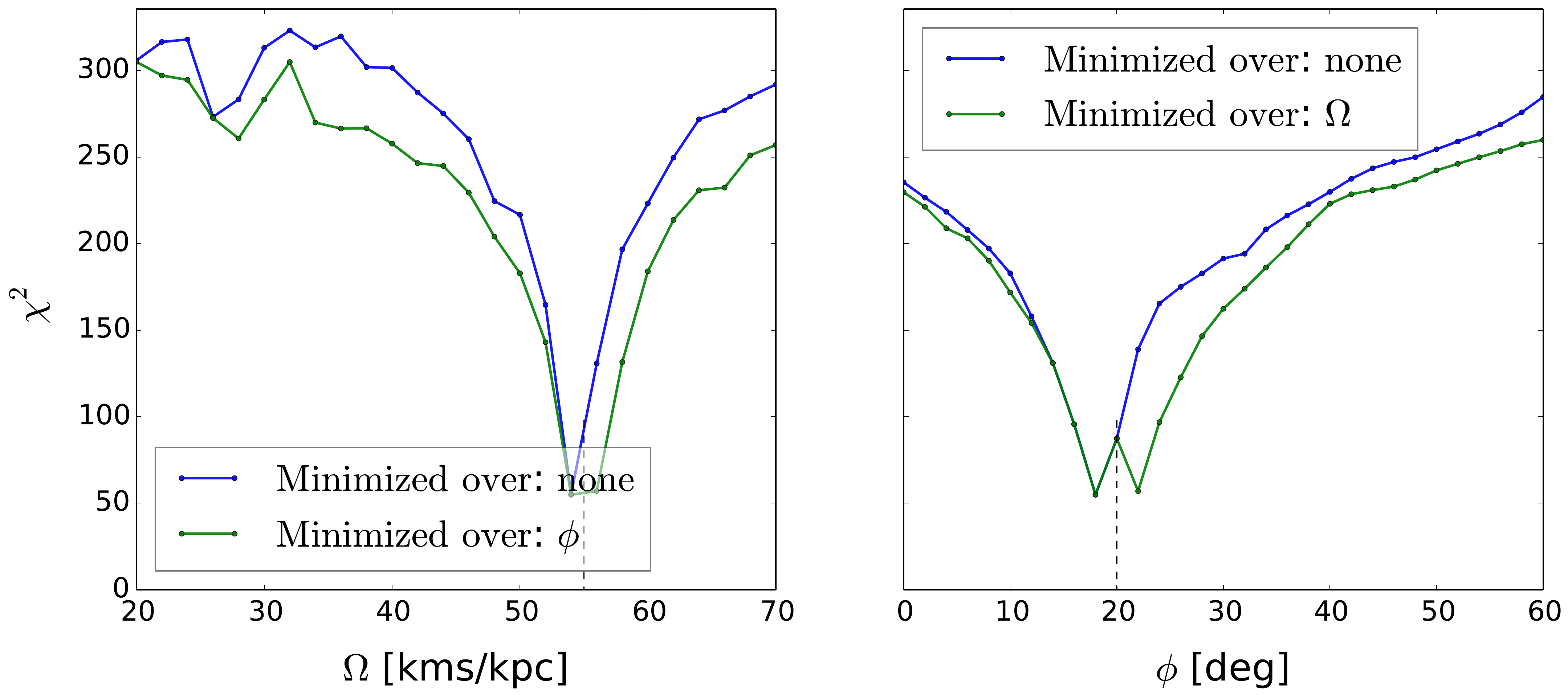} 
  \caption{Variation of SMHD (top), ED (middle) and $\chi^2$ (bottom)
    for the mock dataset constructed using $\alpha=1$ in
    Section~\ref{sec:basictest}.
    The first column of each row shows how the corresponding distance
    varies as a function of the assumed $(\phi,\Omega_{\rm p})$.
    The other two columns plot one-dimensional slices that pass through
    the location of minimum value of the distance.
    For the blue curves in the middle (right) column the slices are
    vertical (horizontal).
    The green curves are constructed by taking the distance of the
    best-fitting $\phi$ as a function of~$\Omega_{\rm p}$ (middle
    column) and the distance of the the best-fitting $\Omega_{\rm p}$ as
    a function of~$\phi$ (right column).
    The vertical lines plot the values of the parameters used to
    construct the mock data.}
  \label{fig:retrieving2}
\end{figure*}
\subsubsection{The basic test}\label{sec:basictest}

The most basic test is given by checking how reliably the model
parameters $(\phi,\Omega_{\rm p})$ are recovered when the models adopt the
correct projection law and the data are uncontaminated by misidentified
features.
The mock dataset used for this test is shown in
Fig.~\ref{fig:retrieving1a}.
Figure~\ref{fig:retrieving2} shows by how much models with assumed parameters
$(\phi,\Omega_{\rm p})$ differ from this mock dataset, as measured by
SMHD, ED, and $\chi^2$.

The plots demonstrate that when the projection law is correct (i.e.,
when there is no uncertainty associated with the chemistry of the ISM or
radiative transfer), then all three distances can be used to locate the
correct model; apart from some noisiness at low values of $\Omega_{\rm
  p}$, they all descend smoothly to a minimum at (or very close to) the
correct values of $(\phi,\Omega_{\rm p})$.
Therefore if we start away from the correct model, e.g.,, with a pattern
speed wrong by $20\, \kms\,\kpc^{-1}$, then all three distances indicate
in which direction we should move to get to the right model.

One point to note from these plots is that knowledge of the envelope
alone suffices to identify the correct model among the restricted family
of models we consider here.
This is not true in general, a point to which we return in
Section~\ref{sec:humanintuition}.

Another point is that the minimum in $\chi^2$ is much sharper than the
minimum in either the SMHD or ED.
$\chi^2$ measures the overlap between the data and model densities.
Adjusting $(\phi,\Omega_{\rm p})$ moves density around the $(l,v)$
plane.
If the model density is similar to the observed one, but shifted in the
$(l,v)$ plane, then the overlap between the two is low and $\chi^2$
tells us that the model is bad, despite the fact that a small adjustment
to the model can lead to a significant improvement.
This happens because $\chi^2$ compares only contents of the same bin,
and does not take into account any cross-bin information.
So, the width of the minimum in the $\chi^2$ is essentially measuring
the width of the features.
We discuss this further in Section~\ref{sec:humanintuition}.
 
None of the distances reach the value of zero because, as we
noted above, the model used to produce the mock data set is not
included, and also the evolutionary time between the two is different.
For the SMHD the value of the minimum is above the value of the plateau
in Fig.\ref{fig:stationarity2}, as are variations produced in the SMHD
by all but the smallest variations of the parameters.
Thus for this particular set of models we are justified in asserting
that the time is a parameter that can be neglected if we are interested
in recovering the value of the pattern speed and of the angle.

\begin{figure}
  \includegraphics[width=0.5\textwidth]{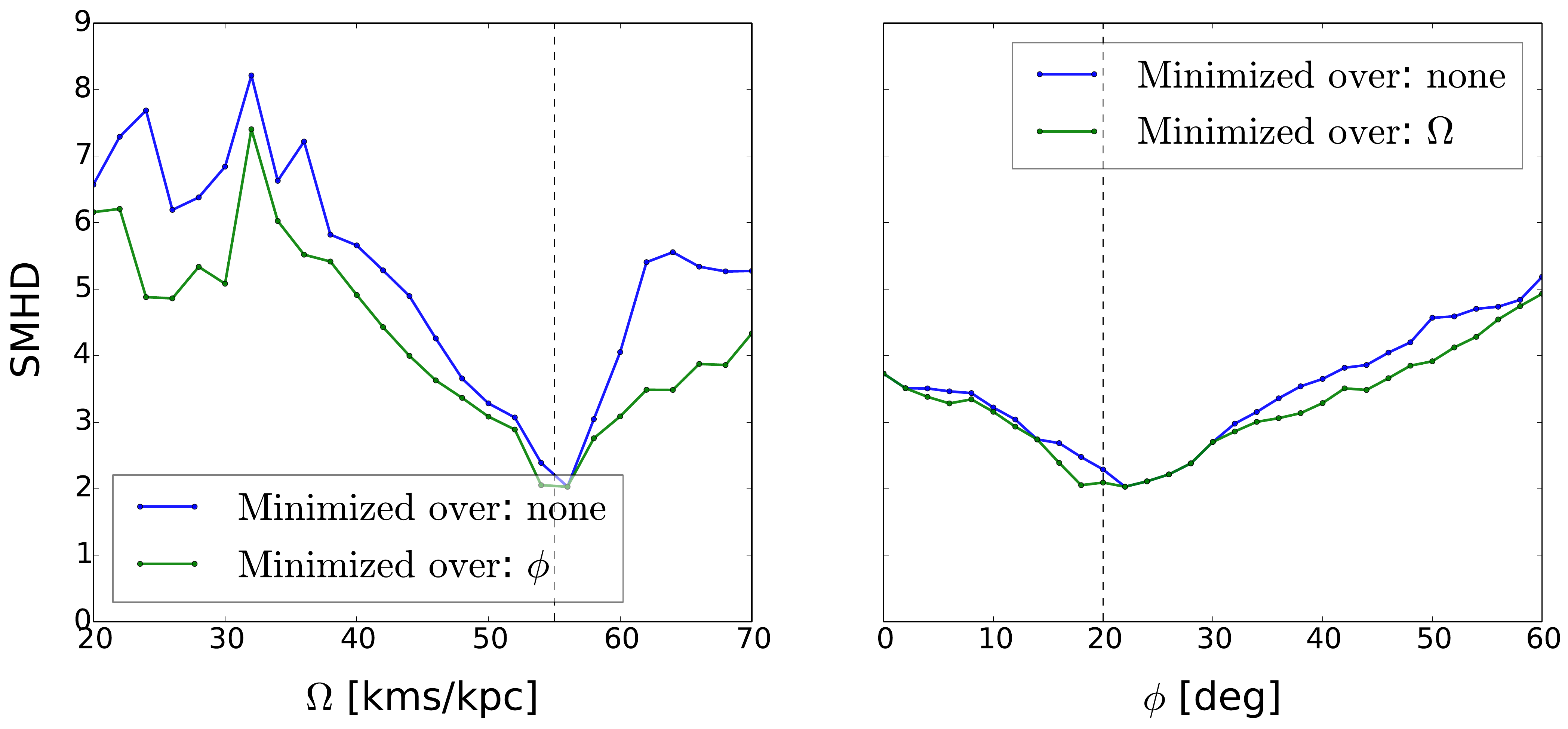} 
  
  \includegraphics[width=0.5\textwidth]{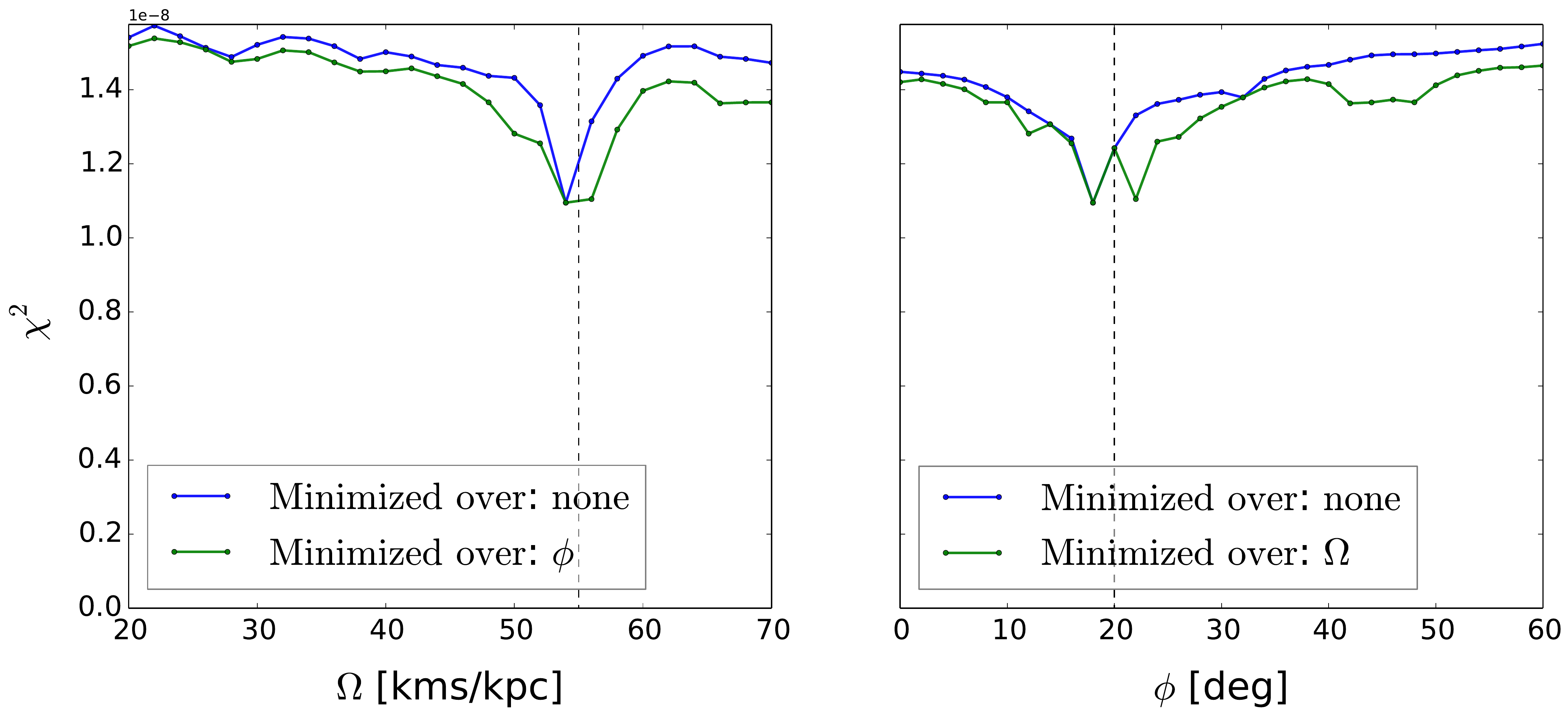} 

\caption{Variation of SMHD (top) and $\chi^2$ (bottom) for the test of
  the projection law described in Section~\ref{sec:testalpha}.  The
  curves have the same meaning as in Fig.~\ref{fig:retrieving2}.}
\label{fig:retrieving3}
\end{figure}
\begin{figure}
\centering
\includegraphics[width=0.23\textwidth]{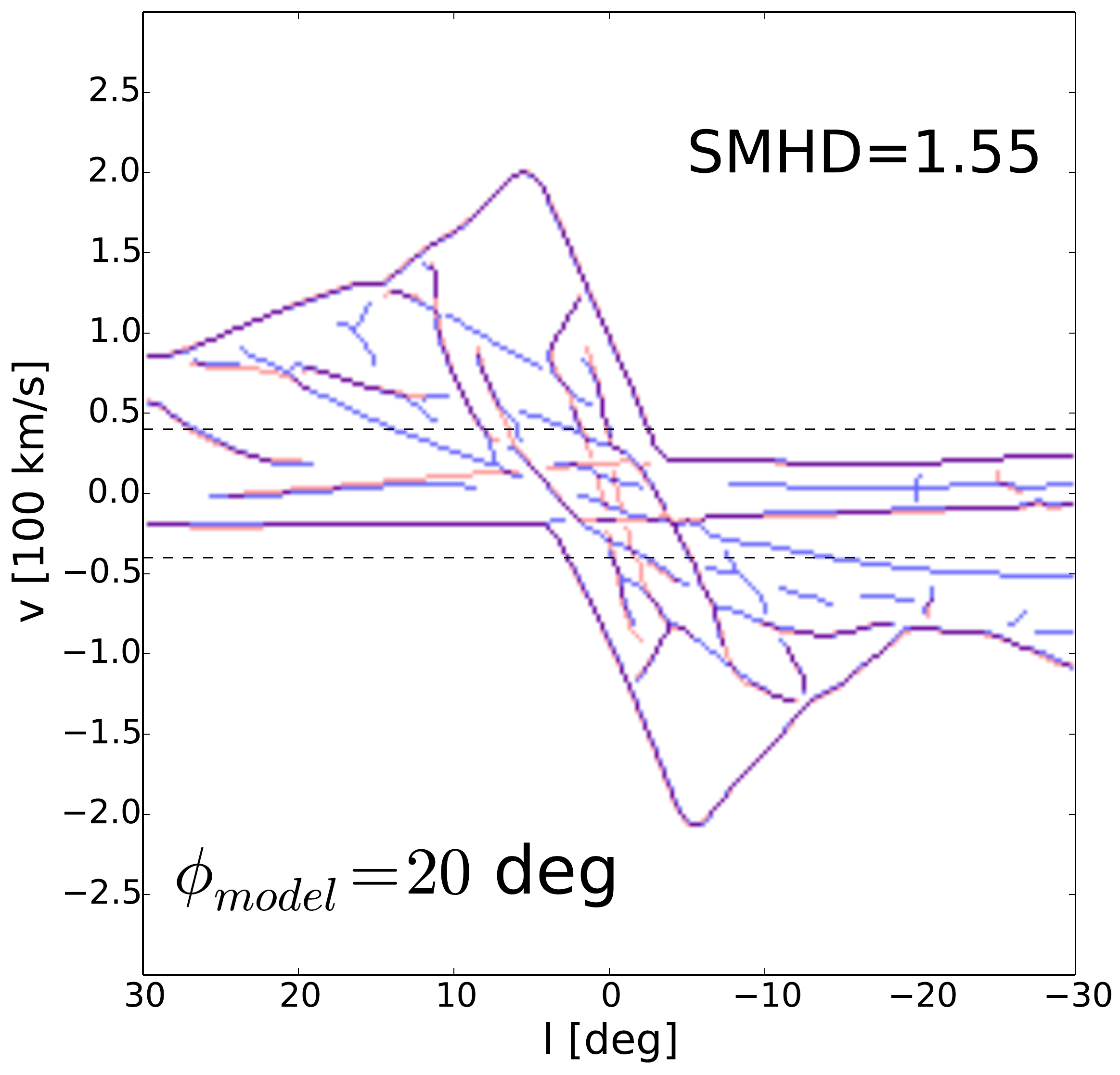}

\includegraphics[width=0.23\textwidth]{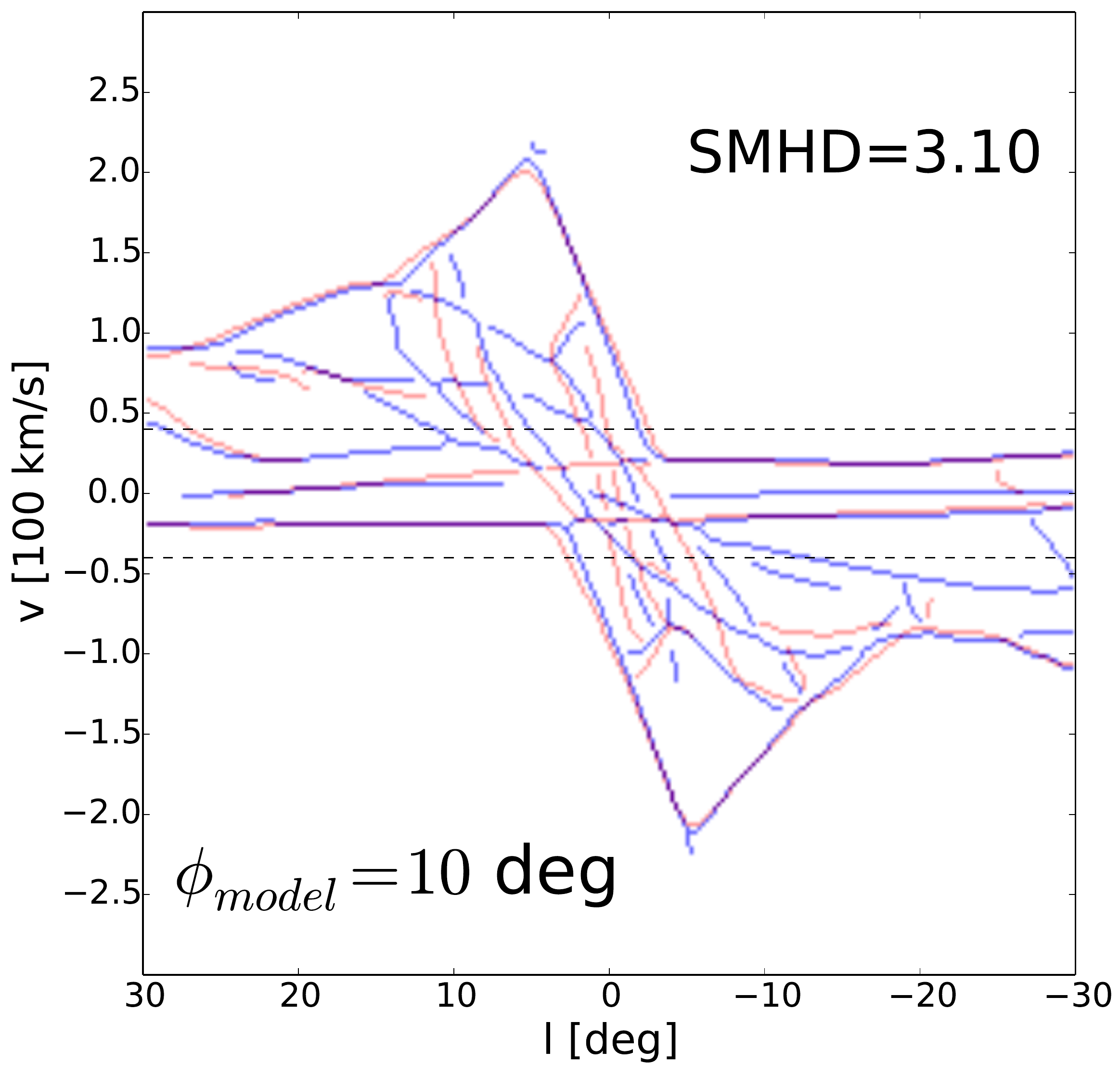}
\includegraphics[width=0.23\textwidth]{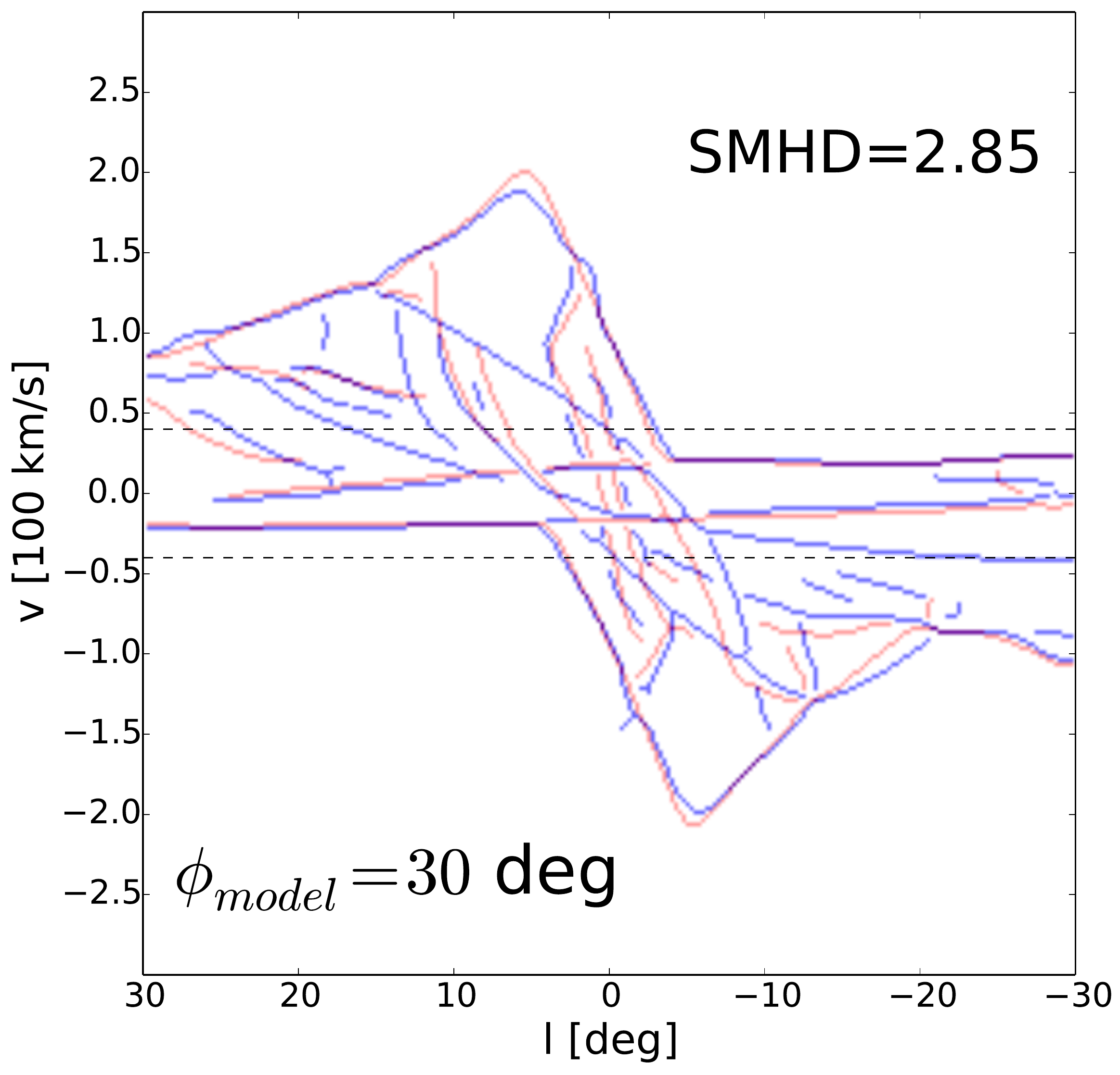}
\caption{Mock data for the alpha test (Figure~\ref{fig:retrieving1},
  panel (b)) overlaid on three models. 
  The red curves show the data features, the blue the data.
  The purpose is to show how the features and the SMHD change when we
  move away from the correct model. The central picture is the model with $\Omega_{\rm p} = 54 \, \kms
  \kpc^{-1}$ and the correct value for the angle $\phi = 20 \degree$.  We see that the features of the
  mock data are almost a subset of the features of the model (but not
  quite). The other two pictures show models with same $\Omega_{\rm p}$
  but $\phi = 10 \degree$ and $\phi = 30 \degree$.  In these, all the
  features are moved slightly in the $(l,v)$ plane  with respect to
  their positions in the middle model. This movement is one of the
  reasons that makes 
  $\chi^2$ unsuited for matching the
  longitude-velocity diagrams. On each figure the values of SMHD between the blue and red models are
  shown. \label{fig:showaway}}
\end{figure}

\subsubsection{Effects of the projection law}
\label{sec:testalpha}
The second test we run is the {\bf alpha test}. In this test, we
change the value of the exponent in the projection
law~\eqref{eq:rhogrid} from $\alpha=1$ to $\alpha=3$ when building our
mock dataset.
All the other parameters retain the values they had in the basic
test. Since a synthetic $(l,v)$ plot depends on how the projection of
the gas is made, so too do the features,  even though, as shown in
Fig. \ref{fig:models3}, they do so only weakly when the projection is
varied within physically plausible limits. 
The purpose of this test is to test whether we can recover the correct
parameters if the models are built using the wrong projection law, 
which we can think of as a crude test of how sensitive the fit is to
assumptions about ISM chemistry and radiative transfer processes.

When $\alpha=3$, the brightness temperature of the mock dataset does not
depend linearly on the total column density, but on its third power
instead.
This is quite an extreme choice, probably well over the edge of the
physically reasonable values.
The mock dataset for this test is shown in \ref{fig:retrieving1b}. We
see that the features change significantly from the basic test.
In particular, some features disappear and are not visible anymore,
while some features are enhanced.
Features that disappear tend to be associated with velocity crowding,
while features that survive tend to be associated with real
overdensities in the gas distribution.
The features for the $\alpha=3$ are almost, but not exactly, a subset of
the features for $\alpha=1$.
Therefore in this test we are trying to match a mock dataset with much
fewer features than our models have.

Fig.\ref{fig:retrieving3} shows the result of the fitting.
Again the model is clearly identified by the minimum SMHD, though the
minimum is not as deep as found in the basic test when the correct
$\alpha$ was adopted.
This happens because the best model now contains extra features that are
not contained in the mock dataset.
Nevertheless, the SMHD still does quite a good job in identifying the
correct model.
This test illustrates the robustness of the method: it shows that if we
use the wrong radiative transfer approximation we should still be able
to retrieve the correct model.

In this test, $\chi^2$ is clearly outperformed by SMHD (compare top and bottom rows in Fig. \ref{fig:retrieving3}).
$\chi^2$ does exhibit a minimum around the parameters of the correct
model, but it is weaker than the minimum of the SMHD.
$\chi^2$ is much flatter than SMHD when we are far from the correct model, 
with minima appearing in $\chi^2$ at $\phi\simeq 45 \degree$ in regions,
while in the same regions SMHD points to the correct model.
The explanation for this is that $\chi^2$, by being heavily 
dependent on the intensities at each point, can fail to recognise
that the overall morphology is similar. 

As a by product of the fact that for $\alpha=3$
the features diminish, this test shows that if for some other reason we
fail to identify some features in the data, then we might still be able
to find the correct model.
In Fig. \ref{fig:showaway} we overlay the mock data of the alpha test
with the correct parameters model and with two models that have one
parameter, the angle, different from the correct value.

We do not plot the results for ED; they are unchanged from
Fig.~\ref{fig:retrieving2}, as one might expect.

\subsubsection{Effects of contamination}
\label{sec:contamination}

\begin{figure}
	\includegraphics[width=0.5\textwidth]{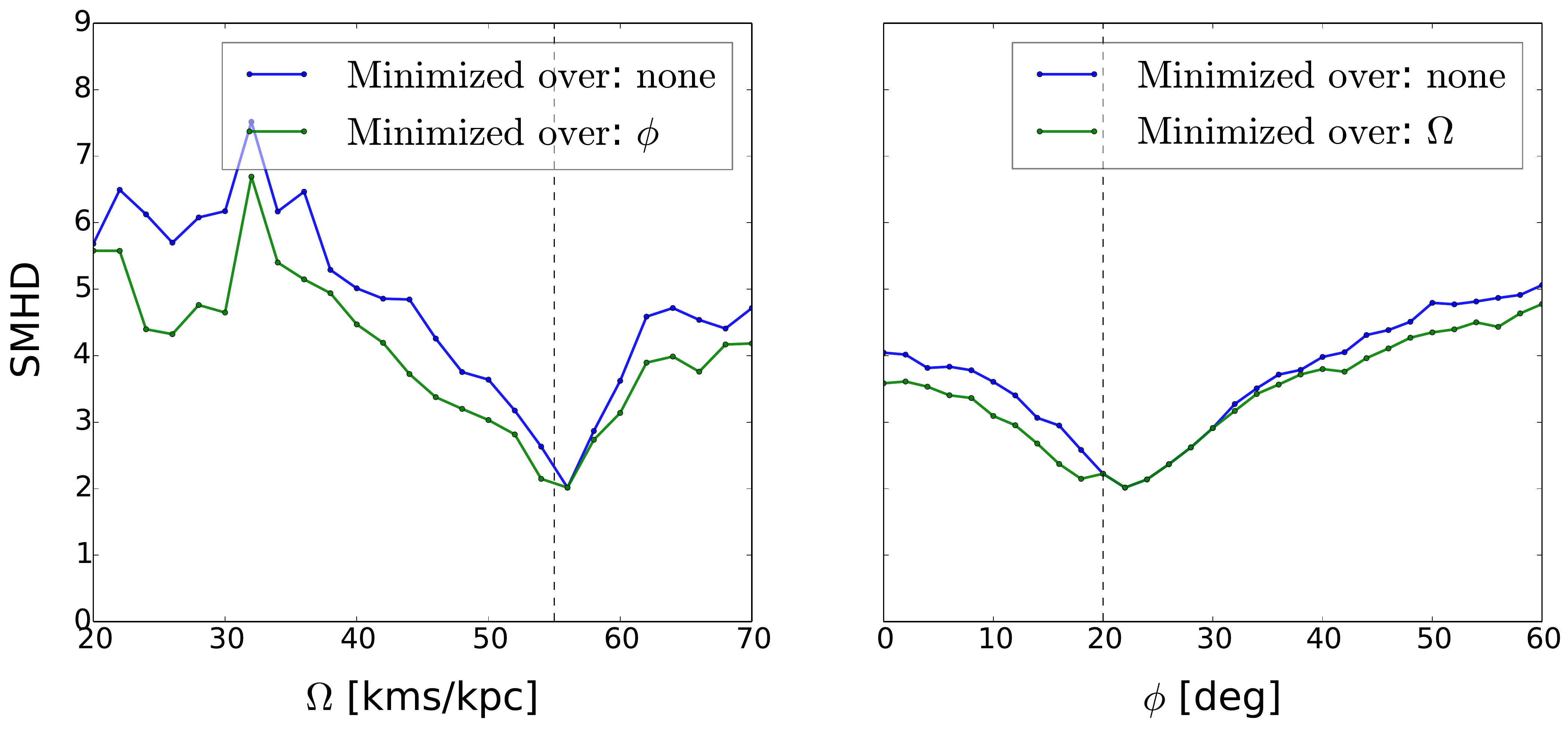}
     \includegraphics[width=0.5\textwidth]{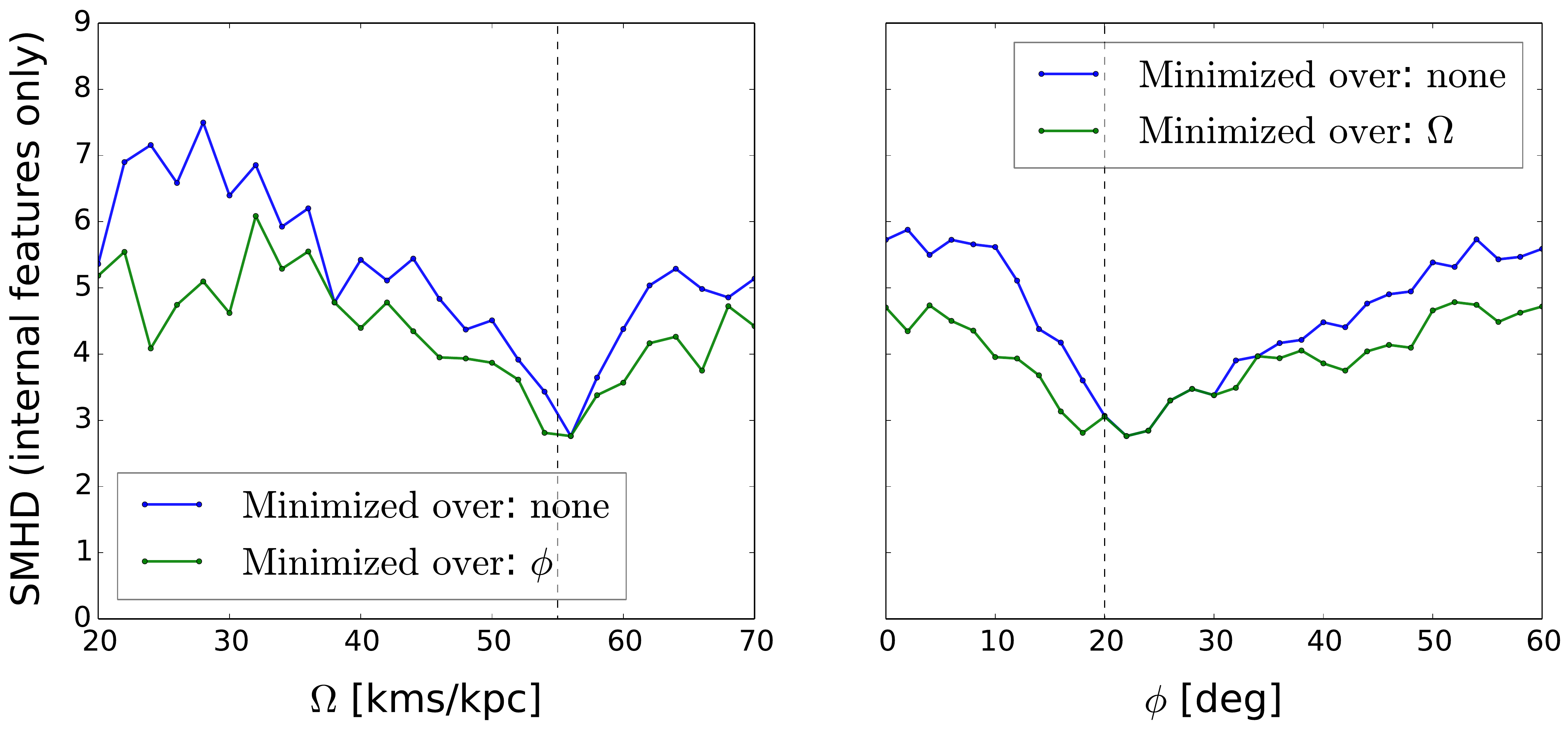}
    \caption{Variation of SMHD in the contamination test
  (Section~\ref{sec:contamination}). In the top row the SMHD is calculated on all features, while in bottom row SMHD is calculated only on internal features (i.e., excluding the envelope). Colours mean the same as in Fig. \ref{fig:retrieving2}.  \label{fig:retrieving4}}
\end{figure}

The last we run is the {\bf contamination test}. 
The mock dataset for this is built by manually adding some extra features
on top of the mock dataset for the basic test and it is shown in
Fig.\ref{fig:retrieving1c}. 
The purpose is to test the robustness of the method against the
presence of spurious features in the data. 
These could represent some features that have been included in the
data but are not really wanted, for example because they are caused by
effects not taken into account by the models
(if the connecting arm were due to magnetic fields and we do not
include these in our simulations, this would count as contamination in
this case). 
Fig. \ref{fig:retrieving4}, top row, shows the results of this test. These are
similar to the results for the alpha test, thus showing robustness of
the method against presence of contamination.
This test can be viewed as adding features as opposed to the alpha
test were we are removing features. Thus the method is robust both
against adding extra features and removing good ones. 

\subsubsection{Variations on SMHD}

In Fig.~\ref{fig:retrieving4}, bottom row, we show the results of using the SMHD
{\it without} the envelope using the contaminated dataset. 
It shows that even if one considers \textit{only} the internal
features, the correct model can still be identified, although the SMHD
becomes more noisy.  So, for this particular class of models, the
internal features alone contain enough information to identify the
correct model, albeit not as well as the envelope alone.

Finally, we have tested how the SMHD performs compares with
unsymmetrized versions of the MHD.
Given a choice between the two possibilities of using the MHD, the
unsymmetrized version performs better than the symmetrized version
either in the alpha test or in the contamination test, but not in both.
At worst, the unsymmetrized MHD displays a shallower minimum, which is
more difficult to identify, and weak secondary minima can appear.
This agrees with our considerations in Sect. \ref{calcdist}.
The symmetrized version provides a compromise able to handle a wider
range of situations.

\subsection{Behaviour for families of models that are far from the
  fiducial model}
\label{sec:humanintuition}
Finally, as a more realistic test, we consider what happens when the
models we search over are very different from the model from which the
data are generated.
We use mock data generated from the model given in
Appendix~\ref{appendix1}, with pattern speed $\Omega_{\rm p}= 48\,
\kms \kpc^{-1}$ and bar angle $\phi=30 \degree$, projected
with~$\alpha=1$.
As the form of the underlying potential of this galaxy model is very
different to the Englmaier \& Gerhard (1999) potentials that we try to
fit to it, we do not expect these parameters to be retrieved
correctly. 
The aim is instead to compare the quality of the best fits to the
mock data according to the SMHD, ED and $\chi^2$ distances.

\begin{figure}
\includegraphics[width=0.25\textwidth]{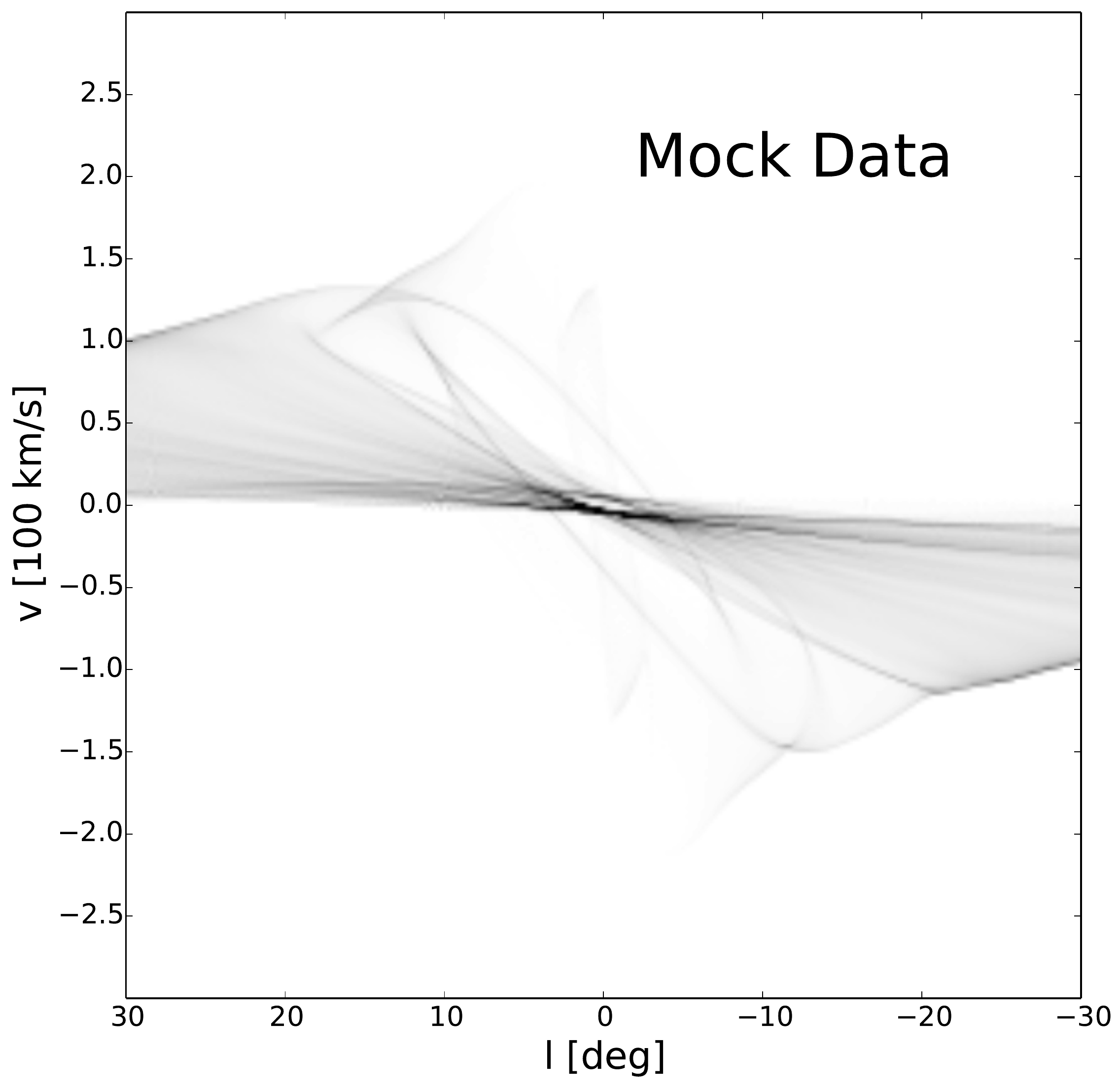}\includegraphics[width=0.25\textwidth]{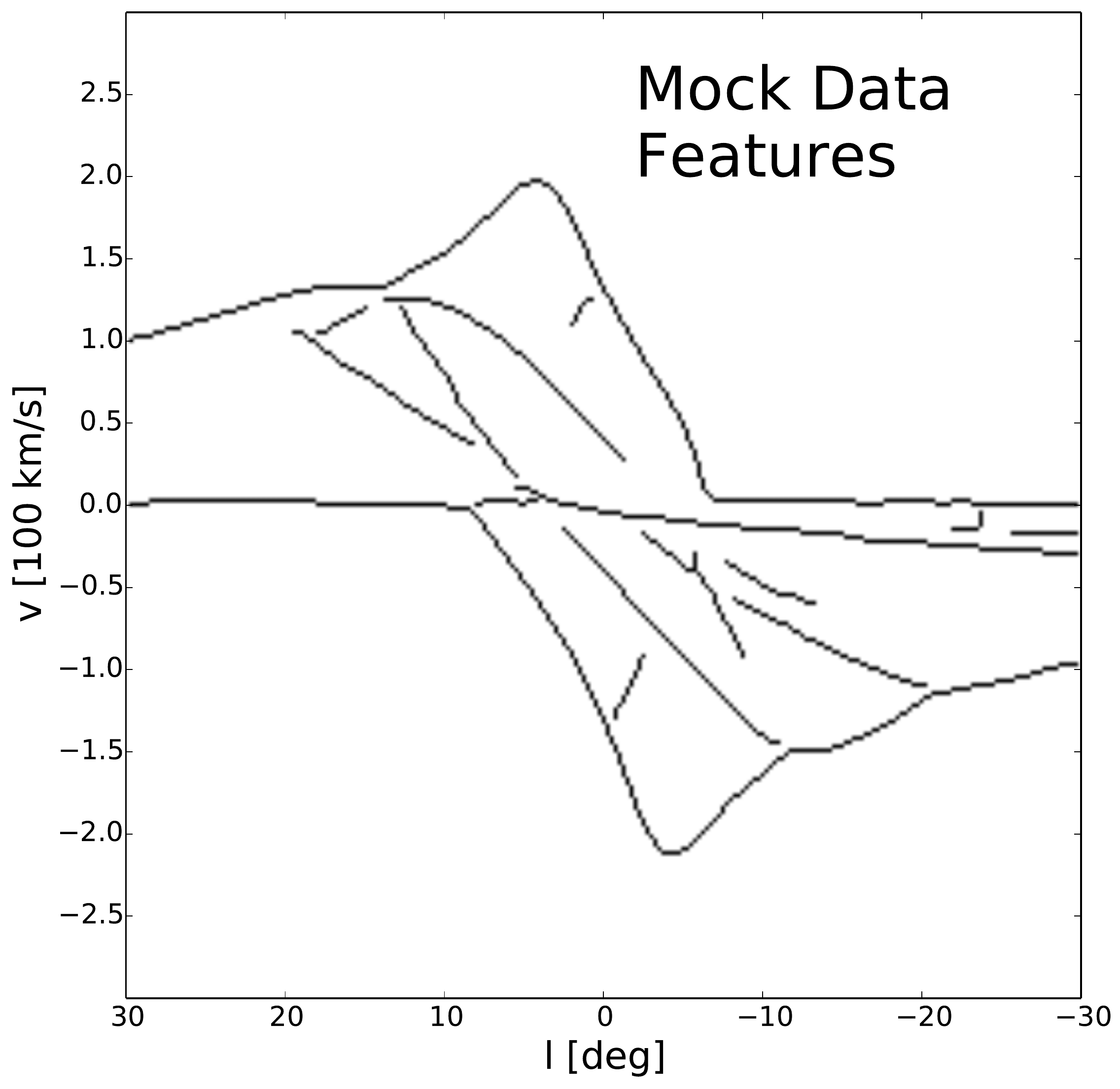}

\includegraphics[width=0.25\textwidth]{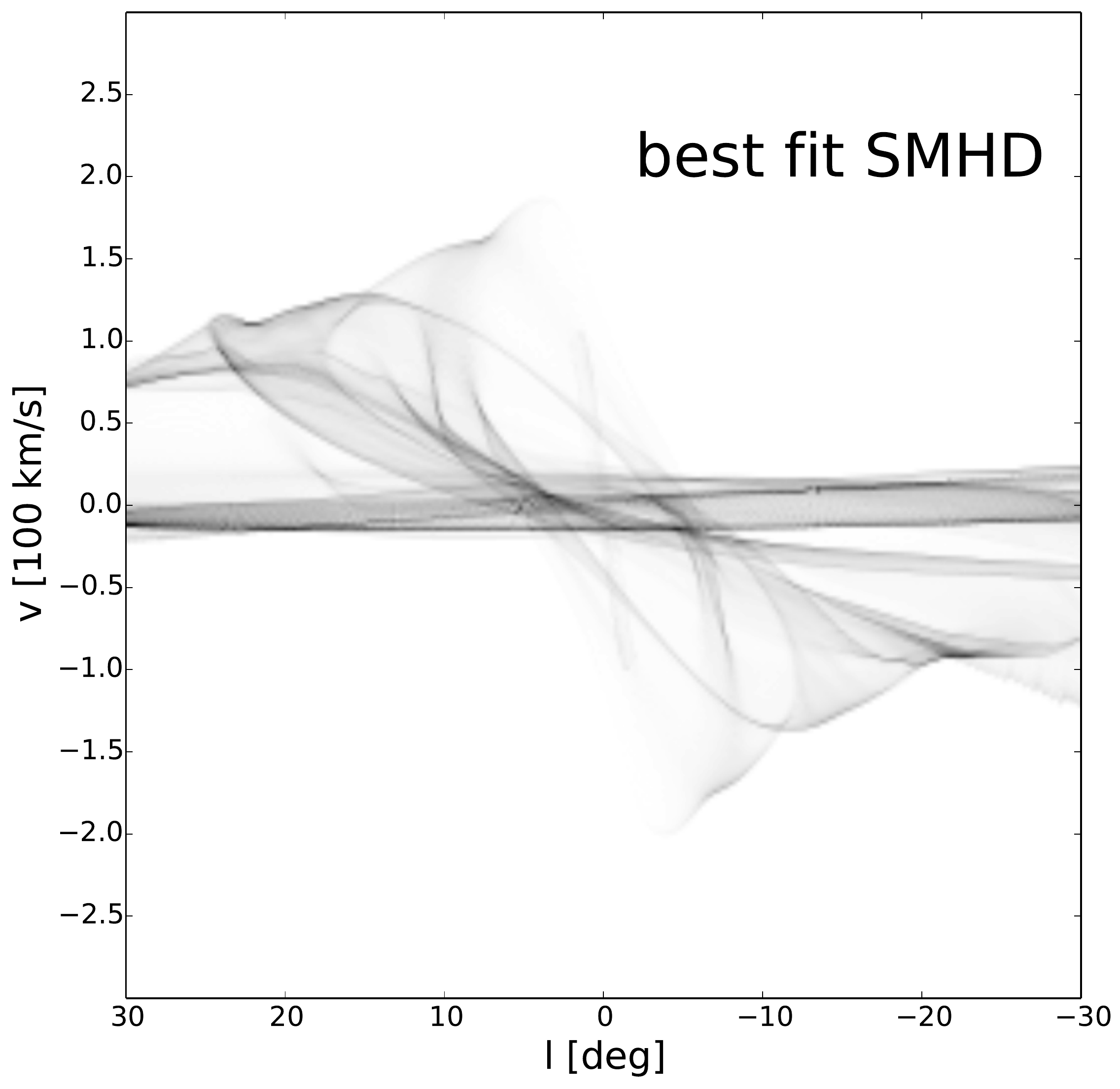}\includegraphics[width=0.25\textwidth]{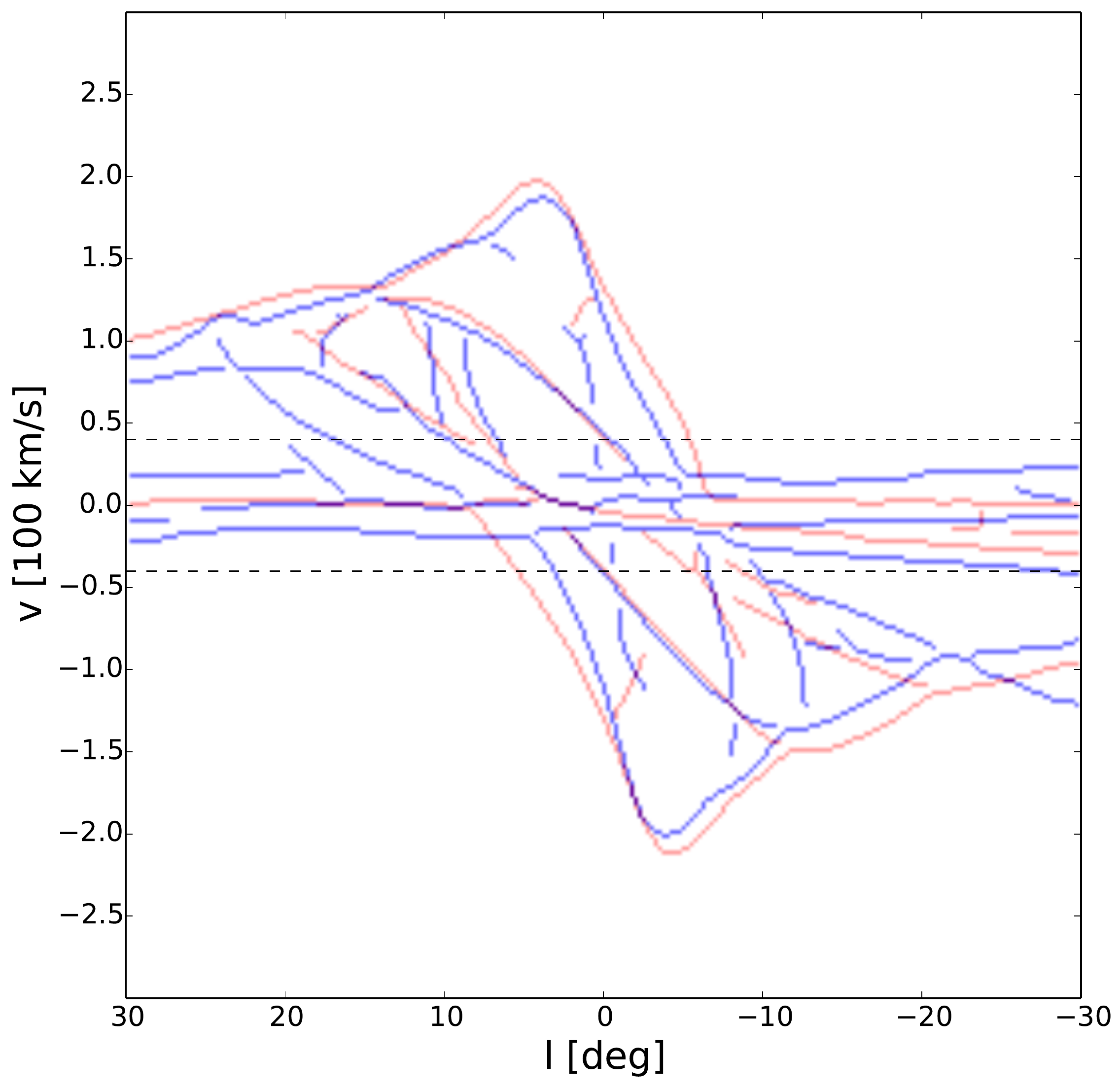}

\includegraphics[width=0.25\textwidth]{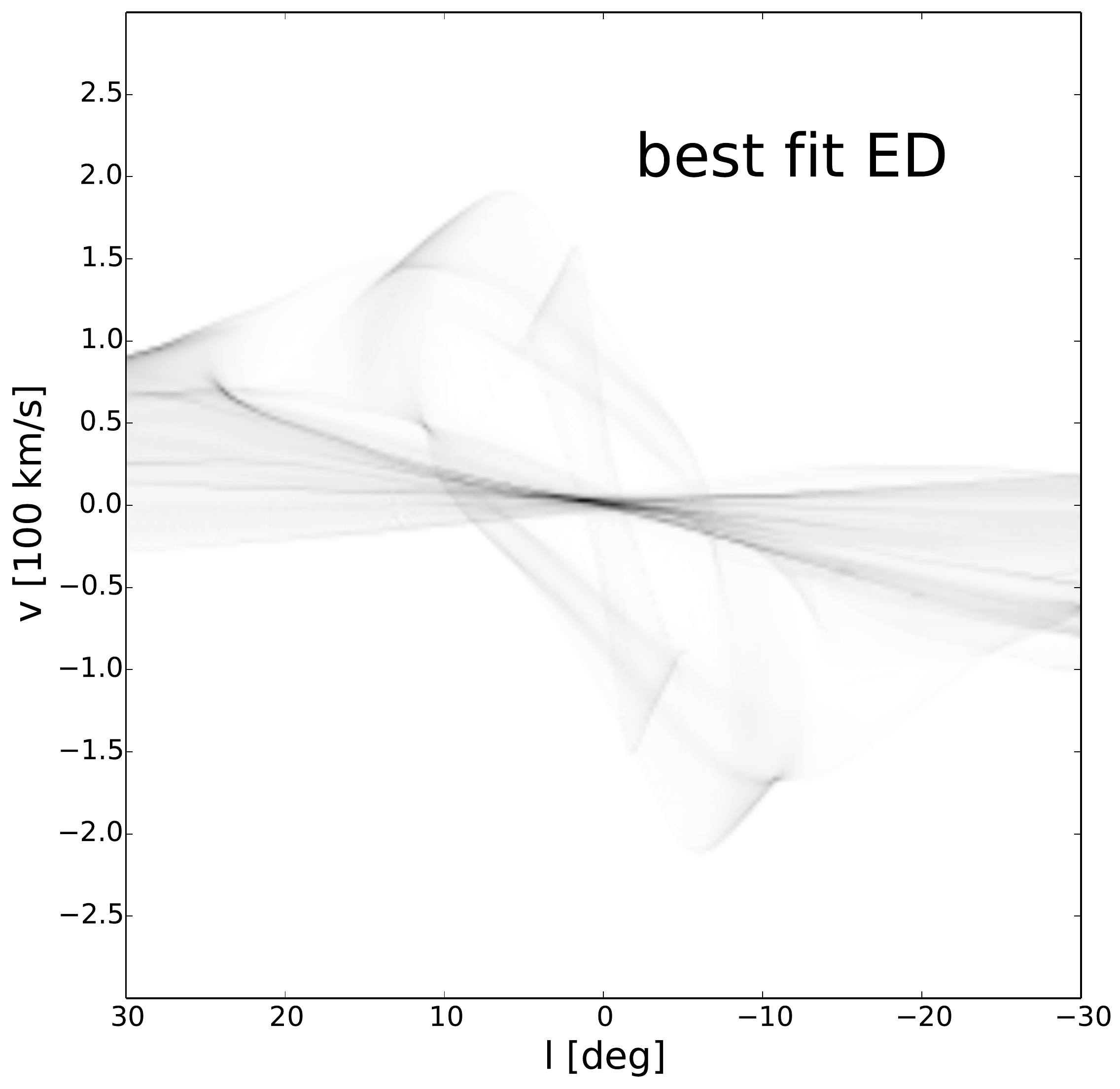}\includegraphics[width=0.25\textwidth]{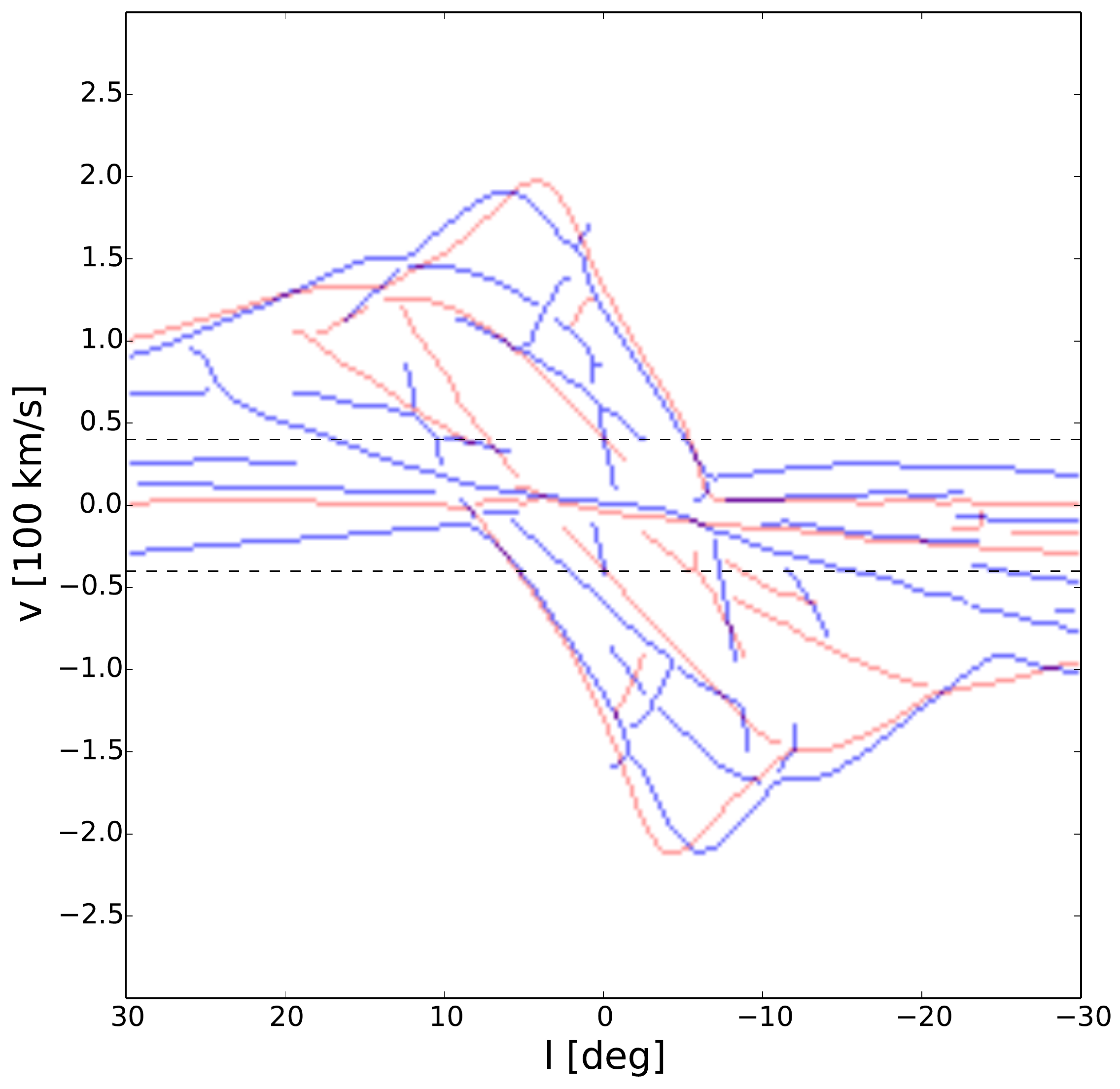}

\includegraphics[width=0.25\textwidth]{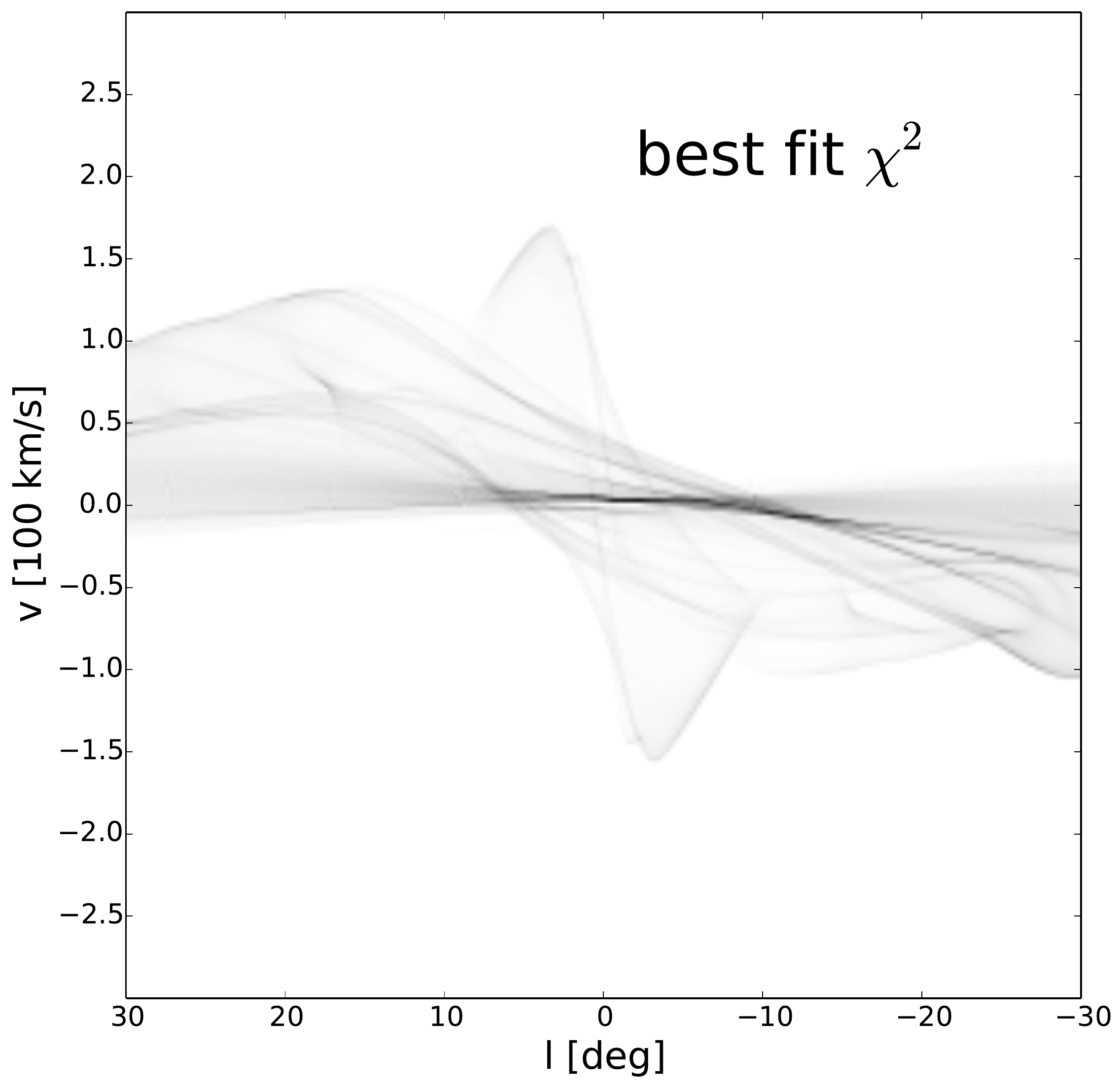}\includegraphics[width=0.25\textwidth]{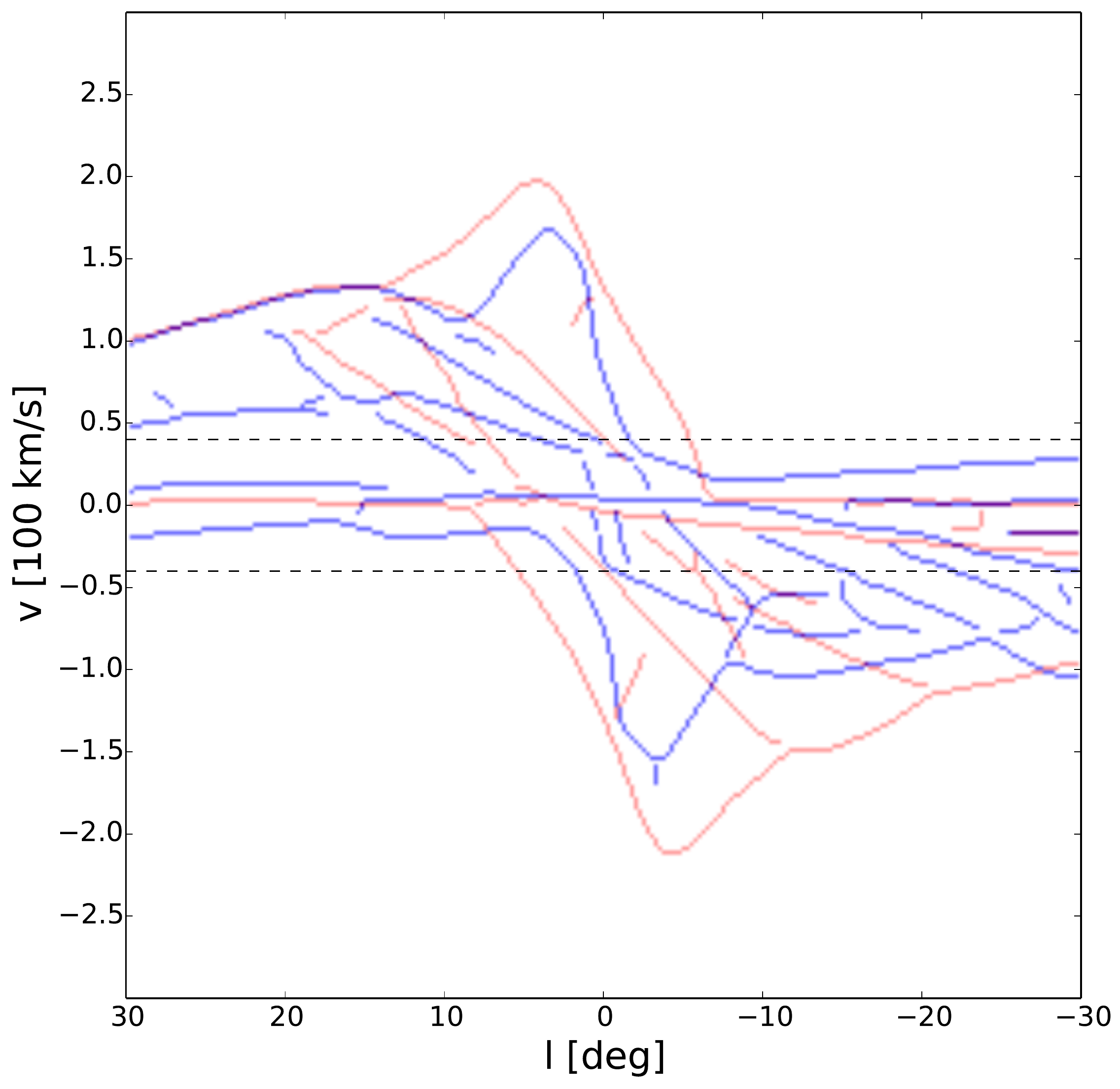}

\caption{A comparison of the best fits obtained by minimising the SMHD,
  $\chi^2$ and the ED for the mock data of
  Section~\ref{sec:humanintuition}.  The top panel shows the mock
  data and its features.  Subsequent panels show the models that
  minimise SMHD, ED and $\chi^2$, respectively. In blue the models and
  in red the data.  The best
  SMHD model matches the mock data features remarkably well, while the
  best $\chi^2$ model looks very different from the mock data. The best
  ED model displays the best-matching envelope, but the internal
  features are not  matched well.  The mock data are drawn from a very
  different potential than the models, and so the purpose of this is
  not to retrieve the correct parameters, but to show that the best
  SMHD model has features better matching the mock data than the best
  Envelope Model and the best $\chi^2$ model.  \label{fig:intuition1}} 
\end{figure}

Fig.~\ref{fig:intuition1} shows the mock data used in this case,
together with the models that minimize SMHD, ED and $\chi^2$.
It is clear that the model that reproduces the features best is,
unsurprisingly, the one that minimizes the SMHD.
The model that minimises the envelope distance matches the envelope very
well, but fails to match the internal features well.
It is evident that the model that minimises $\chi^2$ is entirely
unsatisfactory.

\begin{figure}
\includegraphics[width=0.5\textwidth]{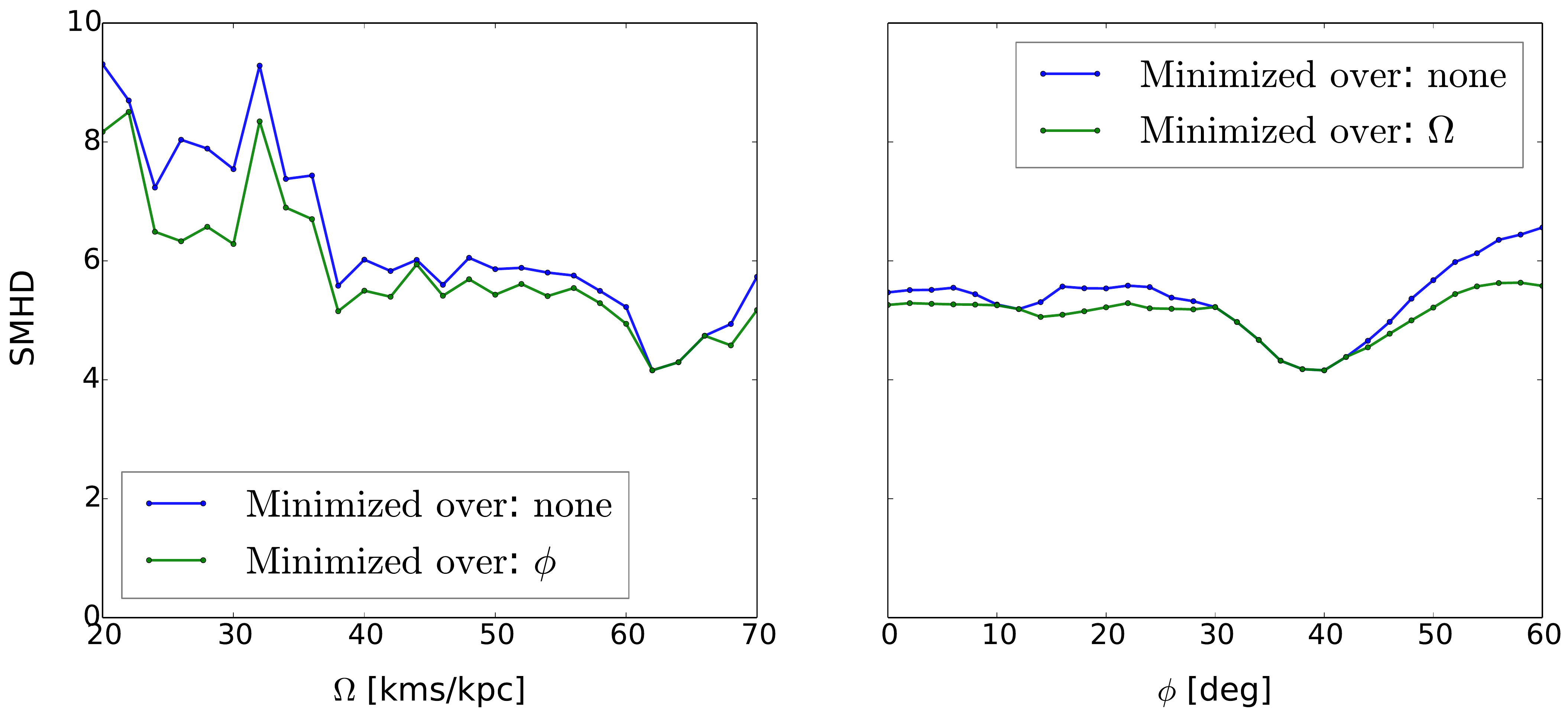}
\includegraphics[width=0.5\textwidth]{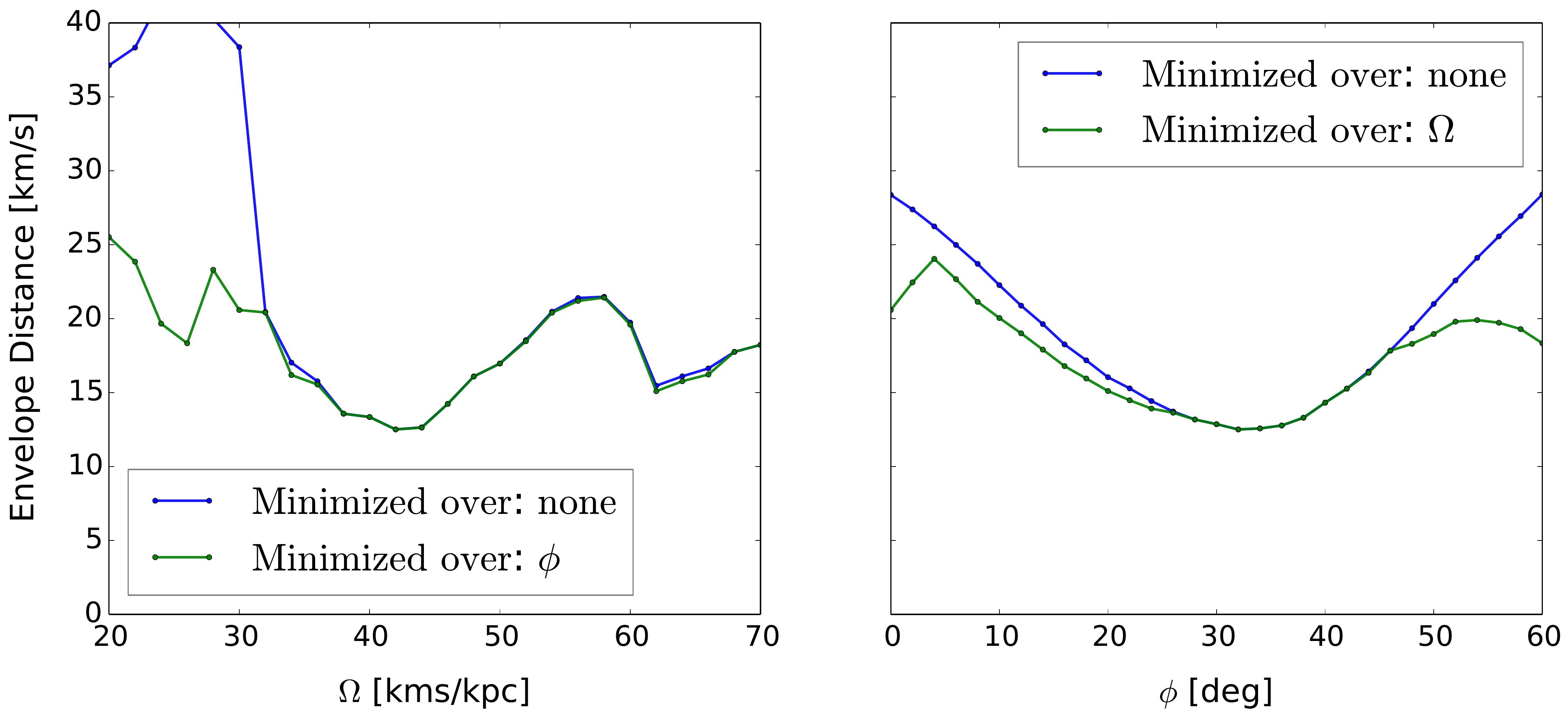}
\includegraphics[width=0.5\textwidth]{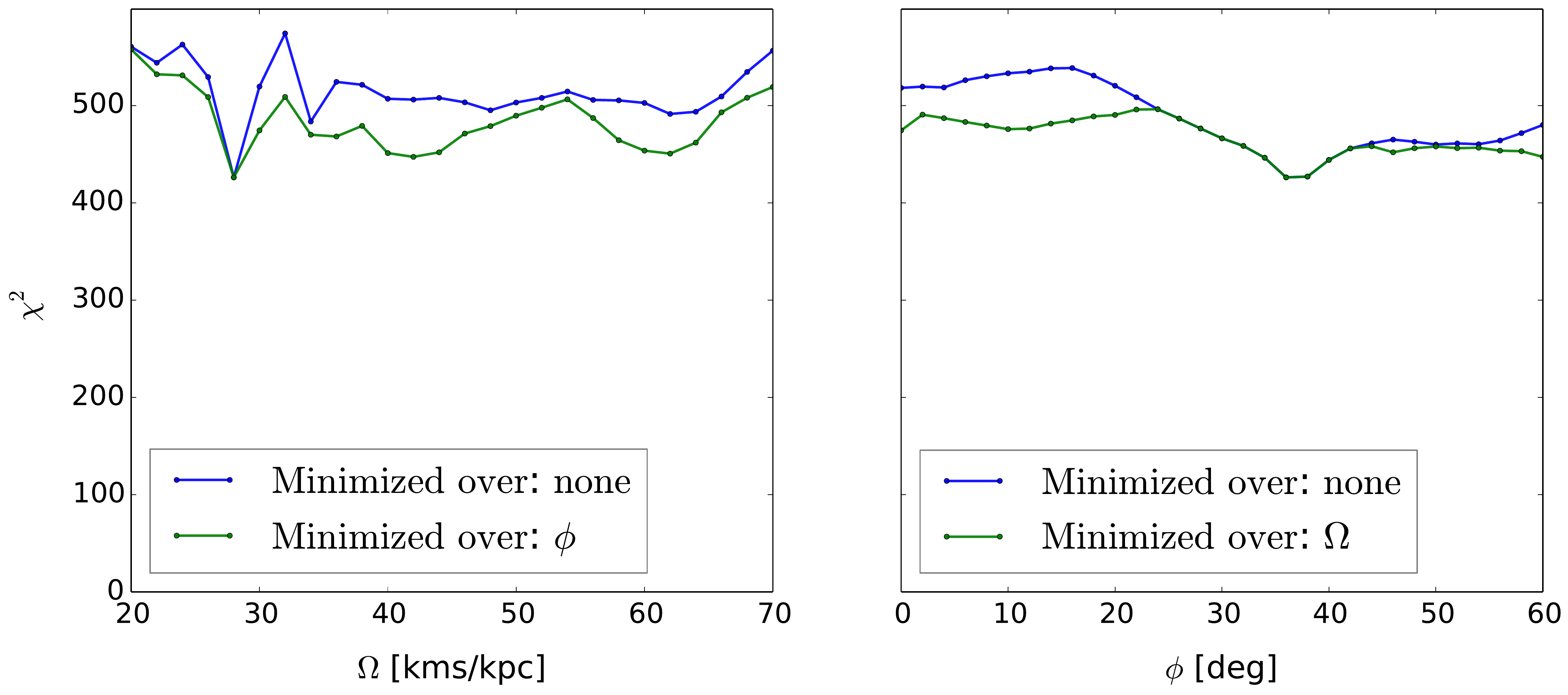}

\caption{Variation of SMHD (top row), envelope distance (middle)
  and $\chi^2$ (bottom row) with model parameters for the mock data of
  Section~\ref{sec:humanintuition}.   Colours
  indicate the same thing as in Fig. \ref{fig:retrieving1}. We see that
  the Envelope is degenerate, displaying two minima. The $\chi^2$ is
  flat and does not indicate clearly the correct direction to the nicer
  model. The SMHD shows a more definite minimum. Moreover, as shown in
  Fig. \ref{fig:intuition1}, the best SMHD is superior to the best
  Envelope and the best $\chi^2$ models. \label{fig:intuition2} }
\end{figure}

In Fig.~\ref{fig:intuition2} we show how the distances vary with model
parameters. The ED exhibits multiple minima, indicating degeneracy.
In fact, it has a secondary minimum at the location of the best SMHD
model, that is weaker than the main minimum.
The $\chi^2$ is more flat, with weak minima here and there.
If we consider that in this test the same approximation of radiative
transfer physics is used for the data and the models, we argue that the
situation would be even more hopeless than the plot indicates if one
were trying to use $\chi^2$ without knowledge of the (unknown) correct
model for the ISM chemistry and radiative transfer.
This indicates that $\chi^2$ is not an appropriate goodness-of-fit
measure, at least not until one understands the latter.
In Sect.~\ref{sec:discussion} we come back to this topic and discuss the
reasons for this behavior.

%% file: RealData.tex
\begin{figure} \subfigure[] {
        \includegraphics[width=0.25\textwidth]{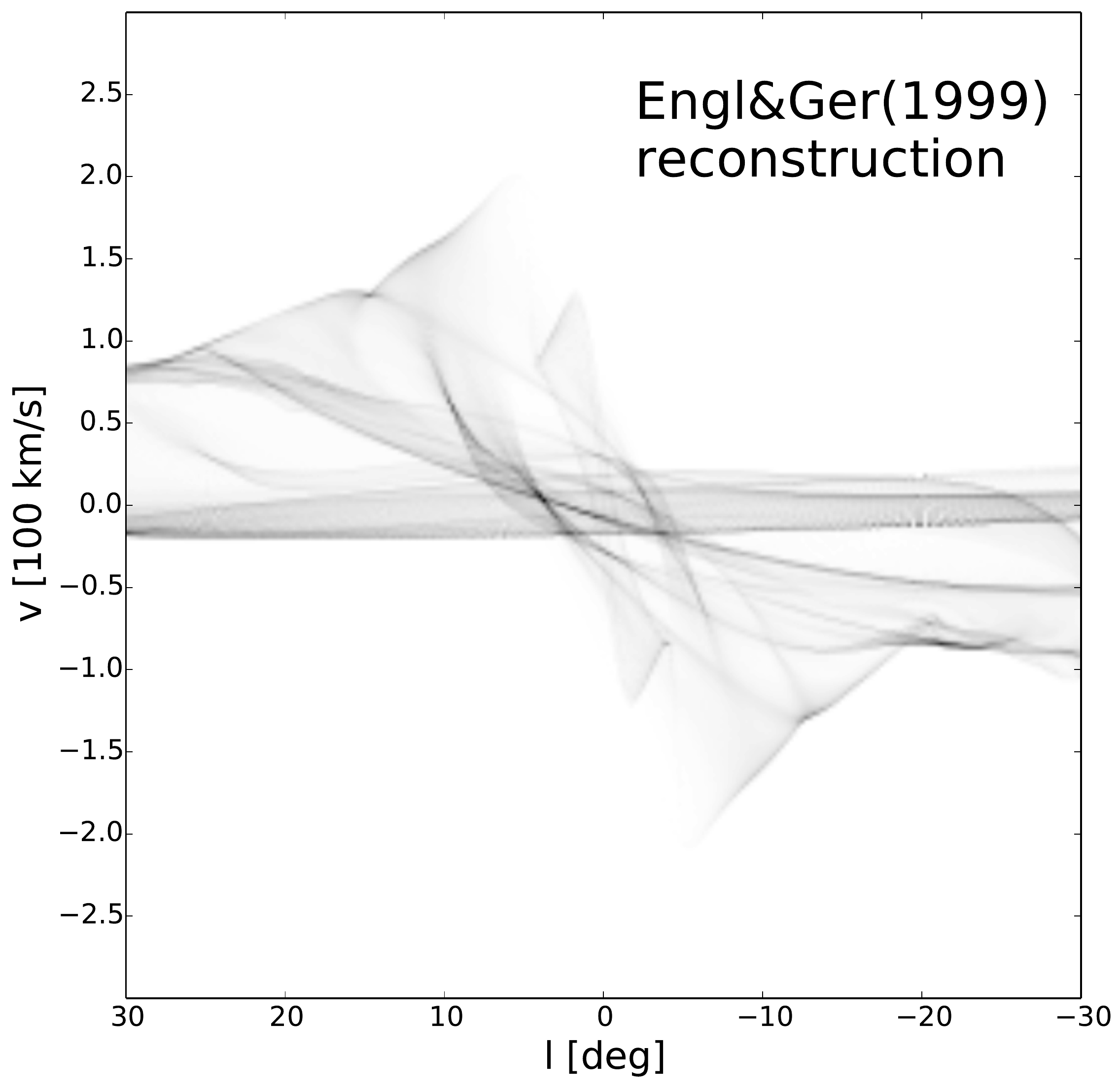}
        \includegraphics[width=0.25\textwidth]{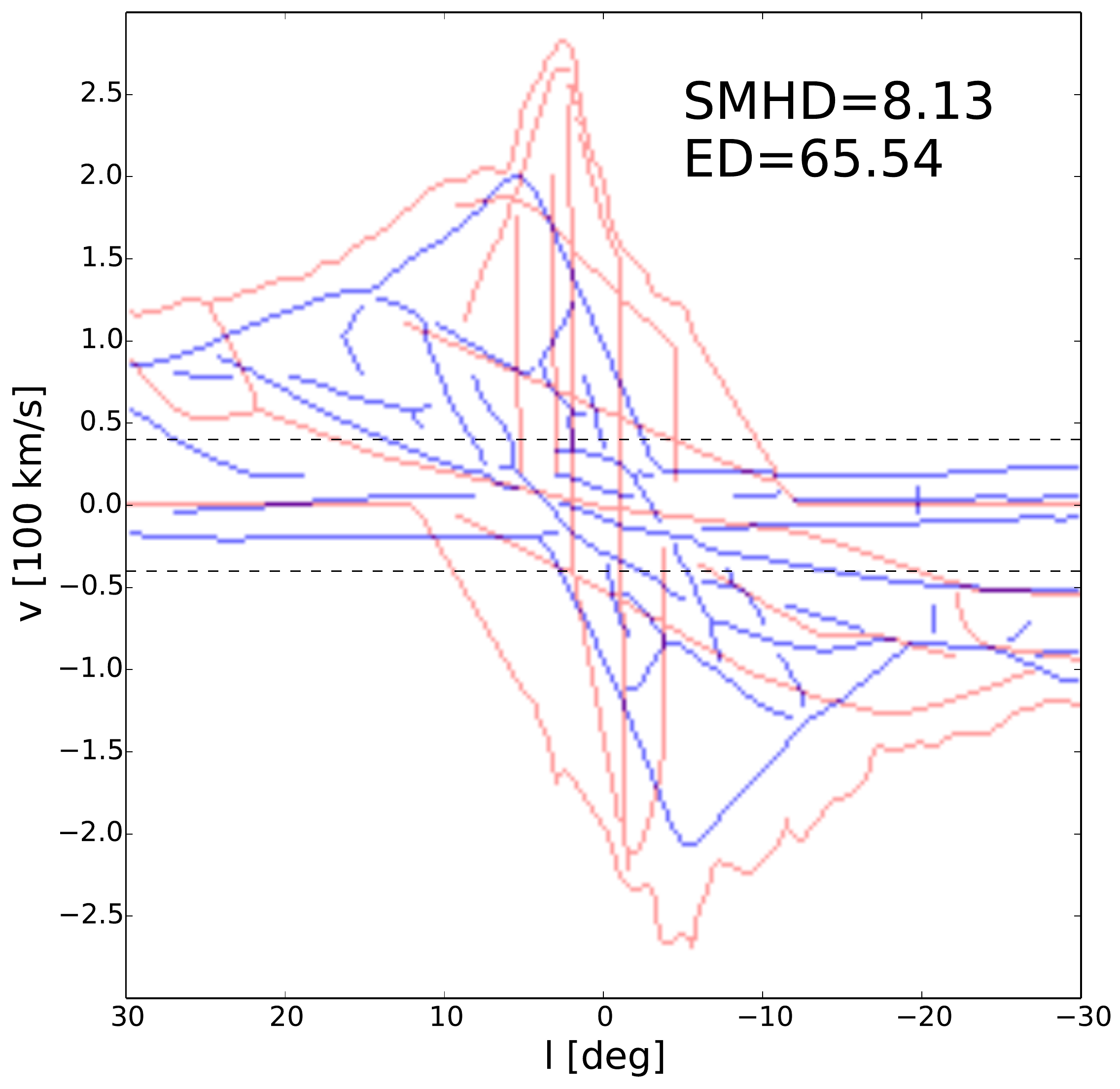}
    }
     \subfigure[]
    {
        \includegraphics[width=0.25\textwidth]{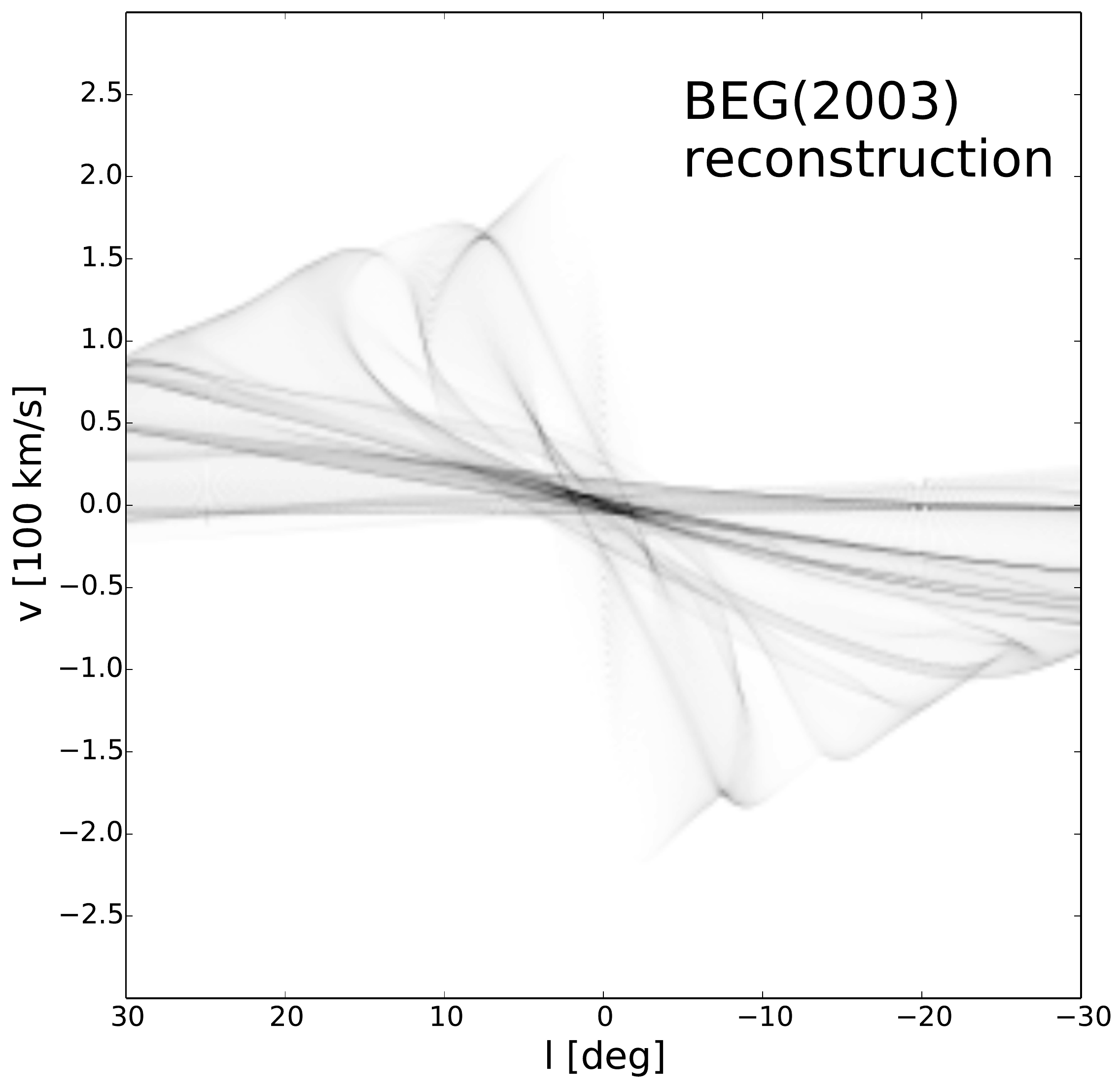}
        \includegraphics[width=0.25\textwidth]{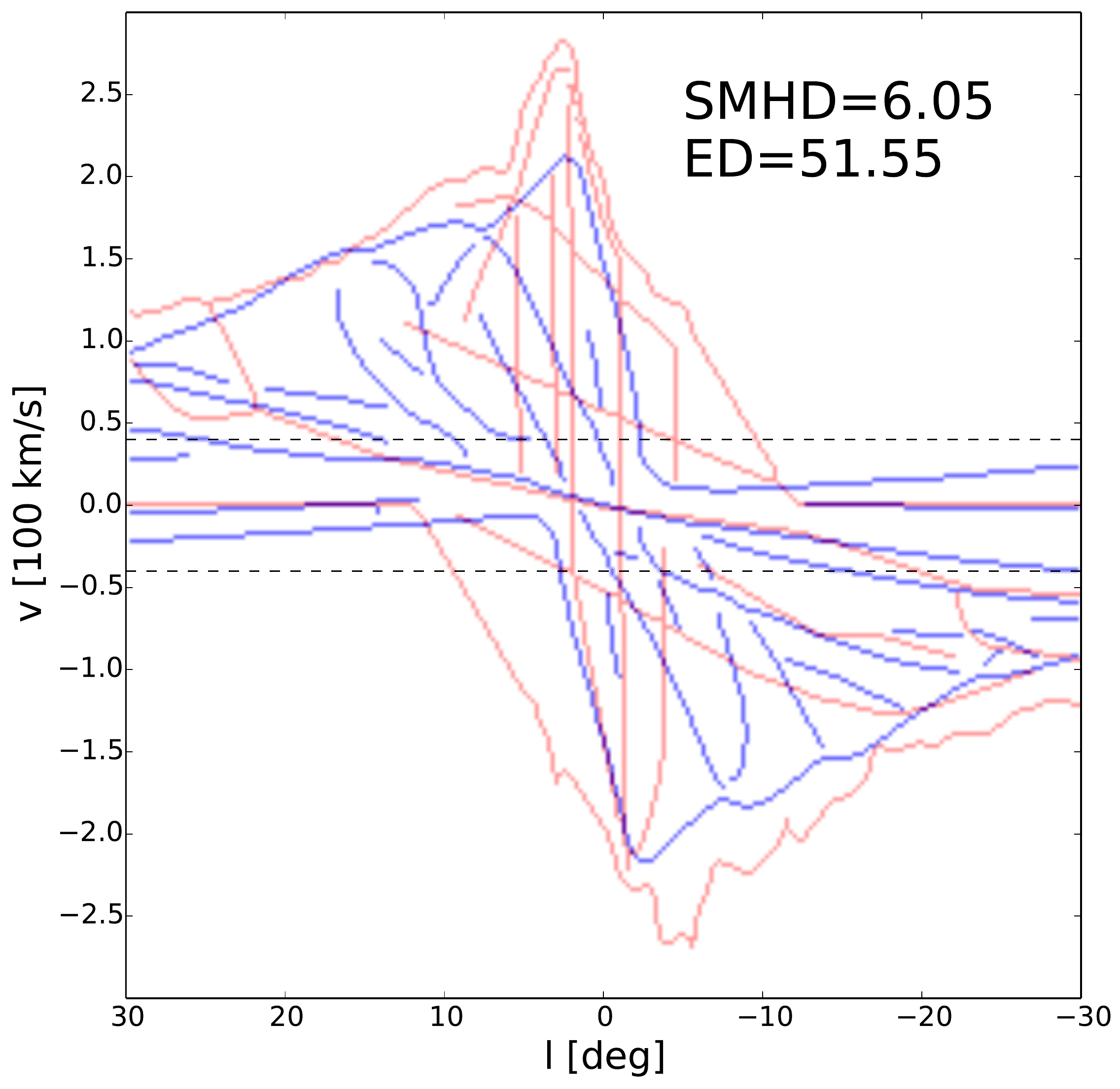}
    }
    \subfigure[]
    {
        \includegraphics[width=0.25\textwidth]{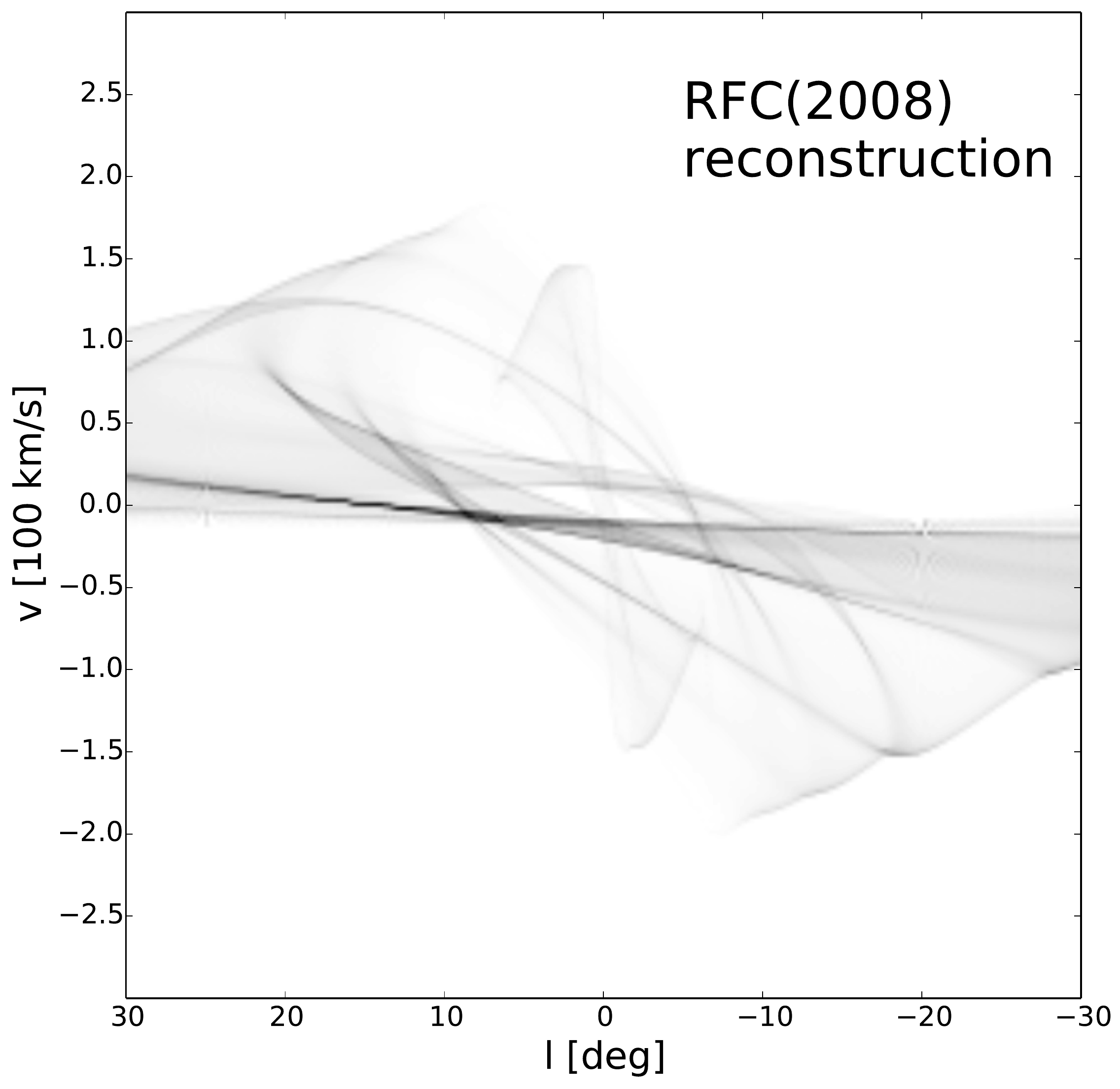}
        \includegraphics[width=0.25\textwidth]{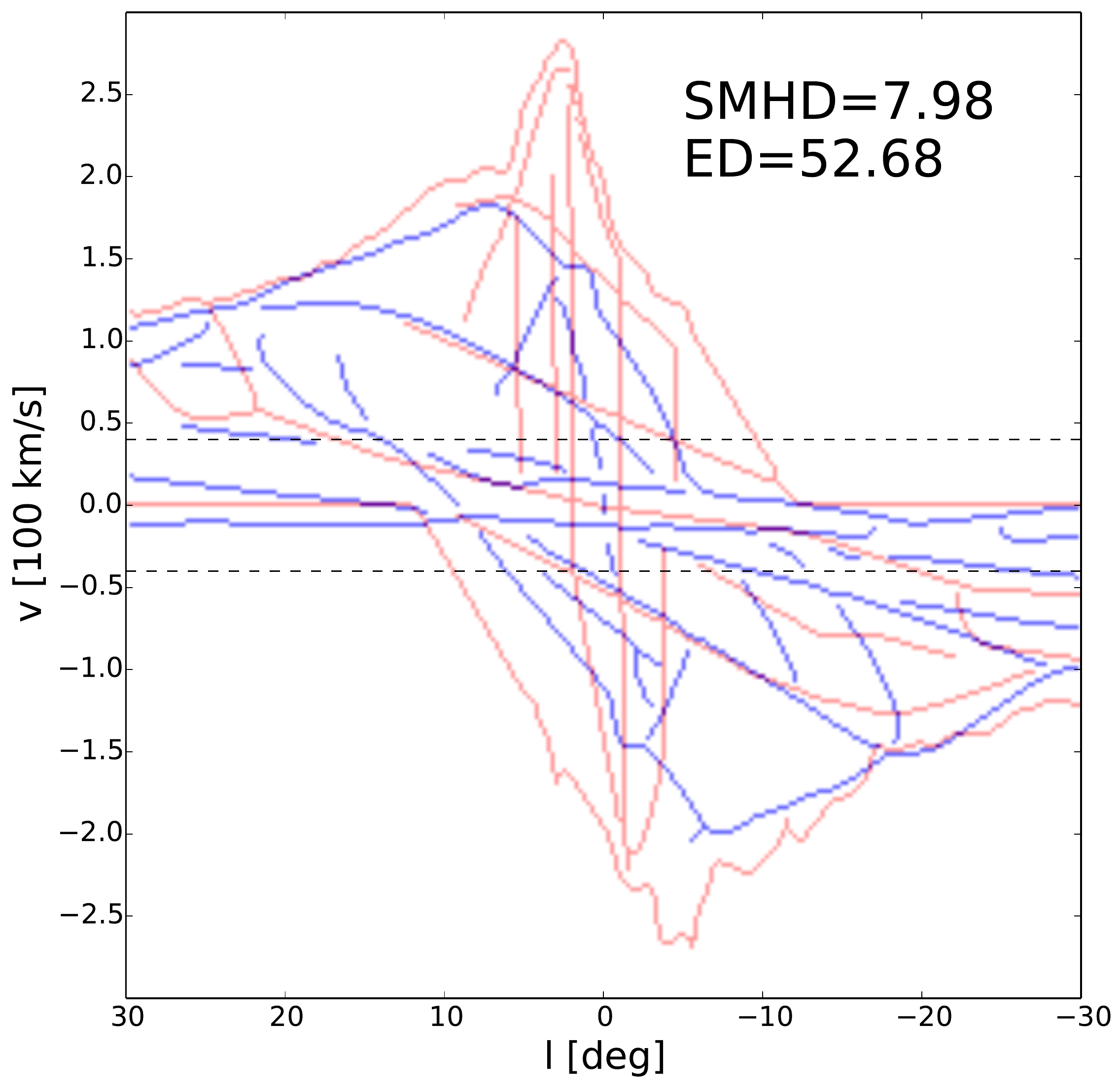}
    }
   \caption{$(l,v)$ plots corresponding to our reconstructions of some of the best models in the literature compared to observations. 
   The corresponding face-on densities are shown in Fig. \ref{fig:models2}. From top to
     bottom, our reconstruction of \protect\cite{englmaiergerhard1999} standard model, \protect\cite{bissantzetal2003} standard model, \protect\cite{combesrodriguez2008} best overall fitting model.
     On the right we overlay the features of the models with observational features discussed in Sect. \ref{sec:observations}. In blue the models and in red the data. All models have a viewing angle (angle between the Sun-Galactic centre line and the major axis of the bar) of $\phi=20\degree$. \label{fig:real3}}
\end{figure}

\begin{figure}
        \includegraphics[width=0.5\textwidth]{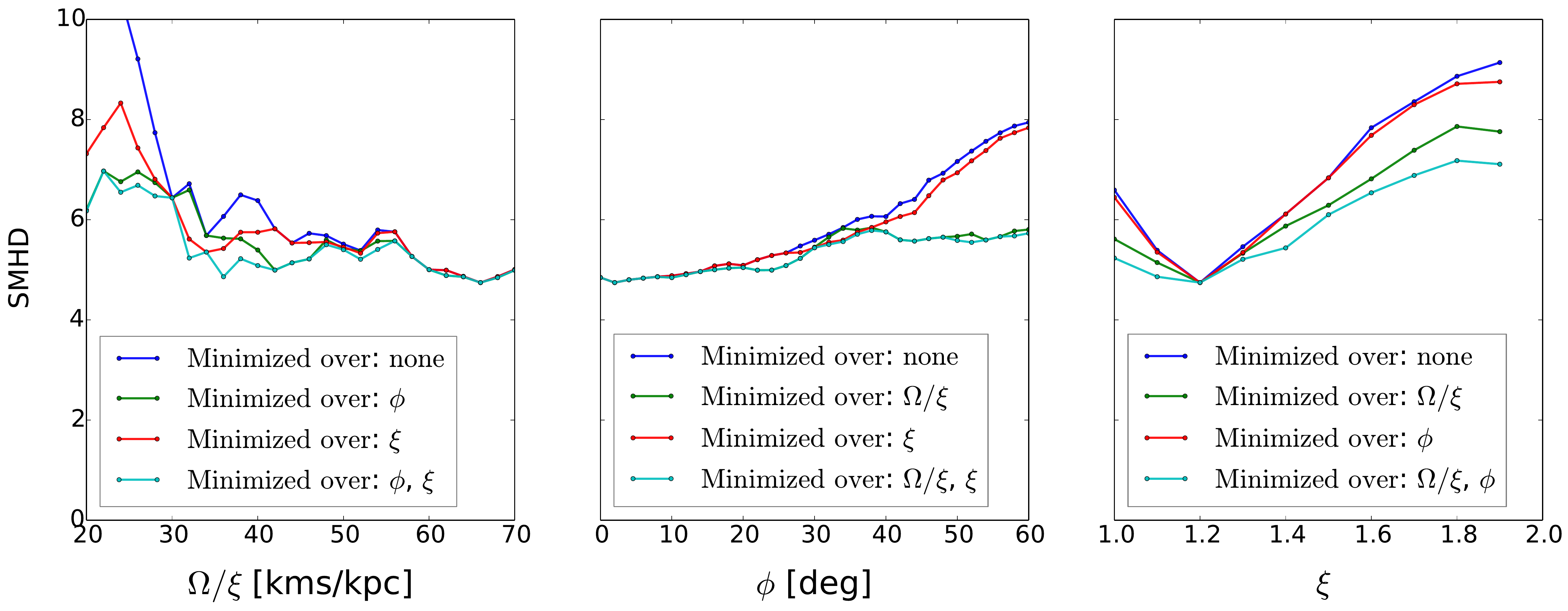}
        \includegraphics[width=0.5\textwidth]{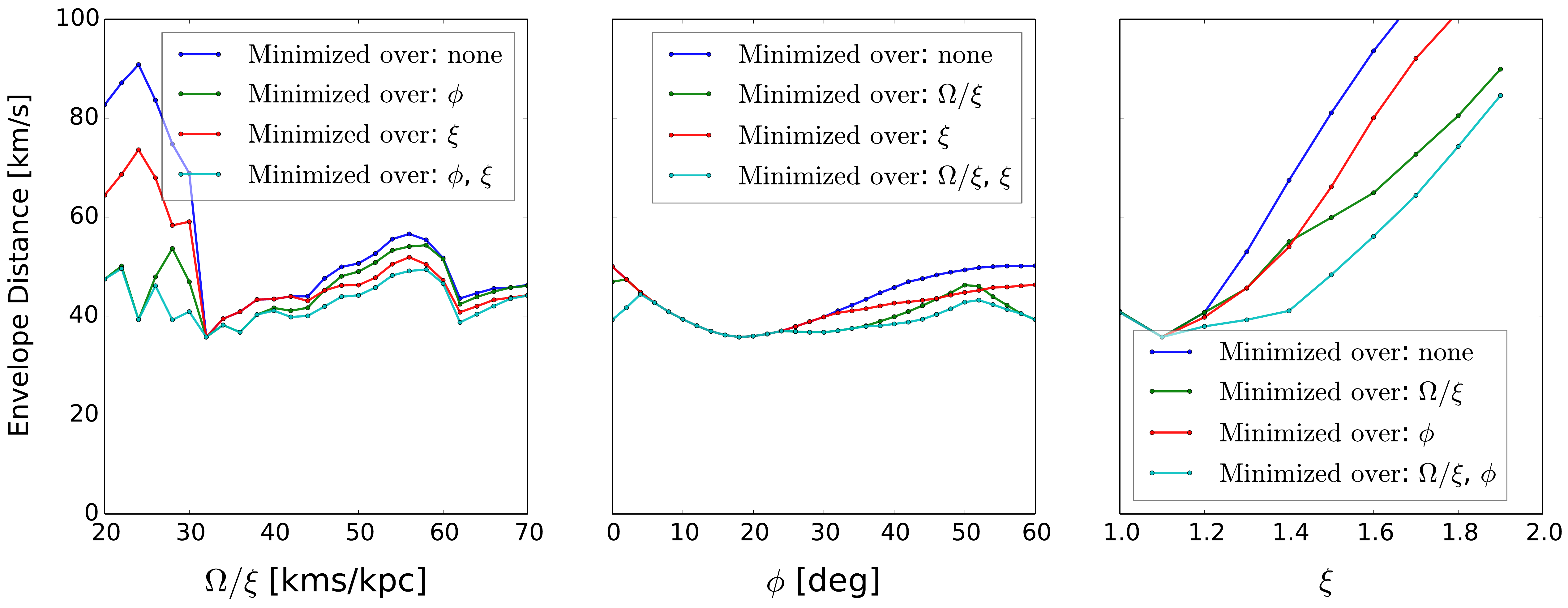}
    \caption{SMHD (top row) and ED (bottom) for
      our reconstructed of the models of \citet{englmaiergerhard1999}
      scaled to fit to features
      in real data (Section~\ref{sec:realdata}).
      The free parameters of the models are the viewing angle~$\phi$,
      the pattern speed~$\Omega$ and the velocity scaling factor~$\xi$.
      As in figure~\ref{fig:retrieving2}, the blue curves show how the
      distances change along straight lines aligned with the
      $(phi,\Omega,\xi)$ coordinate system that pass through the
      location $(\phi,\Omega,\xi)$ of the best fit.  The other curves
      plot the minimum distances when one or both of the other parameters is
      allowed to vary freely.
\label{fig:real1}}

\end{figure}

Having applied the SMHD to mock data, we now test how well it works
when applied to to the features in the real data identified in
Section~\ref{sec:observations}.
We do not produce any new models here, but instead compare the fits
provided by three of the best models from the literature: the standard
model of \cite{englmaiergerhard1999}; that of \cite{bissantzetal2003}
(their Table 1); and the best overall model of
\cite{combesrodriguez2008}, the rotation curve of which is shown in
their Fig.~6.

The model of \cite{englmaiergerhard1999} is stationary in a
frame that corotates with the bar pattern speed of $\Omega=55\, \kms
\, \kpc^{-1}$; there are no spiral arms.
The model of \cite{bissantzetal2003} is nonstationary, as it includes
a bar and a spiral arm component rotating respectively at $\Omega =
58.6 \, \kms \, \kpc^{-1}$ and $\Omega = 19.6 \, \kms \, \kpc^{-1}$.
\cite{combesrodriguez2008} includes two bars, a big bar and a smaller
nuclear bar making a constant angle of $55\degree$ with the first,
both rotating at the same pattern speed $\Omega=30\, \kms \,
\kpc^{-1}$. It has no spiral arms.\footnote{ In order to reproduce the
  rotation curve plotted in Fig.~6 of \cite{combesrodriguez2008} we
  found that we had to replace the exponent of $1/4$ in their
  expression (9) for $r_s$ with an exponent of~$1/2$.
  We assume that this is a typographical error in the paper, even
  though it means that their boxy Gaussian bulge is actually a boxy
  exponential bulge.}

Our reconstructions of snapshots of the density profiles of these
three models are shown in Fig.~\ref{fig:models2}.
The corresponding projected $(l,v)$ distributions for a viewing angle
of $\phi=20\degree$ are shown in Fig.~\ref{fig:real3}.
We have compared our density and $(l,v)$ plots with the appropriate
figures from the original papers and were find that the location of
the features agree very well indeed.
This agreement is surprising, given that our models are based on an
Eulerian grid simulation, whereas
\cite{englmaiergerhard1999} and \cite{bissantzetal2003} used a SPH code while
\cite{combesrodriguez2008} use a sticky-particle code.
It is particularly remarkable that we reproduce the latter so well:
sticky particle simulations in principle solve different fluid
equations than our FS2-based method.
The fact that the three methods give similar results on the scales we
are interested in corroborates our claim that we should first try to
match the observational data with simple models, so as to reproduce
the overall structure, and only later move to more refined models to
reproduce the details.
When the gross structure is not known, it is in general a pointless
exercise to add effects if they turn out to be of secondary
importance.

For each of our three reconstructions, we use the algorithm of
Section~\ref{sec:findfeat} to extract the features.
These are overlaid on the right column of Fig.~\ref{fig:real3} with
the features extracted by eye from the observations
(Section~\ref{sec:observations}).
Unfortunately, these plots highlight the shortcomings of current
models.
The main problems are:
\begin{enumerate}
\item no model is able to reproduce the high velocity peaks at $l
  \simeq \pm 3\degree$ and $l \simeq -4 \degree$. In particular the
  \cite{combesrodriguez2008} model has a very low peak velocity of
  less than $200\, \kms$.
\item no model reproduces the huge forbidden velocities at $(l>0,v<0)$
  and $(l<0,v>0)$. \cite{combesrodriguez2008} do better than the
  others in this, but large uncovered portions remain.
\item broad features, such as the 3-kpc arm, are not reproduced well
  by the \cite{englmaiergerhard1999} or \cite{bissantzetal2003}
  models. The more recent models of \cite{combesrodriguez2008} do a
  better job here, providing a good fit to both the near- and far-side
  3kpc arms.
\item the very complicated central structure, for example the vertical
  features, is not reproduced in any model.
\end{enumerate}  

As a very limited test of whether one could easily improve on this
situation, we fit the same set of models used in Sect. \ref{sec:tests},
based on our reconstruction of the \cite{englmaiergerhard1999}
potential, to the real Galaxy features.
As before the models have $\Omega$ in range $20 \mhyphen 70 \, \kms
\kpc^{-1}$, $\phi$ in range $0 \mhyphen 60 \degree$, all viewed at
evolutionary time $t\simeq 370 \rm Myr$ and projected with $\alpha=1$.
To allow the models some extra freedom we allow an extra parameter
$\xi$ that scales all velocities of the gas.
When velocities are scaled such that $v(r) \to \xi v(r)$, the other
quantities scale in the following way: $\Phi \to \xi^2 \Phi$, $M \to
\xi^2 M$, $c_s \to \xi c_s$, $\Omega \to \xi \Omega$.

\begin{figure}
     \subfigure[]
    {
        \includegraphics[width=0.25\textwidth]{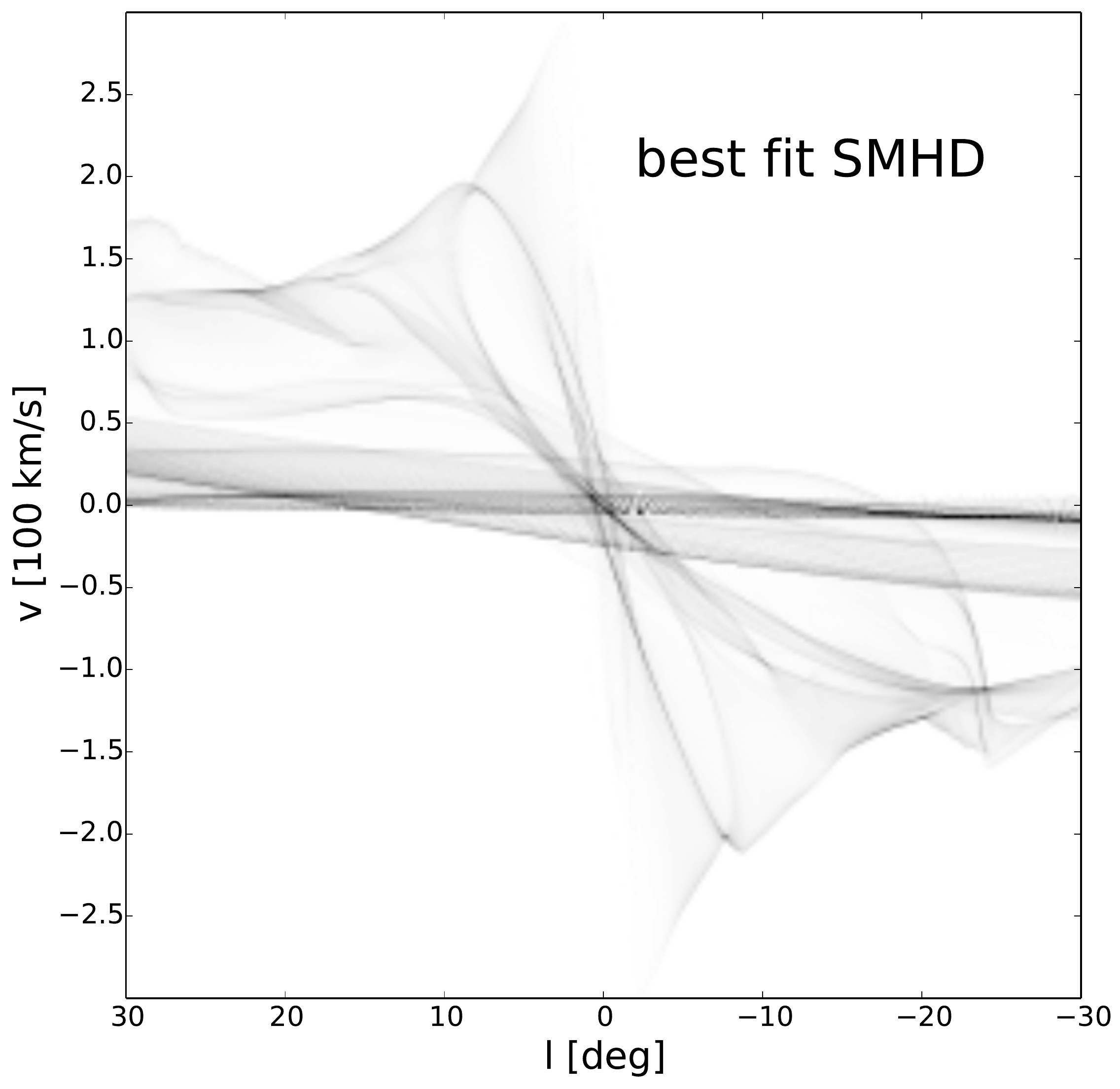}
        \includegraphics[width=0.25\textwidth]{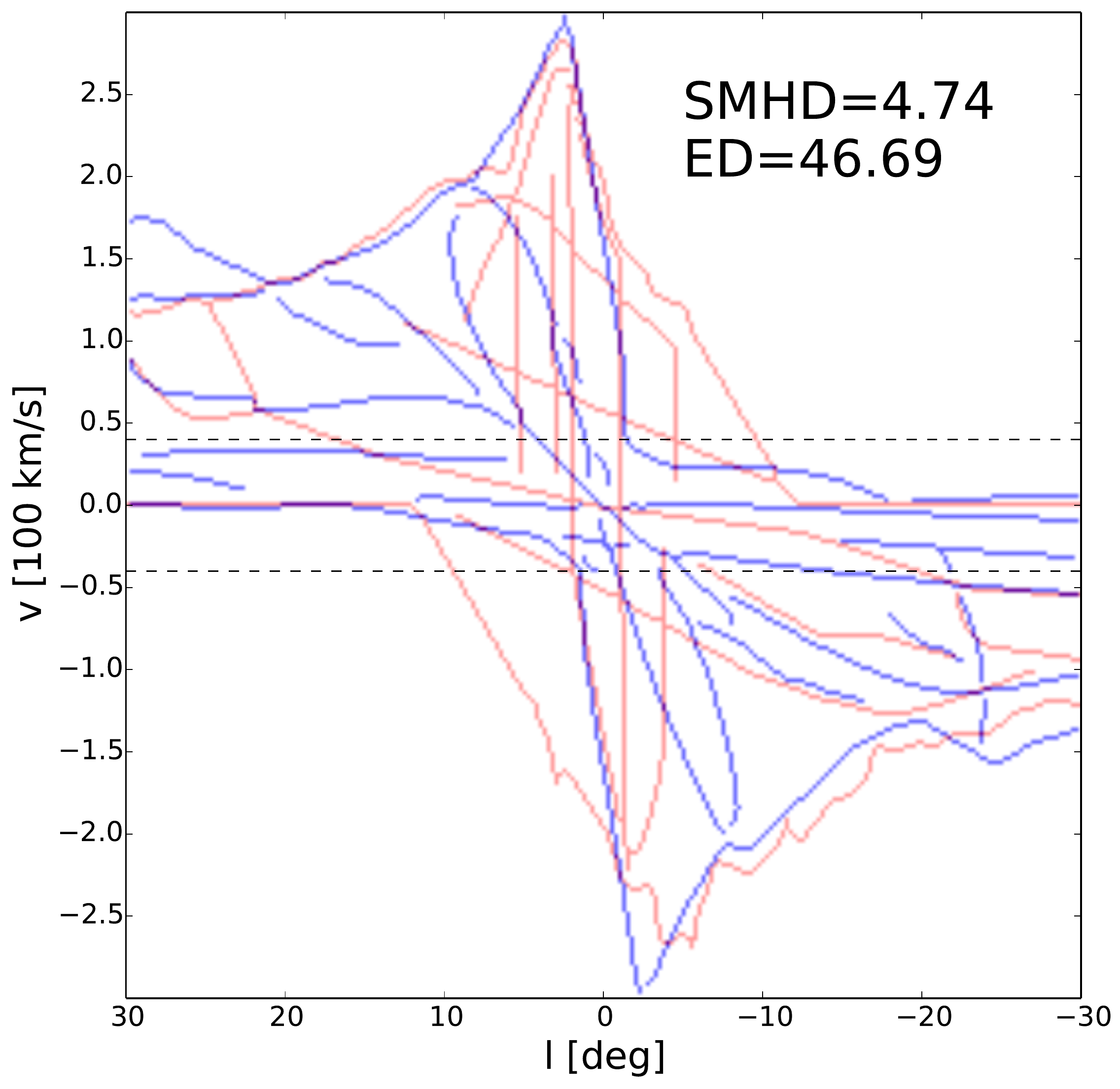}
    }
    \subfigure[]
    {
        \includegraphics[width=0.25\textwidth]{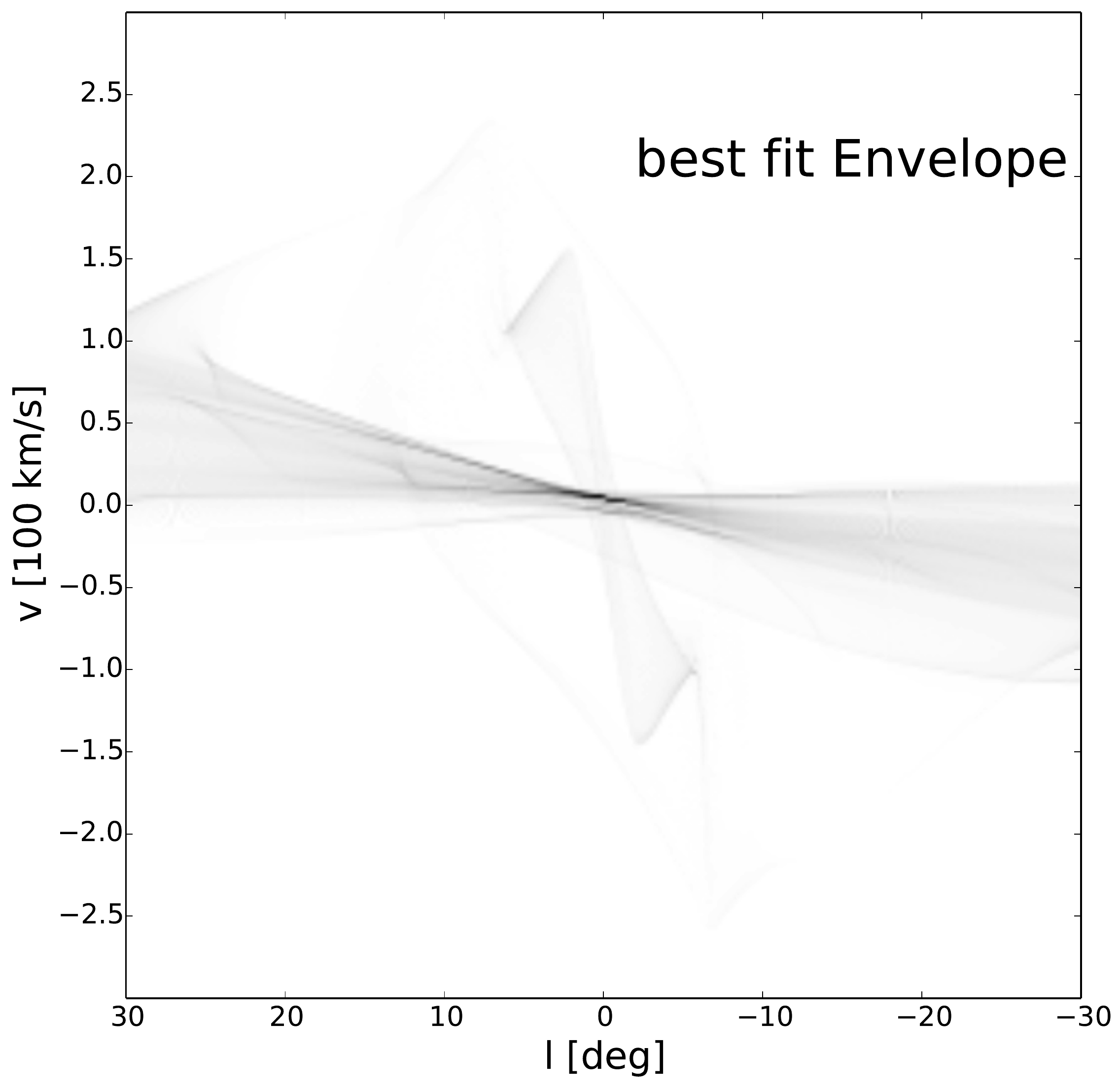}
        \includegraphics[width=0.25\textwidth]{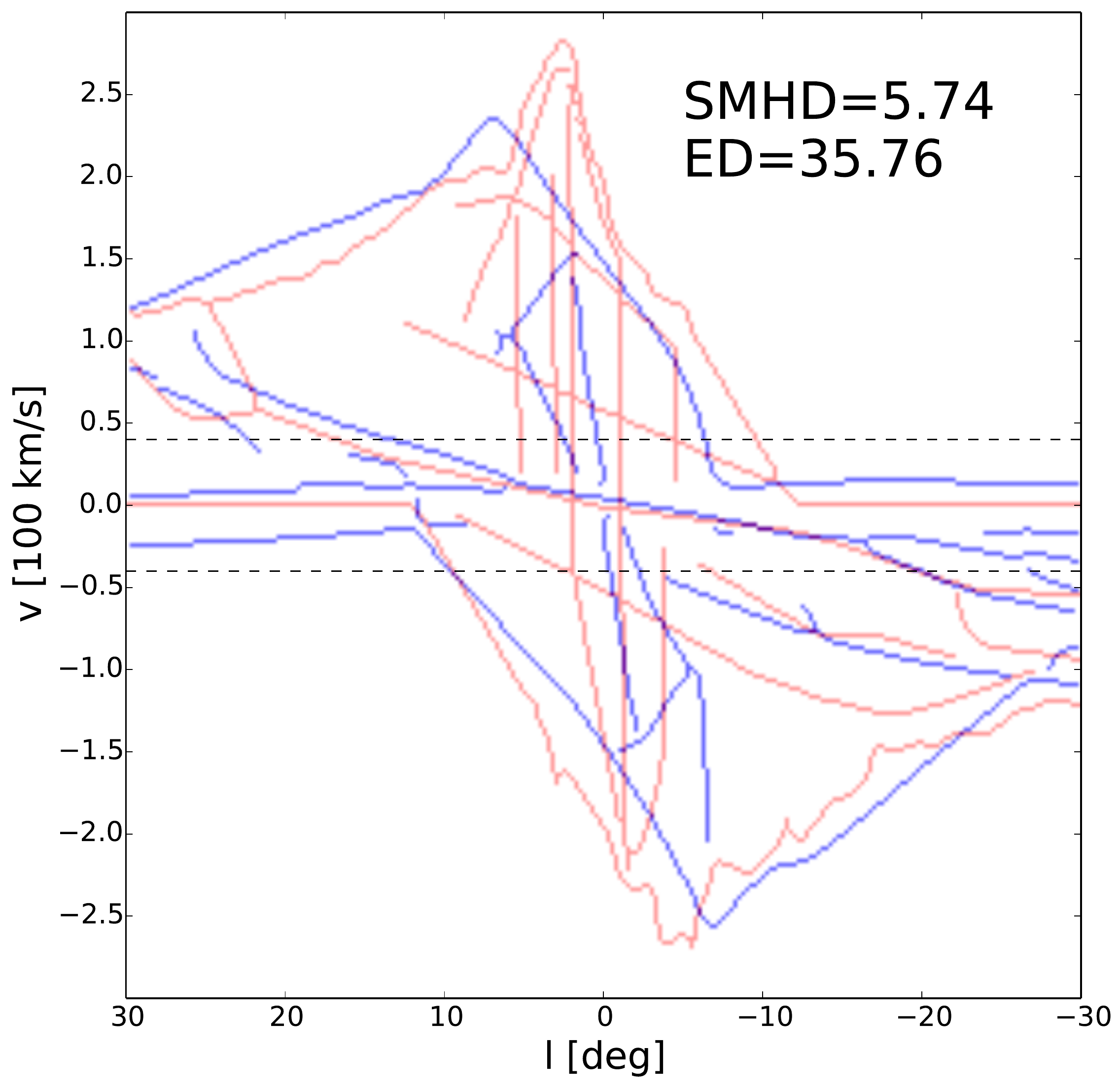}
    }
    \subfigure[]
    {
        \includegraphics[width=0.25\textwidth]{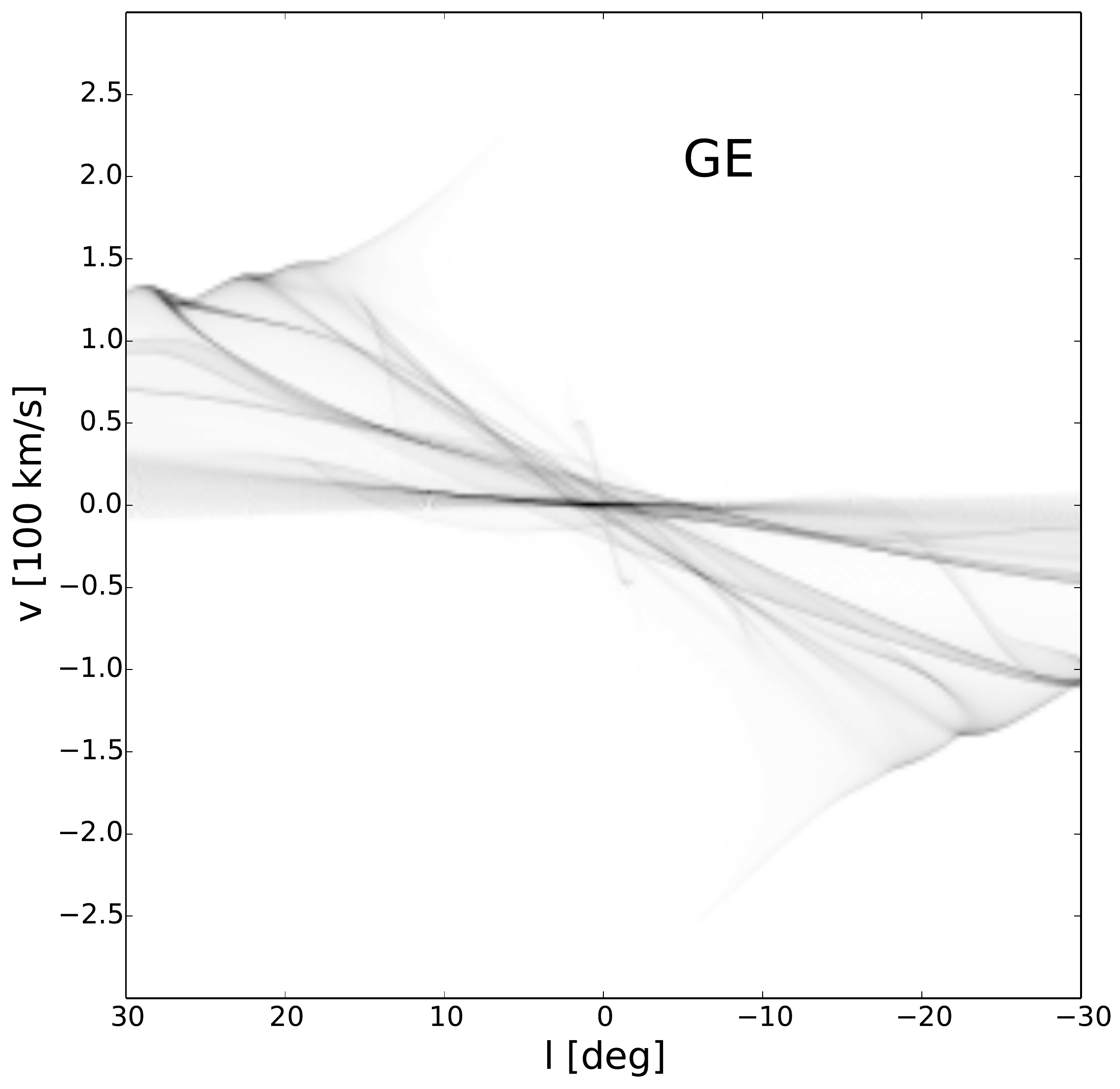}
        \includegraphics[width=0.25\textwidth]{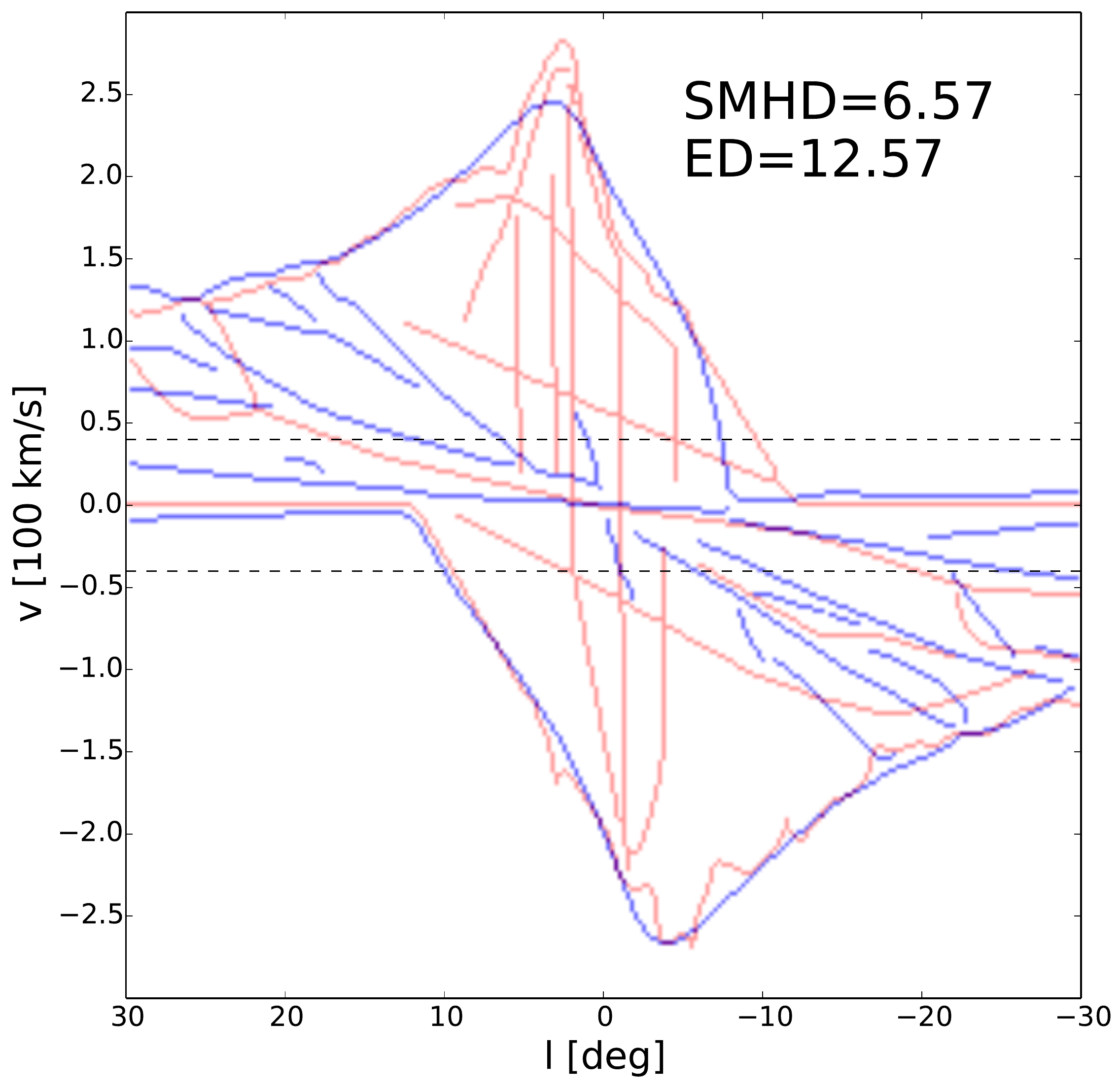}
    }
   \caption{Panels (a) and (b) show the best fit models in our reconstruction of \protect\cite{englmaiergerhard1999} potential, according respectively to SMHD and ED. These two models have values of the parameters corresponding to the values of the minima in Fig. \ref{fig:real2}. Panel (c) shows a model with a very well matching envelope. The simulation underlying this model runs in our reconstruction of \protect\cite{bissantzetal2003} potential, with $\Omega = 30 \, \kms \, \kpc^{-1}$, $\phi=34\degree$, $\xi=1.1$. All models are taken at evolutionary time $t=367 \rm Myr$.
  \label{fig:real2}}
\end{figure}

In Fig. \ref{fig:real1} we show the results of minimising the SMHD
(top row) and ED (bottom row).
We note that the graphs are smooth and not dominated by noise. The
best fits are different according to the two methods.
In Fig. \ref{fig:real2}, panel (a) and (b), we show the best fits
according to SMHD and ED respectively.
The parameter values for these are the overall minima in the graphs of
Fig. \ref{fig:real1}.
These are $\Omega/\xi = 64 \, \kms \, \kpc^{-1}$, $\phi=2\degree$,
$\xi=1.2$ for the best SMHD, and $\Omega/\xi = 32 \, \kms \, \kpc^{-1}$,
$\phi=18\degree$, $\xi=1.1$ for the best ED model.
Unfortunately, we believe that both should be considered
unsatisfactory, as in each important ingredients are missing.
No model is able to reproduce high forbidden velocity and, at the same
time, the high velocity peaks, and features are also not reproduced
well.
Interestingly, in the best ED model a weak vertical feature appears at
negative velocities, approximately at $l\simeq -7\degree$, $-200 < v <
-100 \, \kms$.
To the best of our knowledge, it is the first time that such a feature
appears in a synthetic $(l,v)$ plot and suggests that could explain
this feature if we had the right potential.
In the face-on view of the Galaxy, this feature corresponds to an offset shock lane.

In panel (c) of Fig. \ref{fig:real2} we show a further model, labelled
``GE'', that we found in our reconstruction of \cite{bissantzetal2003}
potential for $\Omega/\xi = 30 \, \kms \, \kpc^{-1}$, $\phi=34\degree$,
$\xi=1.1$.
This model has an envelope that matches the observed one amazingly
well, filling the right forbidden velocities region.
If one were to judge this model only from the envelope, akin to what
\cite{weinersellwood1999} did, this would be considered a very good
one.
We do not consider this to be a particularly good model, however.
The internal features are completely wrong, except close to the
Molecular Ring region.
The very central region is almost featureless, and it exhibits nothing
similar to the 3kpc arm, the connecting arm or the CMZ.
This model illustrates for the real case that the envelope is not
enough to constrain the Galaxy potential, and in this work we argue
that the next piece of information that should be taken into account
is given by the internal features.

%% file: discussion.tex
\subsection{Comparison of SMHD with Envelope Distance and $\chi^2$, and their limitations}

Apart from ``by-eye'' comparisons, the two most widely used ways of
fitting models to observed $(l,v)$ distributions have been some
variations of the Envelope Distance \citep[equ.~\ref{eq:envelope},
e.g.,][]{weinersellwood1999,englmaiergerhard1999} and $\chi^2$
\citep[equ.~\ref{eq:chisq}, e.g.,][]{Pettitt2014}.

While the ED is very robust with respect to changes in radiative
transfer physics, it has the obvious disadvantage of neglecting all
the information coming from internal features; thus, the degeneracy of
the problem is increased.
We know, for example, that it is possible to reproduce {\it any}
terminal velocity curve in the $(l>0,v>0)$ and $(l<0,v<0)$ quadrants
by means of gas moving on purely circular orbits.
In Sect. \ref{sec:humanintuition} we have shown that the Envelope
Distance can be degenerate when the SMHD is not, so the latter can
provide a better-fitting model than the former.
In Sec. \ref{sec:realdata} we have shown a model that fits well the
envelope for the real data, but it is overall unsatisfactory as it
fails to fit the internal features. The internal features must
certainly contain additional information and should be taken into
account when comparing the data with the models.
In Fig. \ref{fig:retrieving4} we give an example of a case in which
internal features alone are sufficient to identify the correct galaxy
model, albeit not quite so securely as when the envelope
is known as well.

In constrast, $\chi^2$ makes full use of all of the available data,
but it suffers from serious drawbacks when one tries to use it to
contrain the geometry and dynamics of the Galaxy.
One of the main concerns is that $\chi^2$ only compares model versus
observed intensities in the same $(l,v)$ bin and does not take into
account any cross-bin information.
If, for example, a model displays features that are very similar to
the observational features, but are slightly misplaced in
the $(l,v)$ plane, the $\chi^2$ distance can suggest that the model
is terrible; in the most extreme case, even a bland, featureless model
might be a formally better fit, while visual inspection would suggest
that the models are actually quite good.
This means that raw $\chi^2$ fails to capture the essence of what is
important in comparing model and data in this case.
This is what happens in the situation of
Sect. \ref{sec:humanintuition}, where we have shown that $\chi^2$
provides unsatisfactory fits even when the same approximation of
radiative transfer physics is used for the data and the models.

$\chi^2$ has other drawbacks.
It is computationally expensive.
As it is entirely dependent on local intensities, its use necessarily
requires detailed modelling of radiative transfer physics, chemistry
and gas dynamics all at the same time, which is very time consuming.
On the other hand, models need to be cheap to test because the space
of possible models is large.
We would like to know not only the pattern speed and orientation of
the bar, but also its mass, length, axis ratio, and possibly more.
Computational expense is the reason why \cite{Pettitt2014}, despite
having a full radiative transfer model, did not use it when fitting
the data and relied on a very simplified radiative transfer model akin
to ours.
$\chi^2$ is clearly not the best choice in situations such as the alpha test in
Sect. \ref{sec:compare}, where everything in the model is correct
except the radiative transfer modelling.

Current models of the gas flow in the Milky Way
(Sect. \ref{sec:realdata}) are unsatisfactory.
We argue that the sensible way of addressing this is to note that, as
we found in Sect. \ref{sec:compare}, the features in the $(l,v)$
distribution give valuable constraints on the Galactic potential that
are largely independent of the details of ISM chemistry or radiative
transfer.
Therefore one should first build models that match well the broad
morphology of the observations to narrow down the potential, and only
later refine this to match the details. As we've shown in Sect. 6, the
envelope is too degenerate for such a task, even in the case of real
data: a model with that reproduces only the envelope well can still be
unsatisfactory.

Thus one should use the SMHD or similar scheme to locate the range of
broadly acceptable potentials and pattern speed(s).
Only when this large-scale structure has been constrained it does make
sense to switch to more sophisticated models that include chemistry
and proper radiative transfer modelling.
We believe that $\chi^2$ (or similar) should play an important role
only in this last step.
An alternative to $\chi^2$ that might be worth considering in this
last step is the ``earth mover distance'' (Appendix \ref{EMD}).

We emphasise that, unlike $\chi^2$, the SMHD is a purely qualitative
measure that cannot sensibly be used to provide formal uncertainties
on the parameters of models that fit the observations.
Given the present ambiguity as to the overall form of the Galactic
potential, we would argue that any such attempt would be misleading.

\subsection{Identification of ``features''}

The SMHD returns a number that quantifies the dissimilarity of two
sets of features in the $(l,v)$ plane: the higher the number, the more
dissimilar the features.  The procedure for identifying features in
data and models requires some remarks.  For models, we have a fully automatic
algorithm (Sect \ref{sec:findfeat}) that, given a model $(l,v)$
distribution returns the features as 1-pixel wide lines.  The features
we identified as being important are bright ridges and the envelope.
Therefore, the algorithm simply detects ridges and envelope given a
model $(l,v)$ distribution.

\subsubsection{Features in data}

It is natural to ask whether the data could be analysed in the same
way as the models to extract features automatically.  Unfortunately,
this turned out to be problematic: unlike our simple, smooth
hydrodynamical models, the real data exhibit substructures due to
clumpiness that are identified as spurious ridges by our
ridge-detection algorithm.
Moreover, the analysis of features in the data often involves looking
at different latitude slices, and each feature may require a special
analysis and considerable work as the example of the Far side 3kpc arm
\citep{Dame2008} shows.
This is clearly beyond our algorithm's capabilities and requires the
skills of experienced astronomers.
For this reason, we rely on human wisdom for the identification
of features in the real data.

\subsubsection{Suitability of models}

Any model for the gas flow comes with a series of implicit
or explicit simplifying assumptions.
In the present paper our gas models are 2D; we neglect the vertical
dimension, which could also play an important role.
We do not include heating and cooling processes due to a variety of
sources, such as supernovae explosions and stellar winds.
Modeling the gas as a smooth fluid means neglecting all the grainy
structure, such as individual clouds with peculiar velocities.
Indeed, the crude modelling with the Euler equation is applicable only
in a coarse-grained sense, and does not take into account explicitly
local turbulence, temperature variations and multiphase nature of the
ISM.
Lastly, we neglect the self-gravity of the gas that could be important
especially near shocks or other structures where gas accumulates.
There is nothing to prevent us from applying the SMHD method to more (or
less) sophisticated models, as long as they produce smooth $(l,v)$
distributions; as discussed above, we argue that for finding the
large-scale structure of the gas distribution, which is currently
poorly understood, the simpler the model the better.

\subsubsection{Possible extensions}
\label{methodcomments}
The method can easily be adjusted to incorporate our beliefs about how
features are generated.
For example, one could argue that the envelope is of different nature
than the internal features, and therefore we should calculate two
separate SMHDs, one that matches only the envelopes and one matching
only the internal features.
Another example is provided by the fact that, because of absorption, for many features we know whether they are
caused by material in front of the Galactic Center (for example the
3kpc arm) or behind it \citep[for example, the $135\kms$ arm, see ][]{Cohen1975}.
It would therefore be natural to match features that we know are in
front of the GC with features that are in front also in the models.
A further example is provided by the tilt of the Inner Galaxy
\citep{burtonbook}: if we believe that a part of the Galaxy -- for
example the inner nuclear disk -- is tilted, then we would like to
match observational tilted features only with features that are
produced within the corresponding region in the models.
This is straightforward to do by fitting multiple SMHDs.

As an illustrative example, to incorporate the information on whether
features are in front or behind the GC, we can proceed as follows. We
divide data features in three sets: $\rm D_f$ are features we know lie
in front of the GC, $\rm D_b$ those that lie behind, $\rm D_u$ those
whose position is unknown. Model features are divided only in two
sets, $\rm M_f$ and $\rm M_b$, as for each model feature we always
have information on its position with respect to the GC. Then a
suitable definition of SMHD that takes into account the new
information (and reduces to the previous definition in absence of new
information) is: {
\begin{equation}
 {\rm SMHD_{new}}(a,b)  =
 				  \frac{A}{2N}  
				 +  \frac{B}{2M} ,
 \end{equation} 
}
where
\begin{equation} A =  {\rm MHD}(D_f,M_f) + {\rm MHD}(D_b,M_b) + {\rm MHD}(D_u,M_f+M_b) \end{equation}
and 
\begin{equation} B = {\rm MHD}(M_f,D_f + D_u) +{\rm MHD}(M_b,D_b + D_u).
\end{equation} Other prior information could be taken into account in
a similar way, dividing the feature in different sets and defining the
rules by which these sets should be matched.


%% file: conclusion.tex
 We have proposed a new way of fitting model $(l,v)$ distributions to
observations.  We have argued that one can separate the effects of the
large-scale dynamics and structure of the Galaxy from those due to
details of radiative transfer and chemistry.  Based on this, our SMHD
provides a way of measuring distances between features in models
versus corrsponding features in the observations.

We have tested the ability of our method to fit models to mock data
generated under a variety of conditions, and have compared it to
alternative methods.
To the best of our knowledge, this is the first time that the ability
of such methods to retrieve model parameters by fitting to $(l,v)$
distributions has been investigated systematically.  We have also
explicitly demonstrated the importance of internal features in the
$(l,v)$ plots, that have been the basis for comparison of $(l,v)$
plots to observations by many authors who have run simulations.

We find that our feature-based SMHD method works well and is much more
robust than other methods.  It works in cases in which the assumed
(crude) radiative transfer model is wrong or when the data are
contaminated.
The Envelope Distance, as expected, is often degenerate in cases
where the SMHD is not, as the ED exploits only a small part of the
information available in the data.
We found this to be true both for mock and real data.
On the other hand, we found $\chi^2$ to be unsuited to the task of
matching longitude-velocity diagrams when this requires exploration of
a huge parameter space.
It works well only in the very vicinity of the correct solution, and
at the price that {\it all} pieces of physics are taken into account
in producing synthetic $(l,v)$ plots, including those that can be
disentangled from the dynamics of the gas.
The main reasons for this behaviour are (i) that $\chi^2$ fails to
take into account cross-bin information, which means that it tends to
favour models that have little structure. (ii) its use is
computationally expensive as it requires modelling the chemistry of
the ISM and carrying out full radiative transfer calculations.
We argued that such calculations are unnecessary if all one wants to
do is to constrain the Galaxy's gravitational potential and the
large-scale distribution of its gas. Given that current dynamical models for the 
Galaxy appear to be far from the truth and that fitting the envelope
alone is not enough,  one should first constrain the
gross morphology found in data using, for example, SMHD, and only
later, once one has almost nailed down the potential, turn on more
details of physics and finally use $\chi^2$ to take advantage of its
statistical interpretation.

Our method is computationally inexpensive because it relies on simple
hydro simulations and avoids the need to model chemistry or to carry
out sophisticated radiative-transfer modelling.  This makes it
suitable for carrying out large, systematic scans of model parameter
space.  It is easily applied to time-dependent simulations.  It can be
used to test the reality of observed features (by comparing fits with
and without the feature present) and can naturally be extended to
test hypotheses, such as ``this feature belongs to a foreground
spiral arm'' or ``that feature is the trace of the $x_1$ orbit'' with
a little extra analysis of the internal dynamics of the models used
for comparison.  It does, however, require some work in that it relies
on features being identified ``by hand'' in the data.  It also
requires that the models used for comparison are smooth enough to
allow the feature extraction algorithm to work; sophisticated models
that produce clumpy structures would probably not be suitable.

We have reconstructed, and reanalysed by applying our method, some of the best models from
the literature that were constructed to fit the \cite{COdata} and
\cite{HIdata} data on the Milky Way, but find that they produce
surprisingly poor fits. We have made an initial attempt to fit the
data, but our family of models was too limited.
We were able to find a model that reproduces the envelope of the
emission accurately, but this fails in explaining the internal
structure of the data; nevertheless, this demonstrated only that the
envelope alone is insufficient to constrain the Galactic potential.
As an interesting by product, we found that the large-scale morphology 
is not very sensitive on the simulation method.
The sticky-particles code used by \citep{combesrodriguez2008} 
gave a large-scale morphology very similar to our grid-based code, despite the fact that 
in principle it solves a different set of hydrodynamical equations. This corroborates our claim that 
to find the gross morphology of the Galaxy one should focus on simple hydrodynamical models.

The problem is now that of finding a sufficiently general class of
model potentials to use in the comparison.
One possibility is to express the first few multipole moments
$\rho_l(r)$ of the Galaxy's mass density distribution in terms of
splines and to develop an automatic scheme for adjusting the spline
weights to minimize the SMHD distance.
A good model should also take into account constraints coming from
different sources, for example infrared data from the 2MASS survey
\citep{2MASS} and the correlations expected between the
three-dimensional distribution of gas and independent results on the
three-dimensional distribution of dust
\citep[e.g.,][]{Marshall+06,Green+14,SaleMagorrian14}.

\section*{Acknowledgments}
We are indebted to James Binney for his helpful insights at all stages
of the development of this work.  MCS thanks Maria Colombo for
suggesting the use of EMD and acknowledges
support from a Clarendon Fund Scholarship.
JM acknowledges support from STFC and ERC.
